\documentclass[a4paper,11pt]{article}
\usepackage{jheppub} % for details on the use of the package, please see the JINST-author-manual
\usepackage{lineno}
\usepackage{breqn}
\usepackage{amsmath}
\usepackage{mathtools}
\usepackage{amssymb}
\usepackage{bbold}
\usepackage{cancel}
\usepackage{soul}
\usepackage{xcolor}
\usepackage{braket} 
\usepackage{hyperref}
\usepackage{cleveref}
\usepackage[presets={abc,vec-cev}]{letterswitharrows}
\usepackage{eso-pic}
\usepackage{graphicx}
\usepackage{subcaption}
\usepackage{mwe}

\def\subsectionautorefname{Sec.}

\newcommand{\eqautoref}[1]{\hyperref[#1]{Eq.\,(\ref*{#1})}}
\title{The Migdal effect in Semiconductors for the Effective Field Theory of Dark Matter Direct Detection}

\author[a,b,c]{Kim V. Berghaus,}
\author[d]{Rouven Essig,}
\author[d,e]{Megan H. McDuffie}
\affiliation[a]{
Institute for Theoretical Physics (ITP), Karlsruhe Institute of Technology (KIT), Wolfgang-Gaede-Str. 1,
76131 Karlsruhe, Germany
}
\affiliation[b]{
Institute for Astroparticle Physics (IAP), Karlsruhe Institute of Technology (KIT),
Hermann-von-Helmholtz-Platz 1, 76344 Eggenstein-Leopoldshafen, Germany}
\affiliation[c]{Walter Burke Institute for Theoretical Physics, California Institute of Technology, Pasadena CA 91125, USA}
\affiliation[d]{C.N. Yang Institute for Theoretical Physics, Stony Brook University, NY 11794, U.S.A}
\affiliation[e]{Physics and Astronomy Department, Stony Brook University, NY 11794, U.S.A}

\emailAdd{kim.berghaus@kit.edu, rouven.essig@stonybrook.edu, megan.mcduffie@stonybrook.edu}

\abstract{The Migdal effect in semiconductors, prompt ionization from a primary nuclear scattering event, can be described across all kinematic regimes using an effective field theory that encodes the complex vibrational and electronic degrees of freedom of the crystal in measurable structure factors. Simultaneously, general dark matter-nucleus interactions can be systematically described using non-relativistic effective field theory operators. We combine these two effective field theory frameworks to calculate the Migdal effect in semiconductors for all ten dimension-six non-relativistic operators. From the effective Hamiltonian, we find that DM-nucleus scattering factorizes from the ionization and vibrational excitation signal as it does in the free-atom case. Using data from EDELWEISS that was taken with a germanium detector, we derive new experimental bounds on each operator and compare these limits to other direct-detection constraints in the literature. We find the accessible parameter space to be disfavored by bounds on heavy mediators contained in simple UV completions that generate the effective operators. }

\begin{document}

\AddToShipoutPictureBG*{%
  \AtPageUpperLeft{%
    \hspace{0.975\paperwidth}%
    \raisebox{-8\baselineskip}{%
      \makebox[0pt][r]{CALT-TH/2026-004}
}}}%

\AddToShipoutPictureBG*{%
  \AtPageUpperLeft{%
    \hspace{0.975\paperwidth}%
    \raisebox{-9\baselineskip}{%
      \makebox[0pt][r]{KA-TP-05-2026}
}}}%

\maketitle
\flushbottom

\section{Introduction} \label{sec:intro}
At astrophysical and cosmological scales, there is compelling observational evidence for the existence of Dark Matter (DM), such as galaxy rotation curves, the bullet cluster, gravitational lensing, and CMB anisotropy measurements. From these observations, many particle DM candidates have been proposed that are consistent on large scales with cold DM---a collisionless, non-relativistic fluid interacting gravitationally with the Standard Model. Discovering non-gravitational DM interactions would revolutionize our understanding of the fundamental properties of DM and help distinguish among different candidates. Numerous Earth-based experiments have been deployed to search for such non-gravitational interactions between DM and ordinary matter. Traditional direct detection experiments search for DM-nucleus interactions that probe $\mathcal{O}(1~{\rm GeV} - 1~{\rm TeV})$ DM masses, for example \cite{CRESST-III-2021, DarkSide-50-2023, SuperCDMS:2023geu, XENONnT2025, LZ:2025igz, PandaX4T_2025}. Studies have extended beyond spin-independent interactions by applying effective field theory approaches~\cite{Fan2010, Fitzpatrick:2013, Anand2013, SuperCDMS:2015lcz, Gazda2016, DarkSide-50:2020swd, LUX:2020oan, Griffin2021, SuperCDMS:2022crd, PhysRevD.111.095033, CDEX:2025mcb}. Generally, the sensitivity of these experiments to elastic DM-nucleus scattering quickly decreases below detector thresholds for DM masses smaller than the proton. Therefore, new strategies are necessary to probe sub-GeV DM, including (1)~DM-electron scattering~\cite{Essig:2011nj}; (2)~using the Migdal effect~\cite{Migdal:1939, Migdal:1978,Ibe:2017yqa} (the inelastic excitation of a bound electron via a DM-nucleus scattering event); and (3)~DM interactions with collective modes~\cite{Knapen:2017ekk}.  

Accordingly, the direct detection landscape has evolved significantly in recent years, with semiconductor targets emerging as powerful probes of DM-electron scattering, particularly sensitive to DM masses from $\sim 100$~keV to $\sim1$~GeV, since detector thresholds equal the band gap of the material $O(\text{eV})$~\cite{Essig:2011nj, Essig2016}. Extensive theoretical~\cite{Essig2016, Griffin2021, DarkELF2021_dielectric, darkELF2022, Singal2023, Giffin:2025hdx} and experimental work~\cite{Tiffenberg:2017aac,Crisler:2018gci,SENSEI:2019ibb,SENSEI:2020dpa,SENSEI:2021hcn,CDEX_electron,SENSEI:2024yyt,SENSEI:2023zdf,SENSEI:2025qvp,DAMIC-M:2023hgj,DAMICm2023,DAMIC-M:2025ltz,DAMIC-M2025,SuperCDMS_electron,SuperCDMS:2020ymb,SuperCDMS2025,SuperCDMS_electron_2025,EDELWEISS2020} over the past decade has explored this DM-electron interaction in crystals. Recent studies extend beyond spin-independent DM-electron interactions~\cite{Catena:2019gfa, Catena2021, Liang:2024ecw,Krnjaic:2024bdd, Giffin:2025hdx} analogously to effective field theory (EFT) studies of DM-nucleon interactions. 

Parallel to this development, the Migdal effect has received  renewed attention for its potential to increase sensitivity to sub-GeV DM interactions with nuclei~\cite{Vergados:2005dpd, Moustakidis:2005gx,Ibe:2017yqa,Essig:2019xkx, Baxter:2019pnz, Li:2022acp, Adams:2022zvg}. Originally proposed by A. Migdal in the late 1930s for studying radioactive nuclei~\cite{Migdal:1939, Migdal:1978}, this inelastic process allows a substantial fraction of the DM kinetic energy to be transferred to atomic electrons, which are more easily detected than nuclear recoils given current experimental thresholds. Importantly, any detector designed for DM-electron interactions inherently possesses sensitivity to DM-nucleus scattering via the Migdal effect~\cite{Ibe:2017yqa,Essig:2019xkx,Baxter:2019pnz}, making this approach immediately accessible to existing experiments. The Migdal effect has already been leveraged to recast ionization measurements in noble liquid~\cite{Dolan2017, Essig:2019xkx, XENON:2019zpr, DarkSide:2022dhx, COSINE-100:2021poy, LZ_Migdal} and semiconductor~\cite{Essig:2019xkx, Knapen2021, SuperCDMS:2022kgp, EDELWEISS:2022ktt, CDEX2022, SENSEI:2023zdf,DAMIC-M2025} detectors, with substantial prior theoretical effort going into quantifying the predicted signal. DM interactions in semiconductors are particularly challenging to calculate because of the complex electronic band structure and lattice vibrations, in contrast to noble liquids where nuclei can be treated as free atoms. However, as discussed previously, the small band gap of semiconductors allows sensitivity to $\mathcal{O}$(MeV) DM masses for DM-nuclear scattering via the Migdal effect. 

Recent work has described the Migdal effect with an EFT framework that captures the semiconductor response in all kinematic regimes for spin-independent DM by encoding the vibrational and electronic degrees of freedom in measurable structure factors~\cite{Berghaus2023} (see also \cite{Esposito:2025iry}).  In this work, we generalize this approach by combining the semiconductor EFT from~\cite{Berghaus2023} with a comprehensive EFT description of general non-relativistic DM-nucleus interactions~\cite{Fitzpatrick:2013, Anand2013}, including couplings to DM spin and/or nucleus spin, as well as angular momentum. We specifically apply our results to forecast the direct detection reach of germanium targets, motivated by experiments such as EDELWEISS~\cite{EDELWEISS_surface, EDELWEISS2020, EDELWEISS:2022ktt}, CDEX~\cite{CDEX:2019hzn,CDEX:2020tkb,  CDEX2022}, and SuperCDMS~\cite{SuperCDMS:2015lcz, SuperCDMS:2022kgp, SuperCDMS:2022crd}. Using existing EDELWEISS data~\cite{EDELWEISS2020}, we derive new experimental bounds on each operator and compare these limits to other direct-detection constraints in the literature. DM-nucleon effective field theory operators have been constrained with the Migdal effect treating the recoiling nucleus as free~\cite{Kang:2018rad,TOMAR2023102851,Liang:2024tef}. We emphasize that careful treatment must be used for low-threshold semiconductor detectors, especially in the region of low DM mass where these detectors have the most competitive reach. Additionally, we present the accompanying phonon signal, which next-generation semiconductor experiments may be sensitive to. A recent complementary study of spin-dependent phonon production from DM can be found in~\cite{Giffin:2025hdx}.

This paper is structured as follows. In~\autoref{sec:NucResp}, we review the DM-nucleus Hamiltonian for a full list of non-relativistic Galilean invariant (i.e., invariant under transformations in inertial reference frames) interactions that are spin 0 or spin-1 mediated. Then, in~\autoref{sec:EFT}, we apply the interaction to the EFT framework describing the Migdal effect in semiconductors. We show that the differential ionization rate factorizes into the ionization probability and the differential (spin-dependent/angular momentum dependent) cross section. In~\autoref{sec:Results}, we recast the EDELWEISS bounds of DM-electron scattering in germanium from~\cite{EDELWEISS2020} to the DM-Migdal effect, and interpret them in the context of other constraints for each non-relativistic operator. Additionally, we will present projections for planned future direct detection experiments. Throughout our analysis, we remain agnostic to any particular choice of UV-completion; however, we briefly discuss implications from general SM mediator bounds, which tend to be very stringent. Finally, we summarize and conclude in~\autoref{sec:summary}. Throughout this work, we will use common definitions found in the literature, which are detailed in~\autoref{app:definitions}. Additionally, in~\autoref{app:velocity} we derive the velocity-dependent effective Hamiltonian relevant for the DM-nucleus interactions we are interested in. Lastly, we simplify the elastic DM-Migdal matrix element in~\autoref{app:Matrix}, which was used to calculate our results.  

\section{DM-Nucleus Scattering Hamiltonian} \label{sec:NucResp}

Elastic DM-nucleus interactions have been well described in the non-relativistic limit in~\cite{Fitzpatrick:2013, Anand2013}. All possible non-relativistic interactions are captured by a complete set of Galilean invariant operators up to dimension-six. These non-relativistic operators can be matched to the relativistic Lagrangian that describe DM-nucleon interactions~\cite{Fan2010, Anand2013, Cirelli2013}. Then in the limit of small momentum transfer compared to the cutoff scale of our theory ($q \ll \Lambda$)\footnote{For the DM-nucleon interactions relevant for the Migdal effect with sub-GeV DM masses, the kinematically relevant regime has momentum transfers $q$ of order $10-100$~keV. Moreover, the cutoff of our theory will be the mass of the particle that mediates the interaction. A discussion regarding choices of UV mediators will be detailed in~\autoref{sec:Results}; however, we will state here that $\Lambda$ can be around, for example $10~\rm{MeV} - 1~\rm{TeV}$, with strong constraints on lower mass mediators. Therefore, we can safely take $q \ll \Lambda$ for all the parameter space we consider.}, the non-relativistic elastic interactions between non-relativistic DM ($\Phi_\chi$) and nucleon ($\Phi_n$) fields are described by effective four-field operators,
\begin{equation} \label{eqn:non-rel Lagrangian}
\mathcal{L} = \sum^{\substack{\text{neutron} \\\text{proton}}}_{n,n'} \sum^{11}_i c^{(n)}_i \mathcal{O}_i \overline{\Phi}_\chi \Phi_\chi \overline{\Phi}_n \Phi_n \: ,
\end{equation}
where the coefficient includes inverse powers of the UV cutoff ($c^{(n)}_i \sim 1/\Lambda^2$ for all operators). Although the EFT coefficients can be written in terms of couplings to iso-spin, we adopt the above notation, which intuitively highlights the coupling to nucleons and can easily be re-written in terms of the DM-nucleon reference cross section, which is commonly constrained by direct detection experiments~\cite{KANG201950}. The dimension-six non-relativistic operators $\mathcal{O}_i$ are constructed from Galilean and Hermitian quantities, for example the spin of the DM particle, $\boldsymbol{S_\chi}$, and the nucleon spin, $\boldsymbol{S_n}$. Additionally, $i \boldsymbol{q}/m_n$, which depends on the three-momentum transferred from the DM, $\boldsymbol{q}$, and the nucleon mass, $m_n$, as well as the transverse velocity, $\boldsymbol{v^\perp} \equiv \boldsymbol{v} + \frac{\boldsymbol{q}}{2 \mu_{\chi N}}$, defined by  $\boldsymbol{v^\perp}{\cdot} \, \boldsymbol{q} = 0$.  We note that $q^2$ is also an invariant quantity and therefore an exhaustive list of non-relativistic operators should include $q^{2m} \mathcal{O}_i $ ($m \in \mathbb{N}$) for all operators $\mathcal{O}_i$. However, it would be difficult to write down a UV model for $q^{2m} \mathcal{O}_i $ without dominant contributions from the leading-order operator $\mathcal{O}_i$. Therefore, in a direct-detection search, these additional operators are momentum suppressed and only appear as a sub-dominant interaction relative to the respective $\mathcal{O}_i$ contribution. Additionally, it is possible to include $\mathcal{O}_2 = (\boldsymbol{v}^\perp)^2$. This operator has a DM-nucleon cross section that is proportional to four powers of velocity, leading to Migdal constraints that scale similar to $\mathcal{O}_3, \mathcal{O}_5$, and $\mathcal{O}_6$. Moreover, $\mathcal{O}_2$ is a coherent operator (similar to $\mathcal{O}_1$) and does not introduce any spin or angular momentum response functions. Lastly, $\mathcal{O}_2$ is difficult to UV complete and is not necessary to describe the non-relativistic reduction of any common DM models. For these reasons we do not include it in our analysis. For simplicity, we continue our discussion with only the leading order non-relativistic operators listed in \autoref{tab:operators}. 
\begin{table}[t!]
\centering
\begin{tabular}{ll}
$\mathcal{O}_1 = \boldsymbol{\mathbb{1}}  $            & $\mathcal{O}_3 = i \boldsymbol{S_n} \cdot \left( \frac{\boldsymbol{q}}{m_n} \times \boldsymbol{v^\perp} \right)  $   \\
$\mathcal{O}_4 = \boldsymbol{S_\chi} \cdot \boldsymbol{S_n}  $  & $\mathcal{O}_5 = i \boldsymbol{S_\chi} \cdot \left( \frac{\boldsymbol{q}}{m_n} \times \boldsymbol{v^\perp} \right)   $ \\
$\mathcal{O}_6 = \left( \boldsymbol{S_\chi} \cdot \frac{\boldsymbol{q}}{m_n} \right) \left( \boldsymbol{S_n} \cdot \frac{\boldsymbol{q}}{m_n} \right) $ & $\mathcal{O}_7 = \boldsymbol{S_n} \cdot \boldsymbol{v^\perp} $ \\
$\mathcal{O}_9 = i \boldsymbol{S_\chi} \cdot \left( \boldsymbol{S_n} \times \frac{\boldsymbol{q}}{m_n} \right)
$ & $\mathcal{O}_8 = \boldsymbol{S_\chi} \cdot \boldsymbol{v^\perp} $ \\
$\mathcal{O}_{10} =  i \left( \boldsymbol{S_n} \cdot \frac{\boldsymbol{q}}{m_n} \right) $  & $\mathcal{O}_{11} = i \left( \boldsymbol{S_\chi} \cdot \frac{\boldsymbol{q}}{m_n} \right) $
\end{tabular}
\caption{All DM-nucleon non-relativistic operators of dimension six described by the interaction in~\eqautoref{eqn:non-rel Lagrangian}. \label{tab:operators}}
\end{table}

In the literature, the terminology \textit{spin-dependent} (SD) is widely used for different subsets of effective operators. In most direct-detection experiments, the SD interaction refers to $\mathcal{O}_4$ and is discussed in contrast to the spin-independent (SI) interaction, $\mathcal{O}_1$ \cite{Tovey200017, Ramani2019, Wang:2021oha, CDEX:2020tkb, XENON:2019zpr, PICASSO2017}. These two operators are not suppressed by momentum or velocity and have therefore been historically considered as dominant contributions in direct detection searches.  In other contexts, the subset of operators $\mathcal{O}_{3}, \mathcal{O}_4, \mathcal{O}_{6}, \mathcal{O}_{7}, \mathcal{O}_{9}, \mathcal{O}_{10}$, are considered to be spin-dependent because they depend on nucleon spin~\cite{AnandProc, Fan:2010gt}. This subset of nucleon spin-dependent operators is distinguishable from the remaining operators, which have coherent effects across the nucleus. Throughout this work we highlight the nucleon and DM spin contributions to these interactions. Moreover, we motivate how various operators result in nuclear responses related to \textit{angular momentum} as well as \textit{total spin and angular momentum}, which are often not explicitly considered in the interpretation of direct detection measurements. 

From the non-relativistic interactions defined in~\eqautoref{eqn:non-rel Lagrangian}, the DM-nucleus Hamiltonian can be constructed, which is necessary for computing the DM-Migdal rate derived in the following section. Accounting for all possible interactions permitted by symmetry, the DM-nucleus Hamiltonian density reads~\cite{Fitzpatrick:2013},
\begin{equation} \label{eqn:DMNucleusHamiltonian}
\begin{aligned}
\mathcal{H}^{(I)}_{\chi N}(\boldsymbol{x}_\chi) &= \sum^A_{\text{nucleons}} l_0 \delta (\boldsymbol{x}_\chi - \boldsymbol{x}_n)
+ \sum^A_{\text{nucleons}} \boldsymbol{l}_5 \cdot \boldsymbol{\sigma} \delta (\boldsymbol{x}_\chi - \boldsymbol{x}_n)  \\
+ & \sum^A_{\text{nucleons}} \frac{l^A_0}{2m_n}  \bigg [-\frac{1}{i} \cev{\nabla}_n \cdot \boldsymbol{\sigma} \delta (\boldsymbol{x}_\chi - \boldsymbol{x}_n)  + \delta (\boldsymbol{x}_\chi - \boldsymbol{x}_n) 
 \boldsymbol{\sigma} \cdot \frac{1}{i} \vec{\nabla}_n  \bigg] \\
+ & \sum^A_{\text{nucleons}} \frac{\boldsymbol{l}_M}{2m_n}  \bigg[-\frac{1}{i} \cev{\nabla}_n  \delta (\boldsymbol{x}_\chi - \boldsymbol{x}_n)  + \delta (\boldsymbol{x}_\chi - \boldsymbol{x}_n)  \frac{1}{i} \vec{\nabla}_n  \bigg] \\
+ & \sum^A_{\text{nucleons}} \frac{\boldsymbol{l}_E}{2m_n}  \bigg[\cev{\nabla}_n \times \boldsymbol{\sigma} \delta (\boldsymbol{x}_\chi - \boldsymbol{x}_n)  + \delta (\boldsymbol{x}_\chi - \boldsymbol{x}_n)  \boldsymbol{\sigma} \times \vec{\nabla}_n  \bigg] \: ,
\end{aligned}
\end{equation}
where $I$ labels the nucleus located in a lattice site of our semiconductor material and $\vec{\nabla}_n$ ($\cev{\nabla}_n$) is a derivative with respect to the position of the nucleon, $\boldsymbol{x}_n$, which acts to the right (left). These derivatives appear for terms that arise from operators dependent on $\boldsymbol{v^\perp} = \boldsymbol{v^T} + \boldsymbol{v^N}$, where $\boldsymbol{v^T}$ acts on the center of mass of the nucleus and $\boldsymbol{v^N}$ acts on the relative distance between the nucleons within the nucleus. This separates the operator into \textit{center of mass} (or \textit{point-like}) and \textit{intrinsic} contributions~\cite{Fitzpatrick:2013, Anand2013}. The EFT coefficients ($l_0, \boldsymbol{l}_5, l^A_0, \boldsymbol{l}_M, \boldsymbol{l}_E$) are functions of $\boldsymbol{S_\chi}$, $\boldsymbol{S_{n}}$, $\boldsymbol{q}/m_n$, $\boldsymbol{v^{\perp}}$ and the DM-nucleon couplings, $c^{(n)}_i$.  We provide the full definitions of the EFT coefficients in~\autoref{app:definitions}. 

From~\eqautoref{eqn:DMNucleusHamiltonian}, the Hamiltonian can be expanded into spherical harmonics, and one can identify the five relevant elastic nuclear response functions. Namely, (1)~the coherent response function, $M$, which contributes to the charge of the overall nucleus, (2)~$\Sigma'$, which arise from nucleon spin-dependent contributions transverse to the transferred momentum, (3)~$\Sigma''$, which describes the longitudinal nucleon spin-dependent response, (4)~$\Delta$, which results from couplings to orbital angular momentum and (5)~$\Phi''$, which captures a combined angular-momentum and spin-dependent response. The Hamiltonian is then~\cite{Fitzpatrick:2013, Anand2013} 
\begin{align}
\label{eq:nucleon_hamiltonian}
H^{(I)}_{\chi N}  = & \sum^{n,p}_N \sum_J \sqrt{4\pi} \sqrt{2J+1} (i^J) \left[ l_0 M_{J0} - \boldsymbol{l}_5 \biggl\{i{\Sigma''}_{J0} \hat{e_0} - \sum_{\beta = \pm 1} i {\Sigma'}_{J1} \frac{\hat{e^*_\beta}}{\sqrt{2}} \biggl\} \right.
\\ \nonumber & \left. - \frac{\boldsymbol{l}_M}{m_n} \biggl\{iq \sum_{\beta = \pm 1} \beta \Delta_{J0} \frac{\hat{e^*_\beta}}{\sqrt{2}} \biggl\} -  \frac{\boldsymbol{l}_E}{m_n} \biggl\{q\Phi''_{J0} \hat{e_0} \biggl\} \right] \:.
\end{align}
Here $J$ is the angular momentum of the material, and we have written $\boldsymbol{q} = |q| \hat{e}$, where $\hat{e}_0 = \hat{q}$ and $\hat{e}_{\pm 1}$ is perpendicular to $\hat{q}$.

The weight of each response is dependent on the target material. Nuclear response form factors for isotopes that have an unpaired neutron or proton have been calculated in~\cite{Fitzpatrick:2013} (fluorine, sodium, germanium, iodine, and xenon) and~\cite{Catena_2015} (hydrogen, helium, aluminum, and nickel). While our analytical results are applicable to any crystal, we focus primarily on germanium semiconductor detectors, motivated by its use in current direct detection experiments. Germanium offers many advantages as a detector material. Regarding the material's sensitivity to the response functions studied in this work, germanium possesses a considerably larger nuclear angular momentum ($J = 9/2$) compared to other materials typically studied in this context, for example from~\cite{Fitzpatrick:2013}. Furthermore, the angular-momentum-and-spin-dependent ($\Phi ''$) contribution in germanium is intermediate, larger than in fluorine and sodium, but smaller than in iodine or xenon. Additionally, since the germanium atom is heavier than other semiconductor materials such as sodium or fluorine, it exhibits a larger coherent response coefficient ($M$). More generally, germanium has a small band gap, which allows for small detector thresholds of order eV. We note, germanium's unpaired neutron enhances both spin ($\Sigma'$, $\Sigma''$) and angular momentum ($\Delta$) contributions for neutron couplings relative to proton couplings.

\section{EFT approach for Migdal effect in semiconductors} \label{sec:EFT}
Having reviewed the theoretical framework for general DM-nucleus interactions, we now turn to the specific challenges of implementing this approach in semiconductor detectors for quantifying the Migdal effect. While the non-relativistic EFT formalism provides a model-independent description of the DM-nucleon interaction, the practical application to crystalline targets requires careful consideration of the many-body electronic structure and lattice dynamics that distinguish semiconductors from gases and liquids. 

The calculation of the  DM Migdal effect in semiconductors was first proposed within the framework of second-order perturbation theory, facilitated by a DM-nucleus contact interaction described by $H_{\chi N}$, and a screened electron-nucleus Hamiltonian $H_{eN}$~\cite{Knapen2021}. This second-order description necessitated a virtual excited nuclear state, amounting to an intermediate lattice mode  mediating the electron ionization, as well as an explicit form of the final wave function of the nucleus. 

One key insight for semiconductor applications is that the complexity of multi-electron responses and vibrational degrees of freedom can be described by measurable quantities. Namely, the material's dielectric function encodes the collective electronic response of the crystal, providing a direct pathway to calculate Migdal electron excitation probabilities~\cite{Knapen2021}. This approach naturally incorporates the complex many-body interactions within the lattice, including screening effects and quantum correlations between electrons. Modern condensed matter methods using density functional theory (DFT) and all-electron reconstruction (AER)~\cite{Griffin2021, Trickle2022, Singal2023} enable accurate computation of these dielectric properties, establishing a robust foundation for predicting Migdal rates in realistic detector materials. On the other hand, the complexity associated with the vibrational degrees of freedom can be related to another measurable material property, the dynamic structure factor. This is most relevant when calculating the Migdal effect for lower DM masses, below 50 MeV, where the final state of the nucleus is poorly approximated by a free ion due to resolving the lattice potential~\cite{Berghaus2023}. 

Subsequently, a newly developed EFT framework exploited the energy separation between the eV scale ionized electron and the $\lesssim \mathcal{O}(100)\,$meV lattice mode that mediates the Migdal electron, showing that the intermediate mode can be integrated out, simplifying the calculation to a first-order effective interaction via an effective Hamiltonian, $H_{eff}$~\cite{Berghaus2023}. Within that EFT it is easy to show that the added complexity due to lattice potentials factorizes in the Migdal calculation with the standard vibrational dynamic structure function encoding the final state of the nucleus. This enabled reliable computations of the Migdal effect in semiconductors within kinematic regimes where the final state of the nucleus is poorly approximated as a free atom due to being subject to the lattice potential. 

The effective Hamiltonian $H_{eff}$ was derived previously in~\cite{Berghaus2023} for the spin-independent ($\mathcal{O}_{1}$) DM-Lattice Hamiltonian.
Here we generalize the framework to general DM-nucleon interactions. At leading order, we find (see~\autoref{app:velocity}) 
\begin{equation} 
\label{eqn:commutationrelation}
H_{eff} = \frac{1}{\omega^2} [H_{\chi L}, [H_L, H_{eL}]] \:, 
\end{equation}
where the DM-lattice Hamiltonian here contains a sum over nuclei $H_{\chi L} = \sum_I H^{(I)}_{\chi N}$ and contains all five nuclear response functions from~\eqautoref{eq:nucleon_hamiltonian}. The electron-lattice Hamiltonian is~\cite{Knapen2021,Berghaus2023},
\begin{equation}
H_{eL} = - \frac{4 \pi \alpha}{V} \sum_{I} \sum_{\boldsymbol{K,K'}} \sum_{\boldsymbol{k}_e} \frac{\epsilon_{\boldsymbol{K,K'}}^{-1} \left( \boldsymbol{k}_e, \omega \right) Z \left( |\boldsymbol{k}_e + \boldsymbol{K'}| \right) }{|\boldsymbol{k}_e + \boldsymbol{K}| |\boldsymbol{k}_e + \boldsymbol{K'}|} e^{i (\boldsymbol{k}_e + \boldsymbol{K}) \cdot \boldsymbol{x_e}} e^{-i (\boldsymbol{k}_e + \boldsymbol{K'}) \cdot \boldsymbol{x}_I} \: ,
\end{equation}
where, $\epsilon_{\boldsymbol{K,K'}}$ is the dielectric function of a given material, which depends on the electron momentum ($\boldsymbol{k}_e$) as well as the electron ionization energy ($\omega$), $\alpha$ is the fine structure constant, and $V$ is the volume of the material. Additionally, the electron-lattice Hamiltonian depends on $Z(k)$, the effective charge of the nuclei, which accounts for shielding effects from tightly bound core electrons.\footnote{For our numerical results in~\autoref{sec:Results}, we use the effective atomic number from~\cite{Knapen2021, Z_effective}.} Finally, the lattice Hamiltonian is 
\begin{equation} \label{eqn:latticeHamiltonian}
H_L = - \sum_I \frac{\nabla^2_I}{2m_N} + U({\boldsymbol{x}_I}) \ ,
\end{equation}
where $\nabla_I$ is the gradient with respect to the lattice position $\boldsymbol{x}_I$, and $m_N$ ($\simeq A m_n$) is the nucleus mass. The inter-atomic potential, $U$, can in general be a complicated function; however, we see in~\autoref{app:velocity} that the potential commutes with the electron-lattice Hamiltonian, so that the effective Hamiltonian does not depend on $U$~\cite{Berghaus2023}. Expanding and simplifying the effective Hamiltonian in~\eqautoref{eqn:commutationrelation} leads to (see~\autoref{app:velocity} for the derivation)
\begin{equation}
\begin{aligned} \label{eqn:H_eff}
H_{eff} &= \frac{1}{\omega^2 m_N} \left(\nabla_I H_{eL} \nabla_I H_{\chi L}\right) \\
& + \frac{1}{\omega^2 m_N} \left[ l^{A}_0  \delta (\boldsymbol{x}_\chi - \boldsymbol{x}_I) 
\left(\boldsymbol{\sigma} \cdot \frac{\vec{\nabla}_I }{i} \right) + \boldsymbol{l}_M \delta (\boldsymbol{x}_\chi - \boldsymbol{x}_n) 
\left(\frac{\vec{\nabla}_I }{i} \right) \right] \left(\frac{-\nabla^2_I H_{eL}}{2m_n} \right) \\
& + \mathcal{O} \left( \frac{1}{\omega^3} \right) \\ 
& \approx \frac{1}{\omega^2 m_N} \left( \nabla_I H_{\chi L} \nabla_I H_{eL}  + \mathcal{O}\left(\frac{|\boldsymbol{k}_e +\boldsymbol{K}|}{m_n} \nabla^2_I H_{eL} \right)\right)\,.
\end{aligned}
\end{equation}
Keeping the leading order term in~\eqautoref{eqn:H_eff}, we now write the matrix element for the DM-lattice Migdal interaction as,
\begin{equation} \label{eqn:MatrixElement}
\begin{aligned} 
\mathcal{M}_{fi} 
    &= \braket{ \chi_f, \boldsymbol{p}_f, n_f, \lambda_f, e_f  \left| \frac{\nabla_I H_{eL} \cdot \nabla_I H_{\chi L}}{\omega^2 m_N} \right| \chi_i, \boldsymbol{p}_i, n_i, \lambda_i, e_f} \: ,
\end{aligned}
\end{equation}
where the DM states depend on the total angular momentum quantum number $j$ and the respective projection quantum number $m$, for example $\bra{\chi_i} = \bra{ j_{(\chi)}, m_{(\chi)}}$ and $\boldsymbol{p}_i$ is the DM momentum. Additionally, the nucleon states depend on its own respective spin quantum numbers, $\bra{n_i} = \bra{j_{(n)}, m_{(n)}}$. We also track the the initial (final) lattice states $\lambda_i$ ($\lambda_f$). Lastly, the electron initial (final) momenta state is $\bra{e_i} = \bra{\boldsymbol{p}_e}$ ($\bra{e_f} = \bra{\boldsymbol{p}_e+\boldsymbol{k}_e}$). In~\autoref{app:Matrix}, we simplify the generalized DM-Migdal matrix element in terms of the EFT coefficients and find,
\begin{equation}
\begin{aligned}
\label{eq:Mfi}
\mathcal{M}_{fi} &= \frac{4 \pi \alpha}{V^2 \omega^2  m_N} \sum_{\boldsymbol{K,K'}} \sum_{\boldsymbol{k}_e} \sum_I  \frac{\boldsymbol{q} \cdot (\boldsymbol{k}_e + \boldsymbol{K'} ) \epsilon_{\boldsymbol{K,K'}}^{-1} (\boldsymbol{k}_e, \omega) Z(|\boldsymbol{k}_e + \boldsymbol{K}|)}{|\boldsymbol{k}_e + \boldsymbol{K}| |\boldsymbol{k}_e + \boldsymbol{K'}|}
\\ & \langle \boldsymbol{p}_e+\boldsymbol{k}_e| e^{i (\boldsymbol{k}_e + \boldsymbol{K}) \cdot \boldsymbol{x_e}} | \boldsymbol{p}_e \rangle  \langle \lambda_f| e^{i (\boldsymbol{k}_e + \boldsymbol{K'} - \boldsymbol{q}) \cdot \boldsymbol{x_I}} | \lambda_i \rangle
\\&  \langle \chi_f, n_f| \sum^{n,p}_N \sum_J \sqrt{4\pi} \sqrt{2J+1} (i^J) \left[ l_0 M_{J0} - \boldsymbol{l}_5 \biggl\{i{\Sigma''}_{J0} \hat{e_0} - \sum_{\beta = \pm 1} i {\Sigma'}_{J1} \frac{\hat{e^*_\beta}}{\sqrt{2}} \biggl\} \right.
\\& \left. - \frac{\boldsymbol{l}_M}{m_n} \biggl\{iq \sum_{\beta = \pm 1} \beta \Delta_{J0} \frac{\hat{e^*_\beta}}{\sqrt{2}} \biggl\} -  \frac{\boldsymbol{l}_E}{m_n} \biggl\{q\Phi''_{J0} \hat{e_0} \biggl\} \right]| \chi_i, n_i \rangle \, ,
\end{aligned}
\end{equation}
where the momentum transferred by the DM is defined as $\boldsymbol{q} = \boldsymbol{p}_i - \boldsymbol{p}_f$. We rewrite the operator from the third and fourth line of DM-Migdal matrix element in~\eqautoref{eq:Mfi} in terms of the effective DM-nucleon couplings $c^{(n)}_i$ from~\eqautoref{eqn:non-rel Lagrangian}, highlighting the relevant nuclear response functions for each EFT operator, 
\begin{equation} \label{eqn:MatrixElementSimplified}
\begin{aligned}
 & \Biggl[ l_0 M_{J0} - \boldsymbol{l}_5 \biggl\{i{\Sigma''}_{J0} \hat{e_0} 
   - \sum_{\beta = \pm 1} i {\Sigma'}_{J1} \frac{\hat{e}^*_\beta}{\sqrt{2}} \biggl\}  
   - \frac{\boldsymbol{l}_M}{m_n} \biggl\{iq \sum_{\beta = \pm 1} \beta \Delta_{J0} \frac{\hat{e}^*_\beta}{\sqrt{2}} \biggl\} 
   - \frac{\boldsymbol{l}_E}{m_n} \biggl\{q\Phi''_{J0} \hat{e_0} \biggl\} \Biggr] 
\\[0.5em]
& = \Biggl[ c^{(n)}_1 M_{J0} 
   + c^{(n)}_3 \biggl\{ \frac{i}{2} \left( \frac{\boldsymbol{q}}{m_n} \times \boldsymbol{v_T}^\perp \right) 
   \sum_{\beta= \pm} i \Sigma'_{J1} \frac{\hat{e}^*_\beta}{\sqrt{2}} 
   - \frac{q^2}{2m^2_n} \Phi''_{J0} \biggr\}  
\\& \quad - c^{(n)}_4 \frac{\boldsymbol{S_\chi}}{2}  
   \left\{ i{\Sigma''}_{J0} \hat{e_0} - \sum_{\beta = \pm 1} i {\Sigma'}_{J1} \frac{\hat{e}^*_\beta}{\sqrt{2}} \right\}
   + c^{(n)}_5 \Biggl\{ i \left( \frac{\boldsymbol{q}}{m_n} \times \boldsymbol{v^\perp_T} \right) \cdot \boldsymbol{S_\chi} M_{J0} 
\\& \qquad \qquad - i \left( \frac{\boldsymbol{q}}{m_n} \times \boldsymbol{S_\chi} \right) \frac{iq}{m_n} 
   \sum_{\beta = \pm 1} \beta {\Delta}_{J0} \frac{\hat{e}^*_\beta}{\sqrt{2}} \Biggr\}
   - c^{(n)}_6 \left( \frac{\boldsymbol{q}}{m_n} \cdot \boldsymbol{S_\chi} \right) \frac{i|q|}{2 m_n} \Sigma''_{J0} 
\\& \quad + c^{(n)}_7 \frac{\boldsymbol{v^\perp_T}}{2} 
   \sum_{\beta= \pm} i \Sigma'_{J1} \frac{\hat{e}^*_\beta}{\sqrt{2}}  
   + c^{(n)}_8 \biggl\{ \left( \boldsymbol{v^\perp_T} \cdot \boldsymbol{S_\chi} \right) M_{J0} 
   + \boldsymbol{S_\chi} \frac{iq}{m_n} \sum_{\beta = \pm 1} \beta {\Delta}_{J0} \frac{\hat{e}^*_\beta}{\sqrt{2}} \biggr\}
\\& \quad + c^{(n)}_9 \frac{i}{2} \left( \frac{\boldsymbol{q}}{m_n} \times \boldsymbol{S_\chi} \right) 
   \cdot \sum_{\beta= \pm} i \Sigma'_{J1} \frac{\hat{e}^*_\beta}{\sqrt{2}}  
   + c^{(n)}_{10} \frac{|q|}{2 m_n} \Sigma''_{J0} 
   + c^{(n)}_{11} i \left( \frac{\boldsymbol{q}}{m_n} \cdot \boldsymbol{S_\chi} \right) M_{J0} \Biggr] \: .
\end{aligned}
\end{equation}
Finally, we derive the DM-Migdal rate using Fermi's Golden Rule,
\begin{equation} \label{eqn:longRate}
\begin{aligned}
\frac{d\Gamma}{d\omega} &\approx \left( \frac{4\pi \alpha}{\omega^2 m_N V^2} \right)^2  \sum_{\boldsymbol{q}} \sum_{\boldsymbol{K},\boldsymbol{Q}} \sum_{\boldsymbol{k}_e} \frac{\boldsymbol{q} \cdot (\boldsymbol{k}_e + \boldsymbol{K})}{|\boldsymbol{k}_e + \boldsymbol{K}|} \frac{\boldsymbol{q} \cdot (\boldsymbol{k}_e + \boldsymbol{Q})}{|\boldsymbol{k}_e + \boldsymbol{Q}|} Z \left( |\boldsymbol{k}_e + \boldsymbol{K}| \right) Z \left( |\boldsymbol{k}_e + \boldsymbol{Q}| \right) \\
& \left( \frac{2\pi V N_T}{4 \pi^2 \alpha} \right) \text{Im} \left( -\epsilon^{-1}_{K,Q} (\boldsymbol{k}_e, \omega) \right) S(\boldsymbol{q} - \boldsymbol{k}_e + \boldsymbol{K}, E_{p_i} - E_{p_f} - \omega) \\
& \sum^{\substack{\text{neutron} \\\text{proton}}}_{n,n'} \Bigg\{ c^{(n)}_1 c^{(n')}_1 F^{(n,n')}_M  + c^{(n)}_3 c^{(n')}_3 \left\{ \frac{q^4}{4m_n^4} F^{(n,n')}_{\Phi''} + \frac{q^2}{8m^2_n} \left(\boldsymbol{v^{\perp}_T} \right)^2 F^{(n,n')}_{\Sigma'} \right\}  \\
& + c^{(n)}_4 c^{(n')}_4  \frac{1}{4}  \left\{ \left\langle \boldsymbol{S_\chi} \cdot \hat{q} \right \rangle^2 F^{(n,n')}_{\Sigma''} + \frac{1}{2} \left( \left\langle \boldsymbol{S_\chi} \right\rangle^2 - \left\langle \boldsymbol{S_\chi} \cdot \hat{q} \right\rangle^2 \right) F^{(n,n')}_{\Sigma'} \right\} \\
& + c^{(n)}_5 c^{(n')}_5 \left\{ \left\langle \boldsymbol{S_\chi} \cdot \left( \frac{\boldsymbol{q}}{m_n} \times \boldsymbol{v^\perp_T} \right)\right\rangle^2 F^{(n,n')}_{M} \right. \\ 
& \qquad \qquad \left. + \frac{q^2}{2m^2_n} \left( \left\langle \boldsymbol{S_\chi} \times \frac{\boldsymbol{q}}{m_n} \right\rangle^2 - \left\langle \left( \boldsymbol{S_\chi} \times \frac{\boldsymbol{q}}{m_n} \right) \cdot \hat{q} \right\rangle^2 \right) F^{(n,n')}_{\Delta} \right\} \\
& + c^{(n)}_6 c^{(n')}_6 \frac{q^2}{4m_n} \left\langle \boldsymbol{S_\chi} \cdot \frac{\boldsymbol{q}}{m_n} \right\rangle^2 F^{(n,n')}_{\Sigma''} + c^{(n)}_7 c^{(n')}_7 \frac{\left(\boldsymbol{v^\perp_T}\right)^2}{8} F^{(n,n')}_{\Sigma'} \\
& + c^{(n)}_8 c^{(n')}_8 \left\{ \left\langle \boldsymbol{S_\chi} \cdot \boldsymbol{v^\perp_T} \right\rangle^2 F^{(n,n')}_M + \frac{q^2}{2m^2_n} \left( \left\langle \boldsymbol{S_\chi} \right\rangle^2 - \left\langle \boldsymbol{S_\chi} \cdot \hat{q} \right\rangle^2 \right) F^{(n,n')}_{\Delta} \right\} \\
& + c^{(n)}_9 c^{(n')}_9 \frac{1}{8} \left\{ \left\langle \boldsymbol{S_\chi} \times \frac{\boldsymbol{q}}{m_n} \right\rangle^2 - \left\langle (\boldsymbol{S_\chi} \times \frac{\boldsymbol{q}}{m_n}) \cdot \hat{q} \right\rangle^2 \right\} F^{(n,n')}_{\Sigma'}  \\
& + c^{(n)}_{10} c^{(n')}_{10} \frac{q^2}{4m^2_n} F^{(n,n')}_{\Sigma''} + c^{(n)}_{11} c^{(n')}_{11} \left\langle \boldsymbol{S_\chi} \cdot \frac{\boldsymbol{q}}{m_n} \right\rangle^2 F^{(n,n')}_M \Bigg\} \: ,
\end{aligned}
\end{equation}
where we have dropped the cross terms with $c_i \neq c_j$, since 
we focus here on constraining each non-relativistic operator independently.\footnote{Cross terms are possible and could be relevant for some UV complete models; we give the relevant form factors in~\autoref{app:definitions}.} The initial electron states can be represented in terms of periodic Bloch wave functions defined in a primitive cell with volume $\Omega$. The initial states, $|\boldsymbol{p}_e]_\Omega$, follow a distribution $f_b(\boldsymbol{p}_e)$, where $b$ labels the electronic branch. Furthermore, we have used the fact that the off-diagonal elements of the dielectric function are small such that $\epsilon^{-1}_{KK'} \approx \epsilon^{-1}_{KK} \delta_{K,K'}$ to write the rate in terms of the electron loss function (ELF) \cite{Knapen2021}, 
\begin{equation}
\begin{aligned}
\text{Im} \left( -\epsilon^{-1}_{\boldsymbol{KQ}} (\boldsymbol{k}_e, \omega) \right) &= \frac{4 \pi^2 \alpha}{V} \sum_{b,b'} \frac{f_b(\boldsymbol{p}_e)}{|\boldsymbol{k}_e+\boldsymbol{K}|} \frac{\epsilon^{-1}_{\boldsymbol{KK'}} \epsilon^{-1 *}_{\boldsymbol{QQ'}}}{|\boldsymbol{k}_e+ \boldsymbol{Q}|} \delta(\omega - E_{b'}, \boldsymbol{p}_e+\boldsymbol{k}_e - E_{b,pe}) \\
&  \left[ \boldsymbol{p}_e \left| e^{-i(\boldsymbol{k}_e + \boldsymbol{Q})\cdot \boldsymbol{x}_e} \right| \boldsymbol{p}_e + \boldsymbol{k}_e \right]_\Omega \left[\boldsymbol{p}_e + \boldsymbol{k}_e \left| e^{i(\boldsymbol{k}_e + \boldsymbol{K})\cdot \boldsymbol{x}_e} \right| \boldsymbol{p}_e \right]_\Omega \:,
\end{aligned}
\end{equation}
and we identify the structure factor,
\begin{equation}
S(\boldsymbol{q'}, E') = \frac{1}{N_T} \sum_{\lambda_f}  \Big| \langle \lambda_f | \sum_I e^{i \boldsymbol{q'} \cdot \boldsymbol{x}_n} | \lambda_i  \rangle \Big|^2 \delta(E_{\lambda_f} - E_{\lambda_i} - E') \:, 
\end{equation}
with $\boldsymbol{q'} \rightarrow \boldsymbol{q} - \boldsymbol{k}_e + \boldsymbol{K}$ and $E' \rightarrow E_{p_i} - E_{p_f} - \omega$, and $N_T$ is the number of atoms in the target. We can then write the rate in terms of the operator form factors, ($F^{(n,n')}_{i,j}$),\footnote{The operator form factors are valid for low momentum transfer in the $q \rightarrow 0$ limit, as pointed out in~\cite{Gresham2014}. In this work, we focus on light DM interactions with $m_\chi \sim 1$~MeV -- $10$~GeV, where this limit is a good approximation. For heavy DM ($10$~GeV $< m_\chi \lesssim 250$~GeV) and higher momentum transfers, the form factors are typically normalized by $F(q=0)$. For more information about this procedure see~\cite{Gresham2014}.} after simplifying the DM spin states, $\langle \boldsymbol{S_\chi} \rangle^2$ (definitions of all form factors and evaluations of the spin states can be found in~\autoref{app:definitions}). Finally, we present the simplified expression for the generalized Migdal rate in a crystal,
\begin{equation} \label{eqn:MigdalRate}
\begin{aligned}
\frac{d\Gamma}{d\omega} &= \frac{8\pi \alpha N_T}{V^3(\omega^2 m_N)^2} \sum_{\boldsymbol{q}} \sum^{\substack{\text{neutron} \\\text{proton}}}_{n,n'} \sum^{11}_{i,j} c^{(n)}_{i} c^{(n')}_j F^{(n,n')}_{i,j} (q,v)  \\
& \sum_{K,Q} \sum_{\boldsymbol{k}_e} \frac{\boldsymbol{q} \cdot (\boldsymbol{k}_e + \boldsymbol{K})}{|\boldsymbol{k}_e + \boldsymbol{K}|} \frac{\boldsymbol{q} \cdot (\boldsymbol{k}_e + \boldsymbol{Q})}{|\boldsymbol{k}_e + \boldsymbol{Q}|} Z(|\boldsymbol{k}_e + \boldsymbol{K}|) Z(|\boldsymbol{k}_e + \boldsymbol{Q}|) \text{Im} \left( -\epsilon^{-1}_{\boldsymbol{K},\boldsymbol{Q}} (\boldsymbol{k}_e, \omega) \right ) \\
& S(\boldsymbol{q} - \boldsymbol{k}_e + \boldsymbol{K}, E_{p_i} - E_{p_f} - \omega) \: .
\end{aligned}
\end{equation}

\section{Results} \label{sec:Results}
\subsection{Rate Spectrum} \label{subsec:RateSpecturm} 
Using the general Migdal rate derived in~\eqautoref{eqn:MigdalRate}, we calculate the direct detection DM-Migdal ionization and phonon spectrum. We divide by the detector mass, $M_T$, and multiply by the number of DM particles that traverse the detector ($n_\chi V$), where $n_\chi$ is the local number density of DM. We then take the infinite volume limit to replace sums from~\eqautoref{eqn:MigdalRate} with integrals as well as integrate over the DM velocity distribution, $f_{\chi}(\boldsymbol{v})$. The total DM-Migdal scattering rate simplifies to
\begin{equation}
\begin{aligned}
R &= \frac{n_\chi}{M_T} \int d \omega \frac{8\pi \alpha N_T}{(\omega^2 m_N)^2} \int d^3 v f_\chi(\boldsymbol{v}) \int \frac{d^3 \boldsymbol{q} }{(2\pi)^3} \underbrace{\sum^{\substack{\text{neutron} \\\text{proton}}}_{n,n'} \sum^{11}_{i,j} c^{(n)}_{i} c^{(n')}_j F^{(n,n')}_{i,j} (q,v)}_{\text{nuclear}}  \\
&  \int \frac{d^3 \boldsymbol{k}_e}{(2\pi)^3} \underbrace{\frac{(\boldsymbol{q} \cdot (\boldsymbol{k}_e + \boldsymbol{K}))^2}{|\boldsymbol{k}_e + \boldsymbol{K}|^2} Z(|\boldsymbol{k}_e + \boldsymbol{K}|)^2 \text{Im} \left (-\epsilon^{-1}_{\boldsymbol{K},\boldsymbol{K}} (\boldsymbol{k}_e, \omega) \right)}_{\text{electronic}}\\
& \underbrace{S(\boldsymbol{q} - \boldsymbol{k}_e + \boldsymbol{K}, \boldsymbol{q}  \boldsymbol{v} - \frac{q^2}{2m_\chi} - \omega)}_{\text{vibrational}} \: ,
\end{aligned}
\end{equation}
highlighting that to leading order the rate factorizes into the nuclear, electronic, and vibrational responses. 
Here we have dropped off-diagonal terms, which effectively sets $\boldsymbol{K} = \boldsymbol{Q}$, and reduces the computational power.

To quantify the differential DM-Migdal phonon spectrum, $dR/dE$, we decompose the three-dimensional velocity integral into its radial and angular components. The isotropic velocity distribution is typically described as a truncated Maxwell-Boltzmann distribution, 
\begin{equation}
f_{\chi}(v) =  \frac{1}{N} \, v \exp{\left(- \frac{v^2+v^2_e}{v^2_0}\right)}
    \begin{cases}
        2\sinh{\left( \frac{2v v_e}{v^2_0} \right) } &  v \leq v_{\rm{esc}} - v_e \, ,\\[0.3em]
        e^{\frac{2vv_e}{v^2_0}} - e^{\frac{-(v^2_{\rm{esc}} - v^2 - v^2_e)}{v^2_0}} &  v_{\rm{esc}}-v_e< v \leq v_{\rm{esc}}+v_e \, , \\[0.3em]
        0 & v> v_{\rm{esc}} + v_e \: ,
    \end{cases}
\end{equation}
where 
\begin{equation}
N = \pi^{1/2} v_e v_0 \left( \text{erf}\left(\frac{v_{\text{esc}}}{v_0}\right) - 2\frac{v_{\text{esc}}}{v_0 \sqrt{\pi}} e^{-\frac{v^2_{\rm{esc}}}{v^2_0} }  \right) \: .
\end{equation}
We use the average Earth velocity $v_{\text{e}}=240$~km/s, the DM escape velocity $v_{\text{esc}}=500$~km/s, and reference velocity $v_0=220$~km/s. Now we can rewrite the integral as a function of phonon energy, $E=qv\cos\theta_{vq}-q^2/(2 m_\chi)-\omega$, resulting in
\begin{equation}
\begin{aligned}
R = & \frac{n_\chi}{M_T} \int d \omega \frac{8\pi \alpha N_T}{(\omega^2 m_N)^2} \int (2 \pi) v^2 dv f_\chi(v) \int \frac{d^3 \boldsymbol{q} }{(2\pi)^3} \sum^{\substack{\text{neutron} \\\text{proton}}}_{n,n'} \sum^{11}_{i,j} c^{(n)}_{i} c^{(n')}_j F^{(n,n')}_{i,j} (q,v)  \\
&\int  \frac{d^3 \boldsymbol{k}_e}{(2\pi)^3}  \frac{(\boldsymbol{q} \cdot (\boldsymbol{k}_e + \boldsymbol{K}))^2}{|\boldsymbol{k}_e + \boldsymbol{K}|^2} Z(|\boldsymbol{k}_e + \boldsymbol{K}|)^2 \text{Im} \left (-\epsilon^{-1}_{\boldsymbol{K},\boldsymbol{K}} (\boldsymbol{k}_e, \omega) \right) \\
&\int^{E_{\rm{max}}}_0 \frac{dE}{qv} S(\boldsymbol{q} - \boldsymbol{k}_e + \boldsymbol{K}, E) \: ,
\end{aligned}
\end{equation}
where $E_{\rm{max}} = qv - q^2/(2 m_\chi) - \omega$. Similarly, we split the integral over $\boldsymbol{q}$ into its angular components. We also assume $q \gg |\boldsymbol{k}_e + \boldsymbol{K}|$ such that the structure factor only depends on the magnitude of the momentum argument, a good approximation for $m_\chi \gtrsim 10$~MeV.\footnote{In the regime where this does not hold, the magnitude of the momentum argument must be kept exact, $\sqrt{q^2 + |\boldsymbol{k}_e + \boldsymbol{K}|^2 - 2q|\boldsymbol{k}_e + \boldsymbol{K}| \cos \theta_{qk}}$, and the integral over $\cos_{qk}$ must be performed explicitly, greatly increasing computational burden.} In this regime, we can evaluate the angular integral $\int d \cos\theta_{qk} \cos^2\theta_{qk}$ analytically. Finally, we restore the full integral over velocity and assume it to be isotropic, which gives the differential DM-Migdal phonon spectrum 
\begin{equation}
\begin{aligned}
\frac{dR}{dE} = & \frac{n_\chi N_T}{M_T} \int d \omega \frac{2 \alpha}{3(\omega^2 m_N \mu_{\chi n})^2} \int_{v_{\text{-}}}^{v_{\text{+}}} d^3v \frac{f_\chi(v)}{v} \int_{q_{\text{-}}}^{q_{\text{+}}} q^3 dq  \sum^{11}_{i,j} \overline{\sigma}^{(i)}_n F^{(n,n)}_{i,j} (q,v)  \\
& \int \frac{d^3 \boldsymbol{k}_e}{(2\pi)^3} Z(|\boldsymbol{k}_e + \boldsymbol{K}|)^2 \text{Im} \left (-\epsilon^{-1}_{\boldsymbol{K},\boldsymbol{K}} (\boldsymbol{k}_e, \omega) \right)  S(q, E)   \: ,
\end{aligned}
\end{equation}
where we have defined the reference cross section for each operator ($\overline{\sigma}^{(i)}_n$) with $(c^{(n)}_i)^2= \pi \overline{\sigma}^{(i)}_n/\mu^2_{\chi n}$ and $\mu_{\chi n}$ is the DM-neutron reduced mass.\footnote{Here, we assume that the DM only couples to the neutron and not the proton, which is approximately true for germanium due to the material's unpaired neutron. Consequently, we have dropped the sum over nucleons from our phonon spectrum. Moreover, for target materials that contain an unpaired proton the differential cross section can be written in terms of the analogous reference cross section relevant for proton scattering, $\overline{\sigma}^{(i)}_p$.} The integration limits for the DM momentum is $q_{\pm} = m_\chi v \pm \sqrt{2m_\chi (m_\chi v^2/2 - \omega - E)}$, and the limits for the DM velocity are 
$v_{\text{-}} = \sqrt{2 (\omega + E) / m_\chi}$ and $v_{\text{+}} = (v_{\text{e}} + v_{\text{esc}})$. 
We use the dielectric function from DarkELF~\cite{DarkELF2021_dielectric}, which is calculated on a grid limited by $\omega \lesssim $ 81 eV and $\boldsymbol{k}_e \lesssim 22$ keV. Although these integration bounds formally go to infinity, the Migdal phonon and ionization spectrum are negligible beyond this range. 

We follow the prescription in~\cite{Berghaus2023} and compute the structure factor $S(q,E)$ for germanium at low momentum with the recursive multi-phonon method, while at higher momenta the structure factor can be approximated using the impulse approximation. We note that the EFT described in~\autoref{sec:EFT} is accurate when there is a separation of scales such that $\omega \gg E$.
Within this regime, the integration bounds $q_\pm$ and $v_{-}$ are not very sensitive to the phonon energy, $E$. Beyond this regime (for larger DM masses and DM momentum transfers), the crystal response approaches that of a free atom with the excitation centered around the nuclear recoil energy, $E_r$, such that $S(q,E) \sim \delta(E-E_r)$  (see ~\cite{Berghaus2023} for more details). 

\begin{figure}[t!]
    \centering
    \begin{subfigure}[b]{0.49\textwidth}   
        \centering 
        \includegraphics[width=\textwidth]{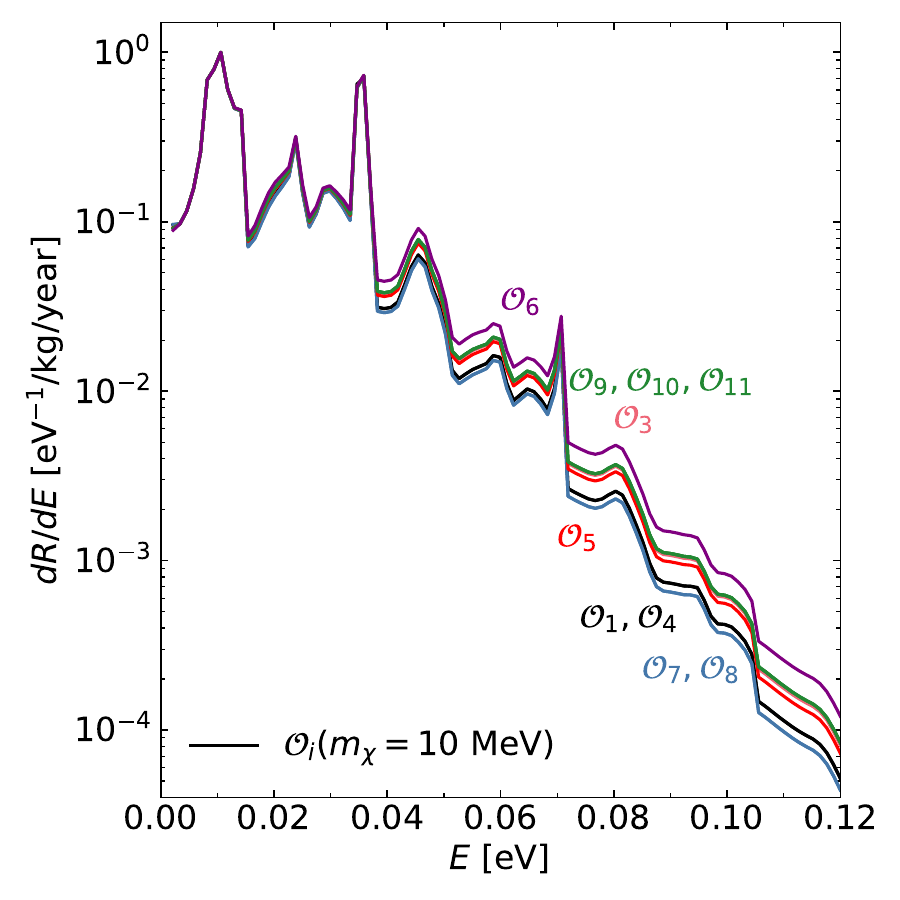}
    \end{subfigure}
    \hfill
    \begin{subfigure}[b]{0.49\textwidth}   
        \centering 
        \includegraphics[width=\textwidth]{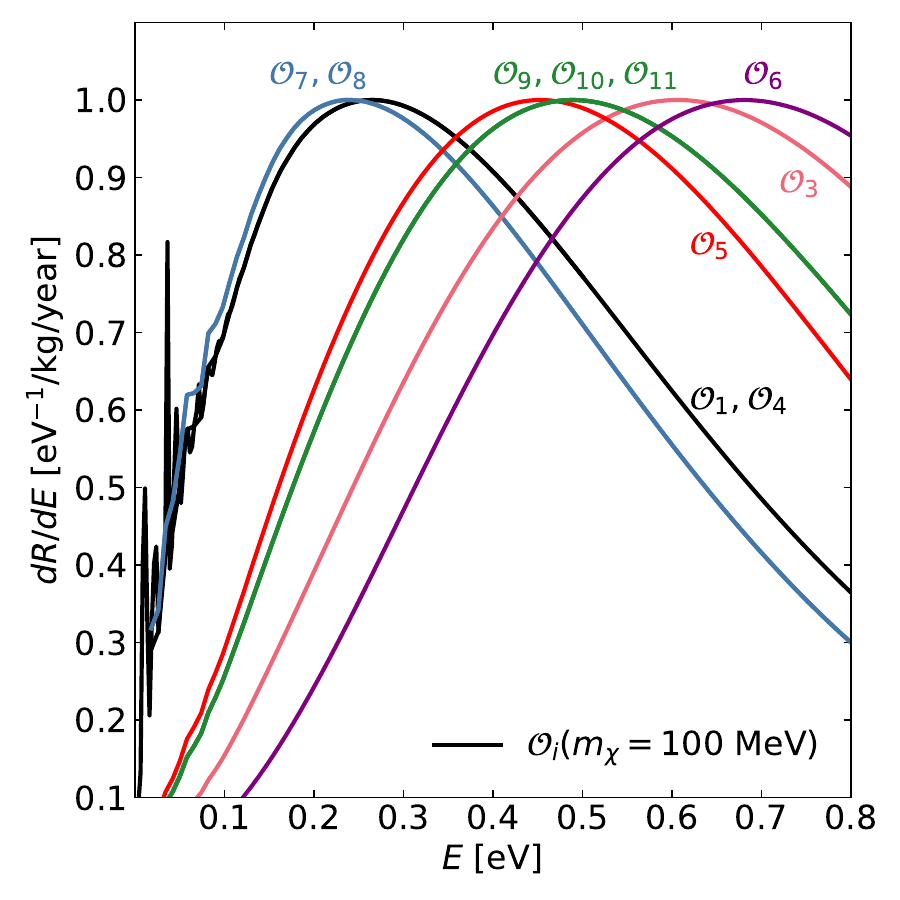}
    \end{subfigure}
    \caption[]
    {DM-Migdal phonon spectrum for a germanium detector (consisting of 7.7\% Ge-73) for DM spin $j_\chi = 1/2$ and DM mass \textbf{(left)} $m_\chi = 10$~MeV and \textbf{(right)} $m_\chi = 100$~MeV. Each curve is normalized to a maximum value of 1~event/(kg-yr-eV), and represents the differential rate with respect to the phonon energy for the reference cross sections given in~\autoref{tab:phonon_scale} for each operator. Note, at $10$~MeV, deviations between operators are so small that we choose to show the spectrum on a log scale. In addition, for 10~MeV, the spectrum for $\mathcal{O}_3$ is similar to the spectra in green. \label{fig:phonon_spectrum}}
\end{figure}

We show the differential Migdal phonon spectrum in~\autoref{fig:phonon_spectrum} assuming an exposure of 1~kg-yr for all operators at $10$~MeV and $100$~MeV. The corresponding reference cross sections are given in \autoref{tab:phonon_scale} for each operator. The nuclear form factors do not explicitly depend on the phonon energy, $E$. As a result, operator form factors that have the same effective velocity and momentum dependence (see~\autoref{app:OpFormFactors}) share the same signal---the integration over the nuclear form factors simply shifts the normalization of the phonon spectrum but reproduces exactly the same shape. At low masses, we see that integrating over the momentum dependence in the structure factor and form factors causes small differences in the phonon spectrum across all operators with different momentum and velocity dependence.
We note $F^{(n,n')}_{3,3}$, effectively has terms proportional to $q^2$ and $q^4$ that are similar in magnitude (due to the weight of the relevant nuclear form factors-- the coherent $F^{(n,n')}_M$ response enhances the $q^4$ contribution). Therefore, for $\mathcal{O}_3$ phonon spectrum, there is a unique dependence on momentum at all DM masses.
For higher DM masses (and higher phonon energies), the dependence on the phonon energy in the integration bounds strengthens. Therefore, operators with the same momentum dependence have distinctly different spectra. 
\begin{table}[h!]
\centering
\begin{tabular}{l|l|l}
Operator & $\overline{\sigma}_n$ [$\rm{cm}^2$] ($m_\chi =10 \,\rm{MeV}$) & $\overline{\sigma}_n$ [$\rm{cm}^2$] ($m_\chi =100 \, \rm{MeV}$) \\ \cline{1-3}
$\mathcal{O}_1 =  \boldsymbol{\mathbb{1}}$                          & $7.7 \times 10^{-40}$                                       & $4.3 \times 10^{-40}$                                       \\
$\mathcal{O}_3 = i \boldsymbol{S_n} \cdot \left( \frac{\boldsymbol{q}}{m_n} \times \boldsymbol{v^\perp} \right) $                                       & $4.2 \times 10^{-19}$                                       & $2.2 \times 10^{-21}$                                       \\
$\mathcal{O}_4 = \boldsymbol{S_\chi} \cdot \boldsymbol{S_n} $                                                                                           & $7.7 \times 10^{-34}$                                       & $4.3 \times 10^{-34}$                                       \\
$\mathcal{O}_5 = i \boldsymbol{S_\chi} \cdot \left( \frac{\boldsymbol{q}}{m_n} \times \boldsymbol{v^\perp} \right) $                                    & $1.2 \times 10^{-22}$                                       & $4.3 \times 10^{-25}$                                       \\
$\mathcal{O}_6 = \left( \boldsymbol{S_\chi} \cdot \frac{\boldsymbol{q}}{m_n} \right) \left( \boldsymbol{S_n} \cdot \frac{\boldsymbol{q}}{m_n} \right) $ & $2.4 \times 10^{-15}$                                       & $2.3 \times 10^{-19}$                                       \\
$\mathcal{O}_7 = \boldsymbol{S_n} \cdot \boldsymbol{v^\perp} $                                                                                          & $4.5 \times 10^{-28}$                                       & $3.4 \times 10^{-28}$                                       \\
$\mathcal{O}_8 = \boldsymbol{S_\chi} \cdot \boldsymbol{v^\perp} $                                                                                       & $9.8 \times 10^{-32}$                                      & $7.6 \times 10^{-32}$                                       \\
$\mathcal{O}_9 = i \boldsymbol{S_\chi} \cdot \left( \boldsymbol{S_n} \times \frac{\boldsymbol{q}}{m_n} \right)$                                         & $1.3 \times 10^{-24}$                                       & $1.0 \times 10^{-26}$                                       \\
$\mathcal{O}_{10} =  i \left( \boldsymbol{S_n} \cdot \frac{\boldsymbol{q}}{m_n} \right) $                                                               & $6.5 \times 10^{-25}$                                       & $5.4 \times 10^{-27}$                                       \\
$\mathcal{O}_{11} = i \left( \boldsymbol{S_\chi} \cdot \frac{\boldsymbol{q}}{m_n} \right) $                                                             & $3.7 \times 10^{-29}$                                       & $1.2 \times 10^{-30}$                                   
\end{tabular}
\caption{Choice of reference cross section for the phonon spectrum of each operator in~\autoref{fig:phonon_spectrum} at two different DM masses. We find that there is a common characteristic shape for operators which have the same momentum and velocity dependence, normalized with a different reference cross section. 
Note, for $\mathcal{O}_{1}$ we assume 100\% of the detector material can interact (rather than 7.7\% for Ge-73) and include the contribution from protons. 
\label{tab:phonon_scale}
} 
\end{table}

Given a future germanium detector with sensitivity to ionization and phonon excitation,~\autoref{fig:phonon_spectrum} quantifies the characteristic shape of the expected phonon spectrum that occurs together with the Migdal ionization. 
Given the large background for low-threshold phonon and multiphonon measurements and possible low-energy backgrounds in ionization-only measurement, the combined ionization and phonon signal of a Migdal event provides a potential discriminator from backgrounds. We note also that the kinematics of the combined phonon+Migdal events differ from the case when phonons are produced in a standard DM-nucleus recoil event without the accompanying Migdal electron; therefore, one could expect the phonon spectrum to be different in Migdal versus non-Migdal events. In \autoref{app:NuclearPhononSpectrum}, we show that for a given operator this is indeed the case for larger DM masses (e.g., 100~MeV), although for lower DM masses (e.g., 10~MeV), the spectra only differ marginally.

The differential ionization spectrum, $dR/d\omega$, which is integrated over phonon energy, is insensitive to the computational method used to model the dynamic structure factor~\cite{Berghaus2023}. We therefore take the simplest form, modeled by the free-ion approximation, $S (\boldsymbol{q}, E) \propto \delta (E-E_r)$, where $E_r$ is the nucleon recoil energy, such that the structure factor can be trivially integrated analytically, and we arrive at
\begin{equation} \label{eqn:differentialRate}
\frac{dR}{d\omega}(\omega) = \frac{n_\chi N_T}{m_N M_T} \int_{q_{\text{-}}}^{q_{\text{+}}}dq \int_{v_{\text{-}}}^{v_{\text{+}}} d^3v v f_\chi \left( v \right) \frac{d\sigma}{dq} \left(q \right) \frac{dP}{d\omega} \left( q,\omega \right) .
\end{equation}
Here we have defined the electron ionization (or shake-off) probability,
\begin{equation}
\frac{dP}{d\omega}(q,\omega) = \frac{2\alpha q^2}{3 \omega^4 m_N} \int \frac{d^3 \boldsymbol{k}_e}{(2\pi)^3} \sum_K Z^2(|\boldsymbol{k}_e + \boldsymbol{K}|) \text{Im}\left( -\epsilon^{-1}_{\boldsymbol{K},\boldsymbol{K}}(\boldsymbol{k}_e, \omega) \right) \: , 
\end{equation}
and the differential DM-nucleus cross section, 
\begin{equation} \label{eqn:differential_xc}
\frac{d\sigma}{dq}(q, v) = \frac{q}{\mu_{\chi n}^2 v^2} \sum_{i,j}^{11} 
\overline{\sigma}^{(i)}_n F_{i,j}^{(n,n)}(q,v) \: .  
\end{equation}
Again, we use the dielectric function from DarkELF~\cite{DarkELF2021_dielectric}. Importantly, the in-medium screening effects, which are included in the dielectric function, suppresses the scattering rate, especially for materials with small bandgaps such as germanium. We note that the dielectric function from DarkELF is only modeled up to $q\sim 6\alpha m_e$, but due to the small momentum transfer to the electrons in the Migdal effect, this is more than sufficient. 

\begin{figure*}[h!]
    \centering
    \begin{subfigure}[b]{\textwidth}
        \centering
        \includegraphics[width=0.5 \textwidth]{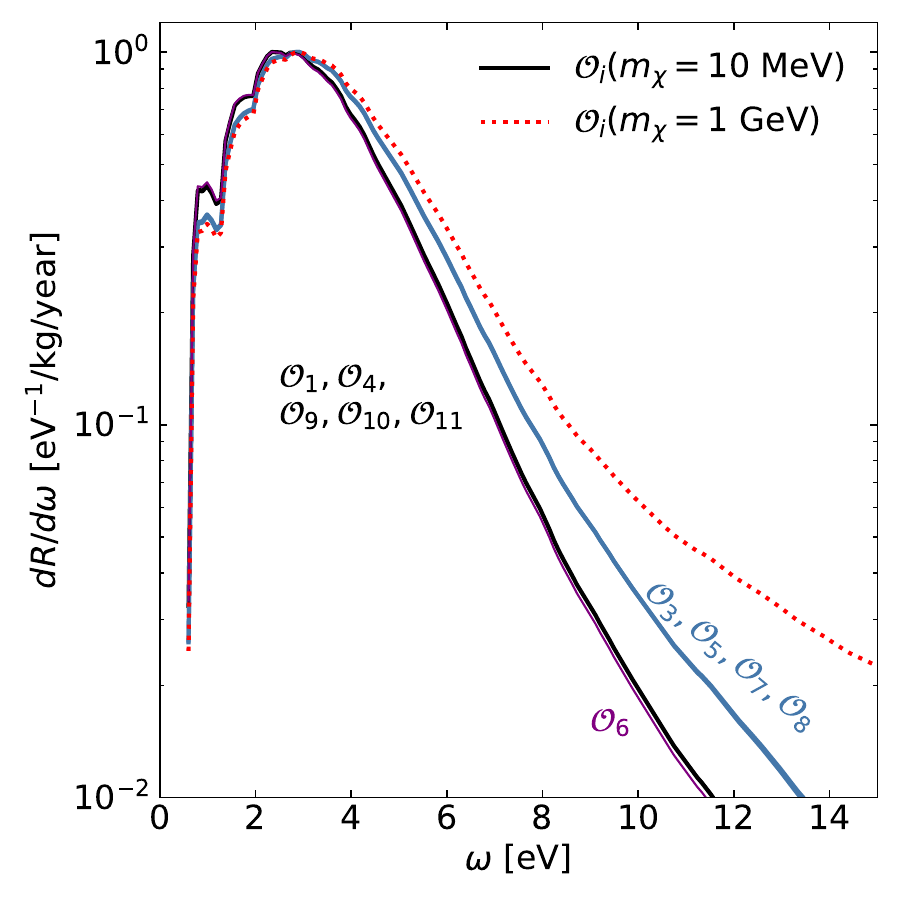}
    \end{subfigure}
    \caption[ ]
    {DM-Migdal ionization spectrum for a germanium detector containing 7.7\% Ge-73, for DM with spin $j_\chi = 1/2$ and for a DM mass of $m_\chi = 10$~MeV (in black, blue, and purple ) and $m_\chi = 1$~GeV (in red). Each curve is normalized to a maximum value of 1~event/(kg-yr-eV), and represents the differential rate with respect to the ionization energy for the reference cross sections given in \autoref{tab:ionization_scale}. For heavier DM masses, the ionization spectrum for each operator \textit{approximately} shares the same characteristic shape. %We detail the reference cross section for each operator in~\autoref{tab:ionization_scale}.
    \label{fig:rate_spectrum}
    } 
\end{figure*}

In~\autoref{fig:rate_spectrum}, we plot~\eqautoref{eqn:differentialRate} for all ten operators  for an EDELWEISS-type germanium detector consisting of 7.7\% Ge-73 for a DM spin of $j_\chi = 1/2$ and a DM mass of $m_\chi = 10$~MeV. Directly from~\eqautoref{eqn:differentialRate} we see that the nuclear form factors are separable from the electron ionization probability. This is a direct consequence of the factorization of the electron-lattice Hamiltonian and DM-lattice Hamiltonian in the leading order term of~\eqautoref{eqn:H_eff}. Therefore, the dominant features of the shape are dictated by the electron loss function. The ionization spectrum peaks around the pair-creation energy---the band gap for germanium is $E_{\rm{gap}} = 0.67$~eV, and the average energy needed to excite additional electron-hole pairs (the pair-creation energy) is $2.9$~eV. Moreover, as the ionization energy increases, the spectrum falls off as $\sim 1/\omega^4$. However, we can see some small difference in the dependence on $\omega$ between non-relativistic operators with different momentum and velocity dependence. Namely, operator form factors depending on $(v^2 - q^2/(4\mu^2_{\chi N}))$ have different structure compared to momentum-independent or $q^2$ and $q^4$ dependent form factors. 
Although the nuclear form factors $F_{i,j}^{n,n'}(q,v)$ do not directly depend on $\omega$, the integration bounds over the DM momentum and velocity depend on the ionization energy. At large DM mass, e.g.~for $\sim 1$~GeV, these minor effects become negligible and all operators share the same characteristic ionization spectrum with different normalization due to the integral over the nuclear form factors (see dotted-gray line in~\autoref{fig:rate_spectrum}).

\begin{table}[h!]
\centering
\begin{tabular}{l|l|l}
Operator     & $\overline{\sigma}_n$ [$\rm{cm^2}$] ($m_\chi =10 \, \rm{MeV}$) & $\overline{\sigma}_n$ [$\rm{cm^2}$] ($m_\chi =1 \,\rm{GeV}$) \\ \cline{1-3}
$\mathcal{O}_1 = \boldsymbol{\mathbb{1}}$                     & $2.3 \times 10^{-38}$   & $4.2 \times 10^{-41}$                                       \\
$\mathcal{O}_3 = i \boldsymbol{S_n} \cdot \left( \frac{\boldsymbol{q}}{m_n} \times \boldsymbol{v^\perp} \right) $                                       & $2.4 \times 10^{-17}$                                        & $2.5 \times 10^{-26}$                                      \\
$\mathcal{O}_4 = \boldsymbol{S_\chi} \cdot \boldsymbol{S_n} $                                                                                           & $2.3 \times 10^{-32}$                                         & $4.2 \times 10^{-35}$                                       \\
$\mathcal{O}_5 = i \boldsymbol{S_\chi} \cdot \left( \frac{\boldsymbol{q}}{m_n} \times \boldsymbol{v^\perp} \right) $                                    & $5.3 \times 10^{-21}$                                         & $1.3 \times 10^{-27}$                                       \\
$\mathcal{O}_6 = \left( \boldsymbol{S_\chi} \cdot \frac{\boldsymbol{q}}{m_n} \right) \left( \boldsymbol{S_n} \cdot \frac{\boldsymbol{q}}{m_n} \right) $ & $9.9 \times 10^{-14}$                                         & $1.8 \times 10^{-24}$                                       \\
$\mathcal{O}_7 = \boldsymbol{S_n} \cdot \boldsymbol{v^\perp} $                                                                                          & $1.6 \times 10^{-26}$                                         & $3.9 \times 10^{-29}$                                       \\
$\mathcal{O}_8 = \boldsymbol{S_\chi} \cdot \boldsymbol{v^\perp} $                                                                                       & $3.5 \times 10^{-30}$                                         & $8.3 \times 10^{-33}$                                     \\
$\mathcal{O}_9 = i \boldsymbol{S_\chi} \cdot \left( \boldsymbol{S_n} \times \frac{\boldsymbol{q}}{m_n} \right)$                                         & $4.8 \times 10^{-23}$                                         & $8.8 \times 10^{-30}$                                       \\
$\mathcal{O}_{10} =  i \left( \boldsymbol{S_n} \cdot \frac{\boldsymbol{q}}{m_n} \right) $                                                               & $2.4 \times 10^{-23}$                                         & $4.4 \times 10^{-30}$                                      \\
$\mathcal{O}_{11} = i \left( \boldsymbol{S_\chi} \cdot \frac{\boldsymbol{q}}{m_n} \right) $                                                             & $5.3 \times 10^{-27}$                                         & $9.6 \times 10^{-34}$                                      
\end{tabular}
\caption{Choice of reference cross section for the ionization spectrum of each operator in~\autoref{fig:rate_spectrum} at two different DM masses. We find that there is a common characteristic shape for operators that have the same momentum and velocity dependence, normalized with a different reference cross section. 
%At high DM mass (i.e., $1\,$GeV) all operators have the same \textit{approximate} shape normalized to a different reference cross section. 
Note, for $\mathcal{O}_{1}$ we assume 100\% of the detector material can interact (rather than 7.7\% for Ge-73) and include the contribution from protons. \label{tab:ionization_scale}}
\end{table}

\subsection{Direct Detection Projections and Constraints}
\begin{figure*} [t!]
    \centering
    \begin{subfigure}[b]{0.49\textwidth}
        \centering
        \includegraphics[width=\textwidth]{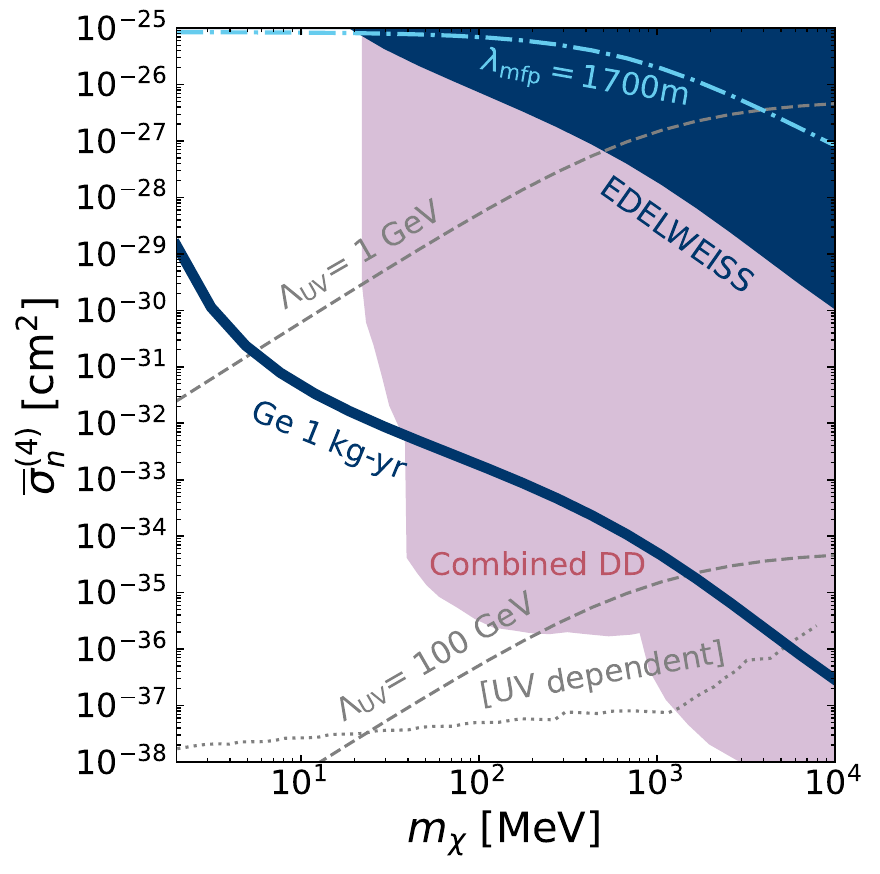}
    \end{subfigure}
    \vskip  \baselineskip
    \begin{subfigure}[b]{0.49\textwidth}   
        \centering 
        \includegraphics[width=\textwidth]{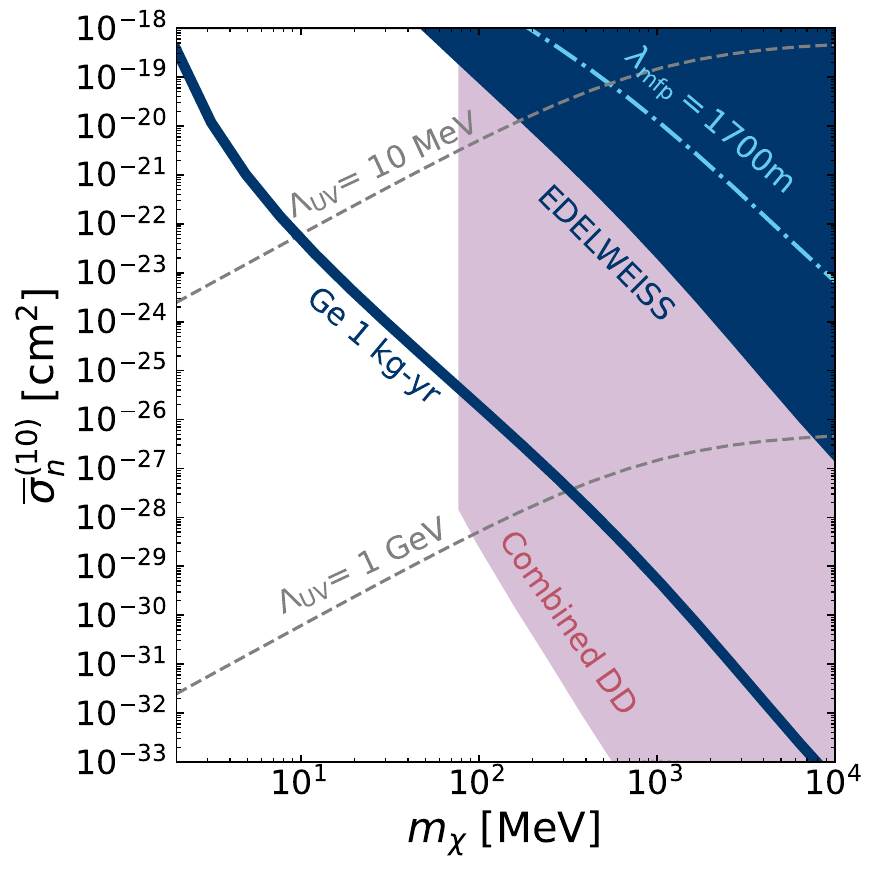}
    \end{subfigure}
    \hfill
    \begin{subfigure}[b]{0.49\textwidth}  
        \centering 
        \includegraphics[width=\textwidth]{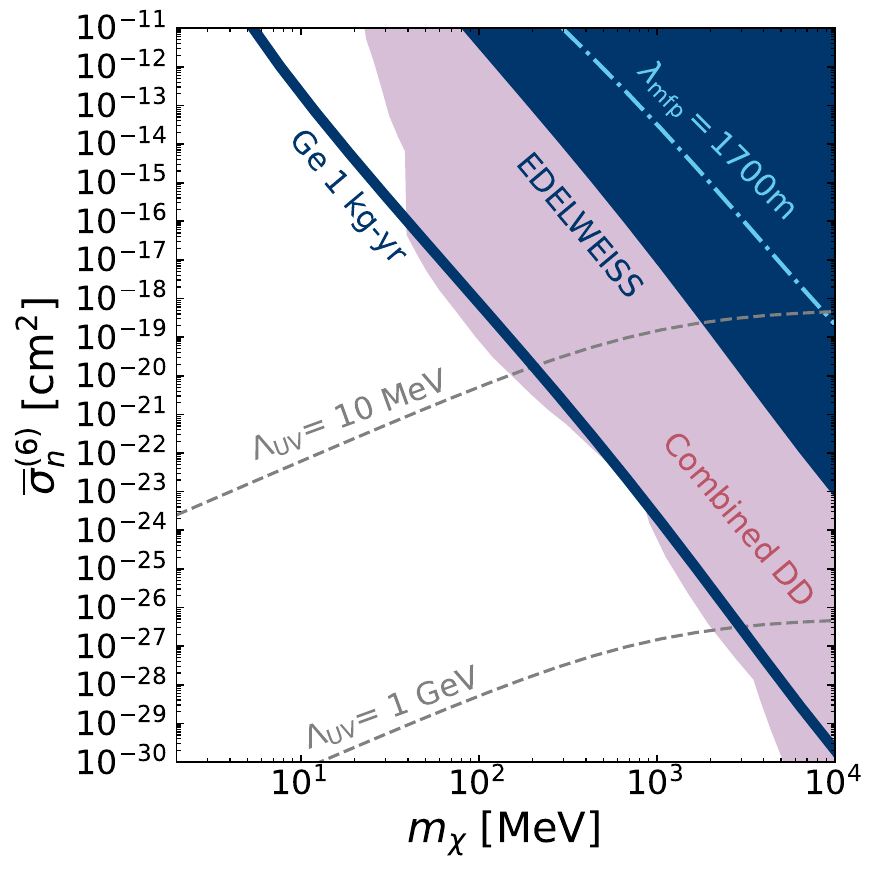}
    \end{subfigure}
    \caption[]
    {We report the projected sensitivity for a 1 kg-year germanium detector (with 7.7\% Ge-73) as a solid dark-blue line for $\mathcal{O}_4$, $\mathcal{O}_{6}$, and $\mathcal{O}_{10}$. Additionally, re-casted EDELWEISS direct detection constraints using data from~\cite{EDELWEISS2020} are shaded in dark-blue. In dot-dashed light-blue, we show the mean free path, above which DM will scatter at least once when traversing 1700~m of the Earth's crust. We include combined direct detection constraints from~\cite{Kang:2018rad} and~\cite{TOMAR2023102851}, which we have re-casted to a DM-nucleon cross section, in pink. Each figure also contains the estimated EFT validity bound in dashed gray, which are projected from the relativistic couplings (see text for details). Work from~\cite{Ramani2019}, which is shown in dotted gray for $\mathcal{O}_4$ and is below the scale of the plot for $\mathcal{O}_6$ and $\mathcal{O}_{10}$, suggests the parameter space shown for these three operators is disfavored due to constraints on heavy mediator couplings to the SM.} 
    \label{fig:ProjectionCrossSection_n}
\end{figure*}

\begin{figure*} [h!]
    \centering
    \begin{subfigure}[b]{0.49\textwidth}   
        \centering 
        \includegraphics[width=\textwidth]{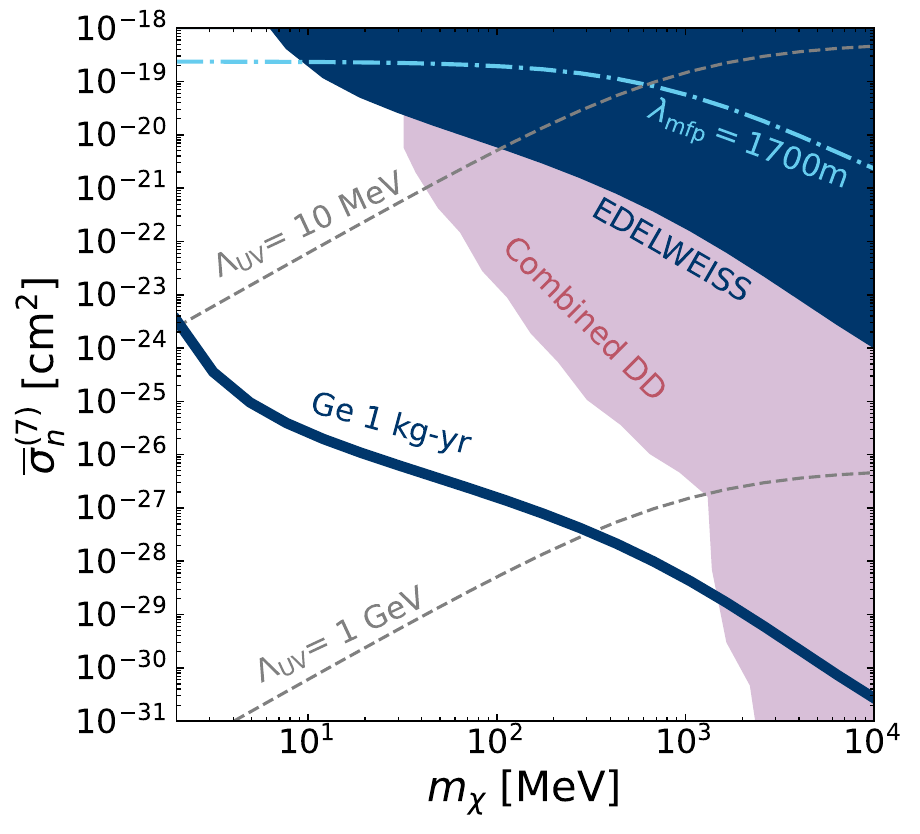}
    \end{subfigure}
    \hfill
    \begin{subfigure}[b]{0.49\textwidth}   
        \centering 
        \includegraphics[width=\textwidth]{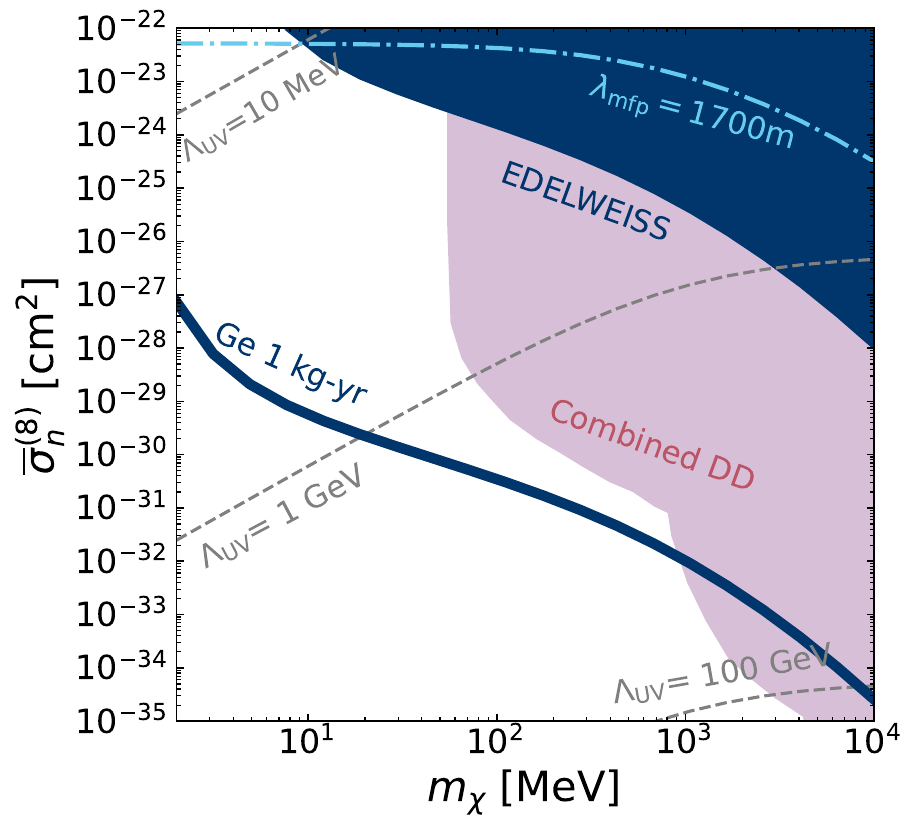}
    \end{subfigure}
    \begin{subfigure}[b]{0.49\textwidth}   
        \centering 
        \includegraphics[width=\textwidth]{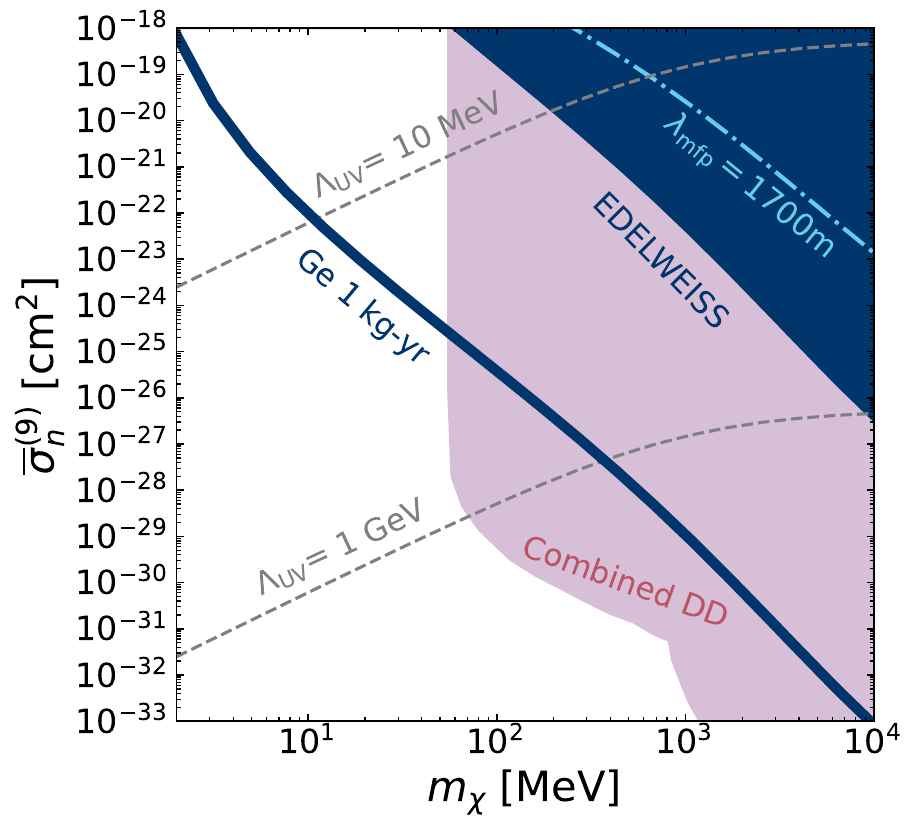}
    \end{subfigure}
    \hfill
    \begin{subfigure}[b]{0.49\textwidth}   
        \centering 
        \includegraphics[width=\textwidth]{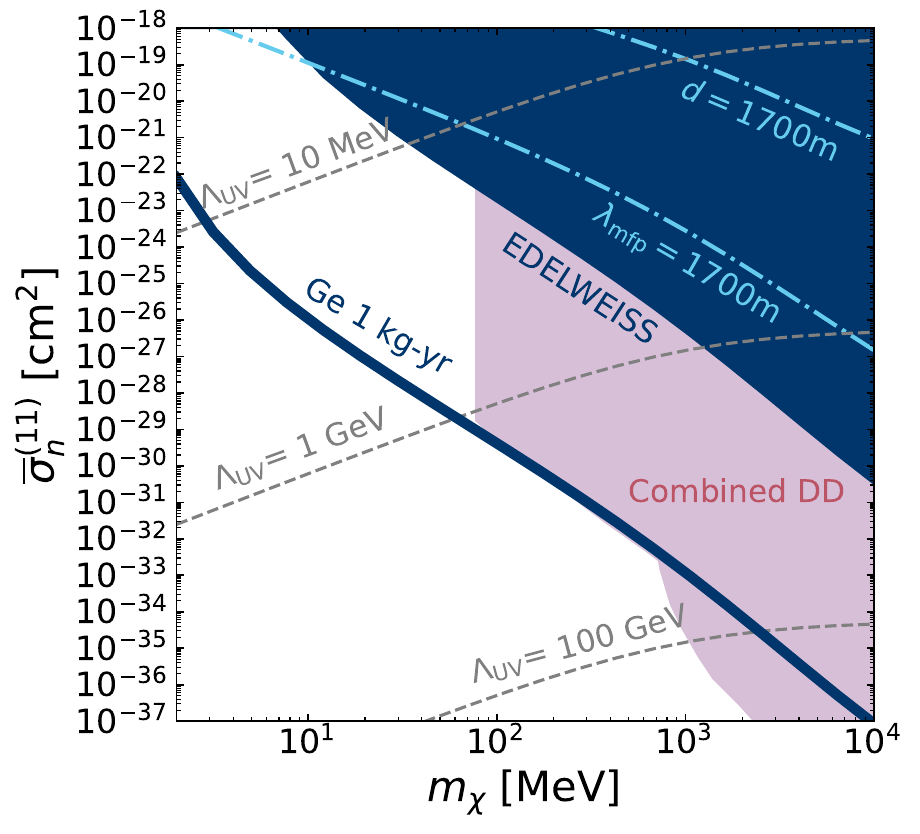}
    \end{subfigure}
    \caption[]
    {We report the projected sensitivity for a 1 kg-year germanium detector (with 7.7\% Ge-73) as a solid dark-blue line for $\mathcal{O}_7$, $\mathcal{O}_8$, $\mathcal{O}_9$, and $\mathcal{O}_{11}$. Additionally, re-casted EDELWEISS direct detection constraints using data from~\cite{EDELWEISS2020} are shaded in dark-blue. In dot-dashed light-blue, we show the mean free path, above which DM will scatter at least once when traversing 1700~m of Earth's crust. Additionally, we include combined direct detection constraints from~\cite{Kang:2018rad} and~\cite{TOMAR2023102851}, which we have re-casted to a DM-nucleon cross section, in pink. Each figure also contains the estimated EFT validity bound, which are projected from the relativistic couplings (see text for details).}
    \label{fig:ProjectionCrossSection_v2_n}
\end{figure*}

\begin{figure*}[h!]
    \centering
    \begin{subfigure}[b]{0.49\textwidth}   
        \centering 
        \includegraphics[width=\textwidth]{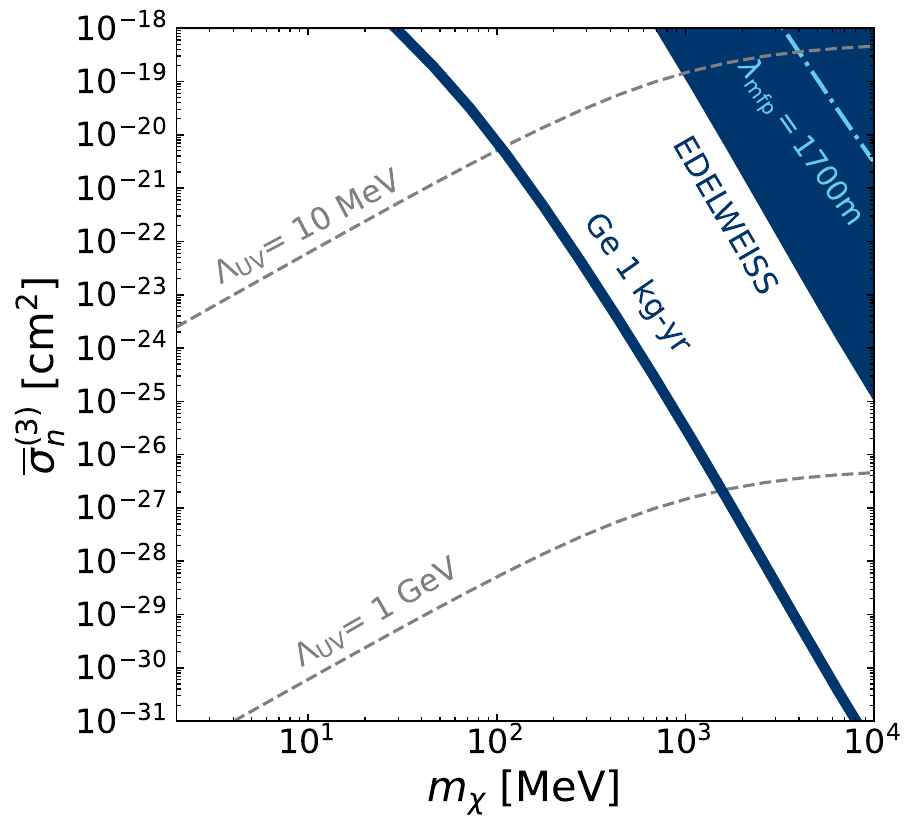}
    \end{subfigure}
    \hfill
    \begin{subfigure}[b]{0.49\textwidth}   
        \centering 
        \includegraphics[width=\textwidth]{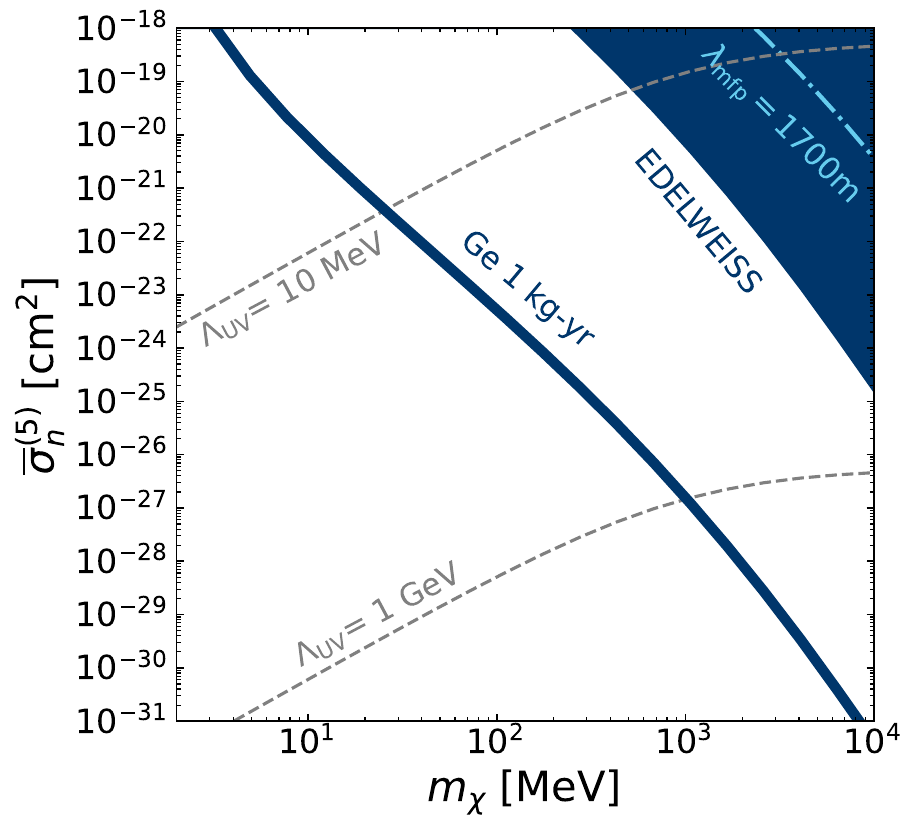}
    \end{subfigure}
    \caption[]
    {We report the projected sensitivity for a 1 kg-year germanium detector (with 7.7\% Ge-73) as a solid dark-blue line for $\mathcal{O}_3$, $\mathcal{O}_5$. Additionally, re-casted EDELWEISS direct detection constraints using data from~\cite{EDELWEISS2020} are shaded in dark-blue. In dot-dashed light-blue, we show the mean free path, above which DM will scatter at least once when traversing 1700~m of Earth's crust. Each figure also contains the estimated EFT validity bound, which are projected from the relativistic couplings (see text for details).} 
    \label{fig:ProjectionCrossSection_v4_n}
\end{figure*}

We recast the DM-electron scattering limits from EDELWEISS~\cite{EDELWEISS2020} to calculate the first DM-Migdal limits for all ten non-relativistic operators given in~\autoref{tab:operators}. We also include projections on DM-Migdal cross sections for neutron couplings for a hypothetical germanium detector (containing 7.7\%~of Ge-73) with 1-kg year exposure, a possible next-generation target. We take the standard choice of setting the DM spin to $j_\chi = 1/2$, and present the results for each of the non-relativistic operators. We show in~\autoref{fig:ProjectionCrossSection_n} constraints on $\mathcal{O}_4$, $\mathcal{O}_{6}$, and $\mathcal{O}_{10}$, which are most often presented as an exemplary sample of velocity-independent, double-velocity dependent, and four-power velocity dependent cross sections.
We show the remaining double-velocity dependent interactions in~\autoref{fig:ProjectionCrossSection_v2_n} and the remaining operators in~\autoref{fig:ProjectionCrossSection_v4_n}. For each of these figures, we show the projection in solid dark-blue line, and the recasted EDELWEISS constraints are shaded in the same dark-blue. 

We have also re-cast Migdal constraints on DM-quark and DM-gluon operators from~\cite{TOMAR2023102851}, which used data from XENON1T~\cite{XENON:2019zpr}, SuperCDMS~\cite{SuperCDMS:2022kgp}, COSINE-100~\cite{COSINE-100:2021poy}, and DarkSide-50~\cite{DarkSide:2022dhx}, to all possible non-relativistic operators. We note that the Migdal constraints from SuperCDMS (which uses a germanium detector) use the free atom approximation rather than the results derived here; however, this approximation is reasonable for the large DM masses being probed in~\cite{SuperCDMS:2022kgp}. Similarly, we re-cast nuclear recoil constraints from~\cite{Kang:2018rad} which used data from XENON1T~\cite{XENON:2018voc}, CDMSlite~\cite{CDMS2018}, SuperCDMS~\cite{SuperCDMS2018}, PICASSO~\cite{PICASSO2017}, PICO-60~\cite{PICO2016, PICO2019}, CRESST-II~\cite{CRESST_II_2016, CRESST_II_data}, DAMA~\cite{DAMA2008}, and DarkSide-50~\cite{DarkSide2018}. We show the combined most stringent bound shaded in pink. To convert from DM-quark and DM-gluon coefficients to DM-nucleon coefficients, we use the convention in~\cite{Bishara:2017pfq}. Note, there is no mapping to the DM-nucleon coefficients for $\mathcal{O}_3$, and the DM-neutron coefficient maps to explicitly zero for $\mathcal{O}_5$ because the neutron charge, $Q_n=0$. 

We highlight that the germanium semiconductor detector performs most competitively at low DM mass, where careful treatment of lattice degrees of freedom is required. For $\mathcal{O}_7$, $\mathcal{O}_8$, $\mathcal{O}_{10}$, and $\mathcal{O}_{11}$, the EDELWEISS constraint presents the most competitive limit, typically for masses below $\sim 10$~MeV or $100$~MeV, depending on the operator. In the case of the EDELWEISS~\cite{EDELWEISS2020} detector, the exposure was relatively small (a $33.4$~g detector and data was collected for $58$~hours). Although other direct detection experiments (shown in pink) may have larger exposure---for example 1 tonne of xenon or 50 tonnes of argon---they are limited by the energy threshold and larger ionization energy. In the case of germanium, the energy threshold is the size of the band-gap, $\mathcal{O}(0.67\,\rm{eV})$. We also compare current direct detection limits with the capability of future germanium detectors with larger exposure, i.e., 1 kg-year. 

\subsection*{Effects from Dark Matter Scattering in Atmosphere or Earth}

For sufficiently large DM-nucleus scattering cross sections, the DM could scatter appreciably in the Earth and change the derived limits and/or sensitivity of projections. We use two methods to estimate the cross section above which these ``Earth-shielding'' effects should be considered more carefully. More generally, we highlight that Earth-shielding effects may need to be considered in the context of constraining non-relativistic operators with direct detection limits. For the first (conservative) estimate, we calculate the mean free path ($\lambda_{\text{mfp}}$) of the DM scattering through Earth's material, before it reaches an experimental detector. This gives an estimate of the cross section for the DM to scatter once within some given distance. The inverse of the mean free path is 
\begin{equation}
\lambda_{\text{mfp}}^{-1} (v) = n_E \int^{2 m_\chi v}_0 dq \frac{d\sigma} {dq} (q,v)\ , %\sigma(v) 
\end{equation}
where $n_E$ is the number density of a given material in the Earth's atmosphere (i.e., for experiments on the surface) or core-mantle-crust system (i.e., for experiments underground). The total DM-nucleus scattering cross section, $\sigma(v)$, is obtain by integrating~\eqautoref{eqn:differential_xc} over the DM momentum. In the case of velocity- and momentum-independent operators (i.e., $\mathcal{O}_1$ and $\mathcal{O}_4$), this cross section (and therefore the mean free path) is velocity independent; for all other operators, we calculate the mean free path at the reference velocity, $v_0$.
The Earth's atmosphere and its core/crust are comprised of multiple materials with different densities. In the case of spin-independent scattering through $\mathcal{O}_1$, the Earth's atmosphere can be well estimated as $\sim 5$ meters of concrete or equivalently $\sim 2$ meters of iron shielding~\cite{Emken:2019tni}. Additionally, the Earth's crust and core can be estimated using a compositions of multiple materials with different densities (for example~\cite{Chen2023} and~\cite{CDEX2022}). In order to study the shielding effects for all ten non-relativistic operators, one must consider DM scattering with a material that has non-zero spin; as an example, N-14 in the Earth's atmosphere, Fe-57 in Earth's core or Al-27 in Earth's crust.\footnote{Nuclear response form factors for N-14 and Al-27 can be found in reference~\cite{Catena_2015}.} For our calculation, we are interested in approximating the shielding due to an EDELWEISS-like experiment, which is approximately $1700\,$m underground. To do so, we estimate the effects of $1700~\rm{m}$ of shielding from a material with non-zero spin and number density, $n_E = 10^{21}~\rm{cm}^{-3}$, similar to that of Fe-57 in the Earth's core (quite dense). The choice of this dense material leads to a more conservative bound on the cross section; we note, that the bound scales inversely proportional to the density. Instead of using a multi-material Earth model, we use only Ge-73 since the form factors for germanium were readily available. In \autoref{fig:ProjectionCrossSection_n}, \autoref{fig:ProjectionCrossSection_v2_n}, and \autoref{fig:ProjectionCrossSection_v4_n} we show a conservative estimate of shielding for a theoretical  underground detector---above the light blue dot-dashed line in these figures, shielding effects must be better estimated to ensure dark matter with large cross sections are able to reach such a detector. 

For a theoretical 1 kg-yr underground germanium detector, the projected reach is far beyond the shielding bound from the mean free path method. However, in the case of $\mathcal{O}_{7}$, $\mathcal{O}_{8}$, and $\mathcal{O}_{11}$, this bound intersects with our re-casted results from the underground EDELWEISS experiment. A better estimate of shielding must be done in these cases. Without using a multi-material Earth model to estimate shielding for this specific experiment, we can perform a better estimate for the same scenario ($1700~\rm{m}$ of shielding from $n_E = 10^{21}~\rm{cm}^{-3}$ of Ge-73). Dark matter that interacts with the Earth material will lose energy but still reach the detector. We calculate the multiple-scattering depth~\cite{Emken:2019tni},
\begin{equation}
d = \int^{E_{d}}_{E_0} dE_\chi \left[ \frac{d \langle E_\chi \rangle}{dx} \right] ^{-1} , 
\end{equation}
where $E_0 = \frac{1}{2}m_\chi v_{\text{esc}}^2$ is the initial kinetic DM energy, and $E_d$ is set by the detector threshold,  approximately $1$~eV for germanium detectors. Furthermore, we have defined the nuclear stopping power,
\begin{equation}
\frac{d \langle E_\chi \rangle}{dx} = n_E \int_0^{2 m_\chi v} dq \frac{q^2}{2 m_E} \frac{d \sigma}{dq} (q,v) \ ,
\end{equation}
where $m_E$ is the mass of the Earth material nucleus. For all non-relativistic operators, the penetration distance gives a looser bound for Earth-shielding. In all cases  other than $\mathcal{O}_{11}$, the multiple-scattering depth suggests Earth shielding effects are important to consider for cross sections that are larger than the range shown in~\autoref{fig:ProjectionCrossSection_n}, \autoref{fig:ProjectionCrossSection_v2_n}, and~\autoref{fig:ProjectionCrossSection_v4_n}. Moreover, these analytical methods can only estimate where Earth shielding effects may become relevant and suggest when a more comprehensive study should be performed. Additionally, for a targeted analysis, each experimental constraint requires its own shielding bound depending on the location underground and the overburden of the given detector. For $\mathcal{O}_1$ and $\mathcal{O}_4$, constraints from collaborations (for example CDEX~\cite{CDEX2022} and XENON1T~\cite{XENON:2019zpr}) include detailed Earth shielding analysis. Additionally, atmospheric shielding was calculated for the EDELWEISS surface experiment~\cite{EDELWEISS_surface} (note, however, that we have analyzed data from the 1700m underground experiment performed in~\cite{EDELWEISS2020}). 

\subsection*{Constraints from the UV}
In this work, we have taken a bottom-up approach, and considered Migdal signatures due to all possible non-relativistic dimension-six EFT operators. Nevertheless, theoretical consistency demands that the interactions arise from a UV complete theory that feature mediators, which are integrated out in the EFT. In~\autoref{fig:ProjectionCrossSection_n}, \autoref{fig:ProjectionCrossSection_v2_n}, and~\autoref{fig:ProjectionCrossSection_v4_n}, we indicate the bounds above which the EFT is not valid for different choices of the UV scale $\Lambda_{\text{UV}} = \{10~\textrm{MeV}, 1~\textrm{GeV}, 100~\textrm{GeV} \}$ in dashed gray, where we impose $c^{(n)}_i < (4 \pi / \Lambda_{\text{UV}}^2)$. 

Other possible UV models, which involve combinations of non-relativistic operators are considered in~\cite{GreshamArXiv}. Additionally, there are direct constraints on pseudoscalar (axial-vector) mediators couplings to the SM and to DM, which are relevant for operators $\mathcal{O}_6$ and $\mathcal{O}_{10}$ ($\mathcal{O}_4$). The SM coupling constraints arise from meson decays, supernova cooling, nuclear decays, and rare $Z$ decays~\cite{BSM_exchanges2017, Ramani2019,Gori:2025jzu}. The most stringent constraints on the DM couplings arise from the bullet cluster~\cite{Ramani2019,Gori:2025jzu}, and are relaxed if only a sub-dominant component of DM interacts via the mediator. These constraints are very strong and disfavor the part of parameter space accessible to direct detection experiments for operators $\mathcal{O}_6$, $\mathcal{O}_{10}$, and $\mathcal{O}_4$. A careful analysis of the constraints on the remaining set of non-relativistic operators would be welcome. In~\autoref{fig:ProjectionCrossSection_n}, we indicate the limits derived in~\cite{Ramani2019} for $\mathcal{O}_4$, noting that more stringent limits have been found recently in~\cite{Gori:2025jzu} for even smaller cross section, which are outside of our shown plot range for the applicable operators. 

\section{Summary and discussion} \label{sec:summary}
As the WIMP DM parameter space becomes more constrained, interest in pushing towards lower DM masses ($\sim 1$~MeV) has required the study of new signals, with important ones being DM-electron and DM-nucleus scattering via the Migdal effect. Although a generalized descriptions of DM-electron~\cite{Liang:2024ecw,Krnjaic:2024bdd} and DM-nucleus~\cite{Fan2010, Fitzpatrick:2013, Anand2013, Griffin2021} interactions have been explored in the literature, there have been fewer comprehensive studies on the general Migdal effect. Additionally, the Migdal effect in semiconductors is computationally more challenging than the Migdal effect in free atoms due to the presence of a lattice potential. 

To derive the generalized DM-Migdal interaction in a crystal, we exploit two separation of scales and construct an effective field theory using non-relativistic DM-nucleon operators. First, we maintain the regime of small DM momentum transfer to the lattice site compared to the UV-cutoff ($q \ll \Lambda$). Second, we utilize the results of~\cite{Berghaus2023}, which highlight the energy separation between the $\mathcal{O}(\rm{eV})$ scale ionized electron and the energy associated with the lattice mode that mediates the Migdal effect, which is $\ll$eV. Using these methods, we are able to probe down to $\sim \rm{MeV}$ DM masses, which corresponds to the lowest energy able to ionize an electron in a semiconductor detector. We show that the resulting rate to leading order factorizes into nuclear, electronic, and vibrational contributions for all operators, generalizing the factorization found in the spin-independent case \cite{Berghaus2023}.

Within this work, we specifically focus on germanium targets. However, similar methodology can be applied to other materials for which the nucleus form factors and electron loss functions have been calculated. Nucleus form factors for a wide set of materials have been calculated in, for example,~\cite{Fitzpatrick:2013} and ~\cite{Catena_2015}. Relevant materials have an unpaired proton or neutron, such that there is a non-zero spin to which the DM spin can couple. Contained within these references are, for example, hydrogen (H-1), helium (He-3), floride (F-19), sodium (Na-23), aluminum (Al-27),  nickel (Ni-59), germanium (Ge-73), iodine (I-127), xenon (Xe-129 and Xe-131). Separately, electron loss functions for crystals such as aluminum (Al), aluminum oxide (Al\textsubscript{2}O\textsubscript{3}), gallium arsenide (GaAs), gallium nitride (GaN), germanium, silicon, silicon dioxide (SiO\textsubscript{2}), and zinc sulfide (ZnS) have been carefully studied using different computational methods by~\cite{darkELF2022, Singal2023, EXCEED_DM} as well as other unconventional materials in~\cite{Hochberg:2025dom}. Comparing the two lists, only Ge-73 and Al-27 is included in both; we therefore highlight Ge-73 which is currently used in direct detection experiments, such as EDELWEISS, CDEX, and SuperCDMS.

We summarize constraints from previous analyses as well as present new constraints from a re-analysis of the EDELWEISS collaboration. We also show projections for a next generation experiment with 1 kg-year exposure. Furthermore, we estimate the cross sections above which ``Earth shielding'' effects are important; in particular, we present the cross sections for each operator where the DM scatters in the Earth's crust (1,700~m of it) at least once before reaching a detector. Detailed simulations would need to be employed to study this more carefully based on a particular experimental setup. 

An important question is which of these operators provide UV completions that are best probed with direct detection experiments, rather than with other data.  In this work, we only showed bounds above which the EFT is not valid for different choices of the UV scale.  A detailed study of UV completions would be useful. 

We highlight that the reanalyzed limits using data from~\cite{EDELWEISS2020} surpass existing direct detection limits at low masses for $\mathcal{O}_{7}$, $\mathcal{O}_{8}$, $\mathcal{O}_{10}$, and $\mathcal{O}_{11}$. In the case of $\mathcal{O}_{3}$ and $\mathcal{O}_{5}$, the constraint shown in this work is the only current direct detection limit. Furthermore, next-generation detectors have competitive reach at low masses for all operators. 

\acknowledgments
We thank Angelo Esposito, Megan Hott, and Aman Singal for useful discussions. KB is funded by the Deutsche Forschungsgemeinschaft (DFG, German Research Foundation) through the Emmy Noether Programme Project No.~548044346.
KB thanks the U.S.~Department of Energy, Office of Science, Office of High Energy Physics, under Award Number DE-SC0011632, and the Walter Burke Institute for Theoretical Physics. 
RE and MHM acknowledge support from DOE Grant DE-SC0025309 and Simons Investigator in Physics Awards 623940 and MPS-SIP-00010469.

\appendix
\renewcommand{\subsectionautorefname}{App.}
\section{Definitions and Identities} \label{app:definitions}

\subsection{Effective Dark Matter-nucleon interaction coefficients} \label{app:l_coefs}
We present the effective field theory coefficients, first introduced in the EFT Hamiltonian in~\eqautoref{eqn:DMNucleusHamiltonian}, in terms of the DM-nucleon coefficients which are relevant to this work:
\begin{equation}
    l_0 = c^{(n)}_1 + i(\frac{\boldsymbol{q}}{m_n} \times \boldsymbol{v^\perp_T}) \cdot \boldsymbol{S}_\chi c^{(n)}_5 + (\boldsymbol{v^\perp_T} \cdot \boldsymbol{S}_\chi)c^{(n)}_8 + i (\frac{\boldsymbol{q}}{m_n} \cdot \boldsymbol{S}_\chi) c^{(n)}_{11}
\end{equation}
\begin{equation}
\begin{aligned}
    l^A_0 &= - \frac{1}{2} c^{(n)}_7 
\end{aligned}
\end{equation}
\begin{equation}
\begin{aligned}
    \boldsymbol{l_5} &= - \frac{1}{2} \Big[i \Big(\frac{\boldsymbol{q}}{m_n} \times \boldsymbol{v^\perp_T}\Big)c^{(n)}_3 + \boldsymbol{S}_\chi c^{(n)}_4 + \frac{\boldsymbol{q}}{m_n} \Big(\frac{\boldsymbol{q}}{m_n} \cdot \boldsymbol{S}_\chi\Big) c^{(n)}_6 \\
    &\qquad + \boldsymbol{v^\perp_T} c^{(n)}_7 + i \Big(\frac{\boldsymbol{q}}{m_n} \times \boldsymbol{S}_\chi\Big) c^{(n)}_9 + i \frac{\boldsymbol{q}}{m_n} c^{(n)}_{10}\Big]
\end{aligned}
\end{equation}
\begin{equation}
    \boldsymbol{l_M} = i (\frac{\boldsymbol{q}}{m_n} \times \boldsymbol{S}_\chi) c^{(n)}_5 - \boldsymbol{S}_\chi c^{(n)}_8
\end{equation}
\begin{equation}
\begin{aligned}
    \boldsymbol{l_E} &= \frac{1}{2} [ \frac{\boldsymbol{q}}{m_n} c^{(n)}_3] \ . 
\end{aligned}
\end{equation}

\subsection{Spherical harmonic identities }
The standard spherical harmonic identity is
\begin{equation}
    e^{i \boldsymbol{q} \cdot \boldsymbol{x}_n} = \sum^\infty_{J=0} \sqrt{4\pi} \sqrt{2J+1} i^J j_J(qx_n) Y_{J0}(\Omega_{x_n}) \: .
\end{equation}
The vector spherical harmonic identity is
\begin{equation}
\hat{e} e^{i \boldsymbol{q} \cdot \boldsymbol{x}_n} = 
\begin{cases} 
\displaystyle\sum_{J=0}^{\infty} \sqrt{4\pi} \sqrt{2J+1}\, i^{J-1} \frac{\vec{\nabla}_n}{q} j_J(qx_n) Y_{J0}(\Omega_{x_n}) 
& \beta = 0 , \\[1.5em]  
\begin{aligned}
&\displaystyle\sum_{J \geq 1}^{\infty} \sqrt{2\pi} \sqrt{2J+1}\, i^{J-2} \Big[\beta j_J(qx_n) \boldsymbol{Y}^\beta_{JJ0}(\Omega_{x_n})\\
&\qquad\qquad + \frac{\vec{\nabla}_n}{q} \times j_J(qx_n) \boldsymbol{Y}^\beta_{JJ0}(\Omega_{x_n})\Big]
\end{aligned}
& \beta = \pm 1 ,
\end{cases}
\end{equation}
where $\hat{e}$ is a unit vector with $\hat{e}_\lambda \parallel \hat{q}$ ($\boldsymbol{q}$ is the DM momentum transfer) and derivatives $\nabla_n$ are with respect to $\boldsymbol{x}_n$, where $\boldsymbol{x}_n$ is the $n$-th nucleon position in the nucleus center of mass frame and $\Omega_{x_n}$ represents azimuthal and polar angles of $\boldsymbol{x}_n$. 
 
\subsection{Nuclear response functions} \label{subsec:NuclearResponseFunctions}
Each of the response functions are built from the Bessel (vector) spherical harmonic $M_{JM}$ ($\boldsymbol{M^M_{JL}}$): 
\begin{equation}
\begin{aligned}
    M_{JM}(q \boldsymbol{x}_n) & \equiv j_J (q \boldsymbol{x}_n) Y_{JM}(\Omega_{x_n}) \\
    \boldsymbol{M^M_{JL}}(q \boldsymbol{x}_n) & \equiv j_J (q \boldsymbol{x}_n) \boldsymbol{Y_{JLM}}(\Omega_{x_n})
\end{aligned}
\end{equation}
\begin{equation}
\begin{aligned}
    M_{JM}(q) & \equiv \sum^A_{\text{nucleons}} M_{JM} (q \boldsymbol{x}_n) \\
    \Delta_{JM}(q) & \equiv \sum^A_{\text{nucleons}} \boldsymbol{M^M_{JJ}}(q \boldsymbol{x}_n) \cdot \frac{1}{q} \vec{\nabla}_n \\
    \Sigma'_{JM}(q) & \equiv -i \sum^A_{\text{nucleons}} \bigg[\frac{1}{q} \vec{\nabla}_n \times \boldsymbol{M^M_{JJ}}(q \boldsymbol{x}_n) \bigg] \cdot \boldsymbol{\sigma} \\
    \Sigma''_{JM}(q) & \equiv \sum^A_{\text{nucleons}} \bigg[\frac{1}{q} \vec{\nabla}_n M_{JM}(q \boldsymbol{x}_n) \bigg] \cdot \boldsymbol{\sigma} \\
    \tilde{\Phi}'_{JM}(q ) & \equiv \sum^A_{\text{nucleons}} \bigg[\frac{1}{q} \vec{\nabla}_n \times \boldsymbol{M^M_{JJ}}(q \boldsymbol{x}_n) \bigg] \cdot \bigg[\boldsymbol{\sigma} \times \frac{1}{q} \vec{\nabla}_n\bigg] + \frac{1}{2} \boldsymbol{M^M_{JJ}}(q \boldsymbol{x}_n) \cdot \boldsymbol{\sigma} \\
    \Phi''_{JM}(q)) & \equiv i\sum^A_{\text{nucleons}}  \bigg[\frac{1}{q} \vec{\nabla}_n M_{JM}(q \boldsymbol{x}_n) \bigg] \cdot \bigg[\boldsymbol{\sigma} \times \frac{1}{q} \vec{\nabla}_n \bigg] \ .
\end{aligned}
\end{equation}
One of the elastic nuclear response, $\tilde{\Phi}'$, arises only in CP non-conserving interactions, and is not generated by any of the non-relativistic operators we consider. Additionally, there are three form factors that only contribute to \textit{inelastic scattering} off the nucleus. These terms do not contribute to the DM-lattice Hamiltonian because of their parity. More specifically, when summing over nuclear spin ($J$) to calculate the respective form factor, the terms are zero for any/all $J$, 
\begin{equation} \label{eqn:inelasticResponses}
\begin{aligned}
    \Sigma_{JM}(q) & \equiv \boldsymbol{M^M_{JJ}}(q \boldsymbol{x}_n) \cdot \boldsymbol{\sigma} \\
    \Delta'_{JM}(q) & \equiv -i \bigg[\frac{1}{q} \vec{\nabla}_n \times \boldsymbol{M^M_{JJ}}(q \boldsymbol{x}_n) \bigg] \cdot \frac{1}{q} \vec{\nabla}_n \\
    \tilde{\Omega}_{JM}(q) & = M_{JM}(q \boldsymbol{x}_n) \boldsymbol{\sigma} \cdot \frac{1}{q} \vec{\nabla}_n + \frac{1}{2} \bigg[ \frac{1}{q} \vec{\nabla}_n M_{JM}(q \boldsymbol{x}_n) \bigg] \cdot \boldsymbol{\sigma}  \\
    & \equiv \Omega_{JM}(q \boldsymbol{x}_n) + \frac{1}{2}\Sigma'_{JM}(q \boldsymbol{x}_n) \ .
\end{aligned}
\end{equation}

\subsection{DM spin states}
To calculate the squared matrix element, the Wigner-Eckart theorem is used to write the expression in terms of the reduced matrix elements. When dealing with the two three-j symbols, which act as coefficients of the reduced matrix elements, orthogonality conditions arise that limit the final spin states ($j_n' = j_n$ and $j_\chi' = j_\chi$). One must use the following simplifications to evaluate the DM spin states: 
\begin{equation}
\begin{aligned}
&\frac{1}{2j_\chi + 1} \sum_{m_i, m_f} \langle j_\chi m_i| 1 |j_\chi m_f \rangle \cdot \langle j_\chi m_f| 1 |j_\chi m_i \rangle = 1 \\
&\frac{1}{2j_\chi + 1} \sum_{m_i, m_f} \langle j_\chi m_i| \boldsymbol{S_\chi} |j_\chi m_f \rangle \cdot \langle j_\chi m_f| \boldsymbol{S_\chi} |j_\chi m_i \rangle = j_\chi (j_\chi + 1) \\
&\frac{1}{2j_\chi + 1} \sum_{m_i, m_f} \langle j_\chi m_i| \boldsymbol{A} \cdot \boldsymbol{S_\chi} |j_\chi m_f \rangle \cdot \langle j_\chi m_f| \boldsymbol{B} \cdot \boldsymbol{S_\chi} |j_\chi m_i \rangle = \frac{\boldsymbol{A} \cdot \boldsymbol{B}}{3}\, j_\chi(j_\chi + 1) \\
&\frac{1}{2j_\chi + 1} \sum_{m_i, m_f} \langle j_\chi m_i| \boldsymbol{A} \times \boldsymbol{S_\chi} |j_\chi m_f \rangle \cdot \langle j_\chi m_f| \boldsymbol{B} \times \boldsymbol{S_\chi} |j_\chi m_i \rangle = \frac{2\boldsymbol{A} \cdot \boldsymbol{B}}{3}\, j_\chi(j_\chi + 1) \\
&\frac{1}{2j_\chi + 1} \sum_{m_i, m_f} \langle j_\chi m_i| \boldsymbol{A} \times \boldsymbol{S_\chi} |j_\chi m_f \rangle \cdot \langle j_\chi m_f| \boldsymbol{S_\chi} |j_\chi m_i \rangle = 0 \ .
\end{aligned} 
\end{equation}

\subsection{Nuclear response form factors}
We use the same notation as~\cite{Fitzpatrick:2013} to define  $F^{(n,n')}_{X}$ where $X$ are the 6 nuclear responses:
\begin{equation}
\begin{aligned}
F^{(n,n')}_M (q^2) &= \frac{4\pi}{2j_n+1} \sum_{J=0,2,...} \langle j_n || M^{(n)}_J (q)|| j_n \rangle \langle j_n || M^{(n')}_J (q)|| j_n \rangle \\
F^{(n,n')}_{\Sigma'}(q^2) &= \frac{4\pi}{2j_n+1} \sum_{J=1,3,...} \langle j_n || {\Sigma'}^{(n)}_J (q)|| j_n \rangle \langle j_n || {\Sigma'}^{(n')}_J (q)|| j_n \rangle \\
F^{(n,n')}_{\Sigma''}(q^2) &= \frac{4\pi}{2j_n+1} \sum_{J=1,3,...} \langle j_n || {\Sigma''}^{(n)}_J (q)|| j_n \rangle \langle j_n || {\Sigma''}^{(n')}_J (q)|| j_n \rangle \\
F^{(n,n')}_{\Delta}(q^2) &= \frac{4\pi}{2j_n+1} \sum_{J=0,2,...} \langle j_n || {\Delta}^{(n)}_J (q)|| j_n \rangle \langle j_n || {\Delta}^{(n')}_J (q)|| j_n \rangle \\
F^{(n,n')}_{\Phi''}(q^2) &= \frac{4\pi}{2j_n+1} \sum_{J=0,2,...} \langle j_n || {\Phi''}^{(n)}_J (q)|| j_n \rangle \langle j_n || {\Phi''}^{(n')}_J (q)|| j_n \rangle \\
F^{(n,n')}_{\tilde{\Phi}'}(q^2) &= \frac{4\pi}{2j_n+1} \sum_{J=0,2,...} \langle j_n || {\tilde{\Phi'}}^{(n)}_J (q)|| j_n \rangle \langle j_n || {\tilde{\Phi'}}^{(n')}_J (q)|| j_n \rangle \\
F^{(n,n')}_{M, \Phi''}(q^2) &= \frac{4\pi}{2j_n+1} \sum_{J=0,2,...} \langle j_n || M^{(n)}_J (q)|| j_n \rangle \langle j_n || {\Phi''}^{(n')}_J (q)|| j_n \rangle \\
F^{(n,n')}_{\Sigma', \Delta}(q^2) &= \frac{4\pi}{2j_n+1} \sum_{J=1,3,...} \langle j_n || {\Sigma'}^{(n)}_J (q)|| j_n \rangle \langle j_n || {\Delta}^{(n')}_J (q)|| j_n \rangle  \ .
\end{aligned} 
\end{equation}
The angular momentum (spin) nuclear response form factor(s) defined at zero momentum transfer $F_{\Delta}(0)$ ($F_{\Sigma '}(0)$ and $F_{\Sigma ''}(0)$) relate to the nucleon angular momentum (spin) expectation value. These expectation values have been calculated using numerous computation methods for other applications and may also be compared to experimental values. As pointed out in~\cite{Gresham2014}, the spin and angular momentum expectation value for germanium that are used by references~\cite{Anand2013, Fitzpatrick:2013} differs from what is used in other literature. Specifically, the nuclear response form factors from Fitzpatrick, et al.~\cite{Fitzpatrick:2013} at zero momentum transfer reproduces $\langle S_n \rangle = 0.475$ and $\langle S_p \rangle = 0.006$ for the neutron and proton spin, as well as $\langle L_n \rangle = 3.832$ and $\langle L_p \rangle = 0.184$ for the neutron and proton angular momentum. In a comprehensive study of state-of-the-art nuclear-shell model calculations, the consequences of including 2-body current effects are examined~\cite{Klos2013, Tovey200017}. This is a sub-leading effect, which couples two nucleons during an interaction via pion exchange and breaks the assumptions that even-numbered protons will have nearly zero effective spin in germanium. Numerous spin expectation values are summarized in~\cite{Klos2013}; however, we will quote the result that is consistent with discussion in~\cite{Gresham2014}, namely, $\langle S_n \rangle = 0.378$ and $\langle S_p \rangle = 0.030$ and $\langle L_n \rangle =  3.732$ and $\langle L_p \rangle = 0.361$, which is taken from calculations performed in~\cite{Dimitrov1995}. To provide an illustrative example, this enhancement in the proton spin can cause the $\mathcal{O}_4$ proton constraints to be an order of magnitude closer to the neutron coupling limits for GeV range DM~\cite{CDMS2018}. Due to the range of accepted values for the proton coupling, we only show results for the neutron, which is the dominant coupling for germanium targets, even if the effects of two-body currents were included. For all computational results, we use the form factors from~\cite{Fitzpatrick:2013}.

\subsection{Operator form factors} \label{app:OpFormFactors}
We use the same notation as~\cite{Fitzpatrick:2013} to define $F^{(n,n')}_{i,j}$ where $i,j \in \{1-11\}$ for the ten non-relativistic operators discussed in this work, and couplings to neutrons and/or protons, 
\begin{equation}
\begin{aligned}
F^{(n,n')}_{1,1} &= F^{(n,n')}_M \\
F^{(n,n')}_{3,3} &=  \frac{q^4}{4m^4_n} F^{(n,n')}_{\Phi''} + \frac{q^2}{8m^2_n} \bigg[ v^2 - \frac{q^2}{4\mu^2_N} \bigg] F^{(n,n')}_{\Sigma'} \\
F^{(n,n')}_{4,4} &=  \frac{C(j_{\chi})}{16} \bigg[ F^{(n,n')}_{\Sigma''} +  F^{(n,n')}_{\Sigma'} \bigg]\\
F^{(n,n')}_{5,5} &=  \frac{C(j_{\chi})}{4} \bigg[ \frac{q^2}{m^2_n} \bigg[ v^2 - \frac{q^2}{4\mu^2_N} \bigg] F^{(n,n')}_M +  \frac{q^4}{m^4_n} F^{(n,n')}_{\Delta} \bigg]  = \frac{q^2}{m^2_n} F^{(n,n')}_{8,8}\\
F^{(n,n')}_{6,6} &=  \frac{C(j_{\chi})}{16} \frac{q^4}{m^4_n} F^{(n,n')}_{\Sigma''}\\
F^{(n,n')}_{7,7} &=  \frac{1}{8} \bigg[ v^2 - \frac{q^2}{4\mu^2_N} \bigg] F^{(n,n')}_{\Sigma'} \\
F^{(n,n')}_{8,8} &=  \frac{C(j_{\chi})}{4} \bigg[ \bigg[ v^2 - \frac{q^2}{4\mu^2_N} \bigg] F^{(n,n')}_M +  \frac{q^2}{m^2_n} F^{(n,n')}_{\Delta} \bigg] \\
F^{(n,n')}_{9,9} &=  C(j_{\chi}) \frac{q^2}{16m^2_n} F^{(n,n')}_{\Sigma'}\\
F^{(n,n')}_{10,10} &=  \frac{q^2}{4m^2_n} F^{(n,n')}_{\Sigma''}\\
F^{(n,n')}_{11,11} &=  C(j_{\chi}) \frac{q^2}{4m^2_n} F^{(n,n')}_M \ . 
\end{aligned}
\end{equation}
In the above equations, we have used that $C(j_{\chi}) = 4j_\chi (j_\chi + 1)/3$. Additionally, we show the form factors for coupled operators, which may be relevant for some UV complete theories (although we do not discuss them in this work),  
\begin{equation}
\begin{aligned}
F^{(n,n')}_{1,3} &=  \frac{q^2}{m^2_n} F^{(n,n')}_{M, \Phi''} \\
F^{(n,n')}_{4,5} &=  -C(j_{\chi}) \frac{q^2}{4 m^2_n} F^{(n,n')}_{\Sigma', \Delta} \\
F^{(n,n')}_{4,6} &=  C(j_{\chi}) \frac{q^2}{16m^2_n} F^{(n,n')}_{\Sigma''} \\
F^{(n,n')}_{8,9} &=  C(j_{\chi}) \frac{q^2}{4 m^2_n} F^{(n,n')}_{\Sigma',\Delta} \ . 
\end{aligned}
\end{equation}

\section{Velocity-dependent effective Hamiltonian} \label{app:velocity}

Starting from the matrix element in~\cite{Berghaus2023}, but including DM-spin states, we have  
\begin{equation}
\begin{aligned}
\mathcal{M}_{fi} &= \sum_\lambda \Bigg[ \frac{1}{\omega + E_{\lambda_f} - E_{\lambda}} \langle \lambda_f, \boldsymbol{p}_e+\boldsymbol{k}_e, n_f | H_{eL} | \lambda, \boldsymbol{p}_e, n_f \rangle \\
&\qquad\qquad \times \langle p_f, \chi_f, \lambda, n_f | H_{\chi L} | p_i, \chi_i, \lambda_i, n_i \rangle \\
&\quad + \frac{1}{E_{\lambda_i} - E_{\lambda} - \omega} \langle p_f, \chi_f, \lambda_f, n_f | H_{\chi L} | p_i, \chi_i, \lambda, n_i \rangle \\
&\qquad\qquad \times \langle \lambda, \boldsymbol{p}_e+\boldsymbol{k}_e, n_i | H_{eL} | \lambda, \boldsymbol{p}_e, n_i \rangle \Bigg] \ ,
\end{aligned} 
\end{equation}
\noindent where $n_i \equiv j_n, m_n$ ($n_f \equiv j_n', m_n'$) are the initial (final) nucleon spin states and $\chi_i \equiv j_\chi, m_\chi$ ($\chi_f \equiv j_\chi', m_\chi'$) are the initial (final) DM spin states. We can approximate the expression to order $O(1/\omega ^2)$ as  
\begin{equation}
\begin{aligned}
\mathcal{M}_{fi} &= \sum_\lambda \Bigg[ \left( \frac{1}{\omega} - \frac{E_{\lambda_f} - E_{\lambda}}{\omega^2} \right) \langle \lambda_f, \boldsymbol{p}_e+\boldsymbol{k}_e, n_f | H_{eL} | \lambda, \boldsymbol{p}_e,n_f \rangle \\
&\qquad\qquad \times \langle p_f, \chi_f, \lambda, n_f | H_{\chi L} | p_i, \chi_i,  \lambda_i, n_i \rangle \\
&\quad - \left( \frac{1}{\omega} - \frac{E_{\lambda} - E_{\lambda_i}}{\omega^2} \right) \langle p_f, \chi_f, \lambda_f,n_f | H_{\chi L} | p_i, \chi_i, \lambda, n_i \rangle \\
&\qquad\qquad \times \langle \lambda, \boldsymbol{p}_e+\boldsymbol{k}_e, n_i | H_{eL} | \lambda, \boldsymbol{p}_e, n_i \rangle \Bigg] \, .
\end{aligned}
\end{equation}
\noindent The $|\lambda \rangle$ states are eigenstates of the unperturbed lattice Hamiltonian, $H_L$, defined in~\eqautoref{eqn:latticeHamiltonian}. We can use the completeness relations to find a similar result as in~\cite{Berghaus2023}, namely, 
\begin{equation}
\begin{aligned}
\mathcal{M}_{fi} &= \frac{1}{\omega^2} \bigg[ \langle p_f, \chi_f, \lambda_f, \boldsymbol{p}_e+\boldsymbol{k}_e, n_f| \left( H_{eL} H_L - H_L H_{eL} \right) H_{\chi L} \\
& - H_{\chi L} \left( H_{eL} H_L - H_L H_{eL} \right)| p_i, \chi_i, \lambda, \boldsymbol{p}_e, n_i, \rangle  \bigg] \\
&=  \langle p_f, \chi_f, \lambda_f, \boldsymbol{p}_e+\boldsymbol{k}_e, n_f | \frac{1}{\omega^2} \left[ H_{\chi L}, \left[ H_L, H_{e L}\right] \right]| p_i, \chi_i \lambda, \boldsymbol{p}_e, n_i \rangle \ ,
\end{aligned} 
\end{equation}
To continue, we calculate the commutation relation $\left[ H_L, H_{e L}\right]$,
\begin{equation} 
\left[ H_L, H_{e L}\right] = - \frac{\nabla_I}{m_N} H_{eL}\nabla_I - \frac{\nabla_I^2}{2m_N} H_{eL} \cdot \mathbb{1} \: ,
\end{equation}
where the derivatives are with respect to lattice cite, $I$. Using  test functions, $f$ and $g$, we can find the full commutation relation, which breaks up into four terms,
\begin{equation}
\begin{aligned} \label{eqn:commutation}
g \left[ H_{\chi L}, \left[ H_L, H_{e L}\right] \right] f &=  - g H_{\chi L} \frac{\nabla_I H_{eL}}{m_N} \nabla_I f + gH_{\chi L} \left(-\frac{\nabla_I^2 H_{eL}}{2m_N}  \right) f \\
& + g \frac{\nabla_I H_{eL}}{m_N} \nabla_I \left( H_{\chi L} f \right) +  g \left( \frac{\nabla_I^2 H_{eL}}{2m_N}  \right) H_{\chi L}f \:.
\end{aligned}
\end{equation}
Plugging in the DM-lattice Hamiltonian, we can simplify each term. To do so, we define $H_{\chi L}^c$ as the velocity-independent (or "constant") terms in the DM-Hamiltonian, namely terms proportional to $l_0$ and $l_5$. We also rename the coefficients $l/2M \rightarrow l'$ to not confuse where the factors of mass are coming from. Lastly, we only look at $H^{(I)}_{\chi L}$, such that we can trivially drop the sums in front of every term (for convenience, we also drop the sum over nucleons):
\begin{equation}
\begin{aligned}
g H^{(I)}_{\chi L} \frac{-\nabla_I H_{eL}}{m_N} \nabla_I f   & = g H_{\chi L}^c \left(\frac{-\nabla_I H_{eL}}{m_N} \cdot \nabla_I f \right) \\
& + g \left[l^{'A}_0 \left(-\frac{\cev{\nabla}_n }{i} \cdot \boldsymbol{\sigma}\right) + \boldsymbol{l}_M' \left(-\frac{\cev{\nabla}_n}{i}\right) + \boldsymbol{l}_E' \left(\cev{\nabla}_n \times \boldsymbol{\sigma}\right) \right] \\
&\qquad \times \delta (\boldsymbol{x}_\chi - \boldsymbol{x}_n) \left(\frac{-\nabla_I H_{eL}}{m_N} \cdot \nabla_I f \right)  \\
& + g \delta (\boldsymbol{x}_\chi - \boldsymbol{x}_n) \left[ l^{'A}_0 \left( \boldsymbol{\sigma} \cdot \frac{\vec{\nabla}_n}{i}\right) + \boldsymbol{l}_M' \left(\frac{\vec{\nabla}_n}{i}\right) + \boldsymbol{l}_E' \left(\boldsymbol{\sigma} \times \vec{\nabla}_n \right) \right] \\
&\qquad \times \left(\frac{-\nabla_I H_{eL}}{m_N} \cdot \nabla_I f \right)
\end{aligned} 
\end{equation}
\begin{equation}
\begin{aligned}
g H^{(I)}_{\chi L} \left(-\frac{\nabla_I^2 H_{eL}}{2m_N} \right) f & = g H_{\chi L}^c \left(-\frac{\nabla_I^2 H_{eL}}{2m_N} \right)f \\
& + g \left[ l^{'A}_0 \left(-\frac{\cev{\nabla}_n}{i} \cdot \boldsymbol{\sigma}\right) + \boldsymbol{l}_M' \left(-\frac{\cev{\nabla}_n}{i}\right) + \boldsymbol{l}_E' \left(\cev{\nabla}_n \times \boldsymbol{\sigma}\right) \right] \\
&\qquad \times \delta (\boldsymbol{x}_\chi - \boldsymbol{x}_n) \left(-\frac{\nabla_I^2 H_{eL}}{2m_N} \right)f  \\
& + g \delta (\boldsymbol{x}_\chi - \boldsymbol{x}_n) \left[ l^{'A}_0 \left( \boldsymbol{\sigma} \cdot \frac{\vec{\nabla}_n}{i}\right) + \boldsymbol{l}_M' \left(\frac{\vec{\nabla}_n}{i}\right) + \boldsymbol{l}_E' \left(\boldsymbol{\sigma} \times \vec{\nabla}_n\right) \right] \\
&\qquad \times \left(-\frac{\nabla_I^2 H_{eL}}{2m_N} \right)f
\end{aligned} 
\end{equation}
\begin{equation}
\begin{aligned}
g \frac{\nabla_I H_{eL}}{m_N} \nabla_I (H^{(I)}_{\chi L} f) & = g \frac{\nabla_I H_{eL}}{m_N} \left(\nabla_I  H^{c}_{\chi L} f +  H^{c}_{\chi L} \nabla_I f \right) \\
& + g \left(\frac{\nabla_I H_{eL}}{m_N}\right) \nabla_I \left\{ \left[l^{'A}_0 \left(\frac{\cev{\nabla}_n }{i} \cdot \boldsymbol{\sigma}\right) + \boldsymbol{l}_M' \left(-\frac{\cev{\nabla}_n}{i}\right) \right. \right. \\
&\qquad\qquad \left. \left. + \boldsymbol{l}_E' \left(\cev{\nabla}_n \times \boldsymbol{\sigma}\right) \right]  \delta (\boldsymbol{x}_\chi - \boldsymbol{x}_n) \right\} f \\
& + g \left(\frac{\nabla_I H_{eL}}{m_N}\right) \left[l^{'A}_0 \left(\frac{\cev{\nabla}_n }{i} \cdot \boldsymbol{\sigma}\right) + \boldsymbol{l}_M' \left(-\frac{\cev{\nabla}_n}{i}\right) \right. \\
&\qquad\qquad \left. + \boldsymbol{l}_E' \left(\cev{\nabla}_n \times \boldsymbol{\sigma}\right) \right]  \delta (\boldsymbol{x}_\chi - \boldsymbol{x}_n) \nabla_I f  \\
& + g \left(\frac{\nabla_I H_{eL}}{m_N}\right) \nabla_I \left\{ \delta (\boldsymbol{x}_\chi - \boldsymbol{x}_n) \left[ l^{'A}_0 \left( \boldsymbol{\sigma} \cdot \frac{\vec{\nabla}_n}{i}\right) \right. \right. \\
&\qquad\qquad \left. \left. + \boldsymbol{l}_M' \left(\frac{\vec{\nabla}_n}{i}\right) + \boldsymbol{l}_E' \left(\boldsymbol{\sigma} \times \vec{\nabla}_n\right) \right] \right\} f \\
& + g \left(\frac{\nabla_I H_{eL}}{m_N}\right) \delta (\boldsymbol{x}_\chi - \boldsymbol{x}_n) \left[ l^{'A}_0 \left( \boldsymbol{\sigma} \cdot \frac{\vec{\nabla}_n}{i}\right) \right. \\
&\qquad\qquad \left. + \boldsymbol{l}_M' \left(\frac{\vec{\nabla}_n}{i}\right) + \boldsymbol{l}_E' \left(\boldsymbol{\sigma} \times \vec{\nabla}_n\right) \right] \nabla_I f
\end{aligned} 
\end{equation}
\begin{equation}
\begin{aligned}
g \left( \frac{\nabla_I^2 H_{eL}}{2m_N}  \right) H^{(I)}_{\chi L} f  & = g \left(\frac{\nabla_I^2 H_{eL}}{2m_N} \right) H_{\chi L}^c f \\
& + g \left(\frac{\nabla_I^2 H_{eL}}{2m_N} \right) \left[ l^{'A}_0 \left(-\frac{\cev{\nabla}_n}{i} \cdot \boldsymbol{\sigma}\right) + \boldsymbol{l}_M' \left(-\frac{\cev{\nabla}_n}{i}\right) \right. \\
&\qquad \left. + \boldsymbol{l}_E' \left(\cev{\nabla}_n \times \boldsymbol{\sigma}\right) \right] \delta (\boldsymbol{x}_\chi - \boldsymbol{x}_n) f  \\
& + g \left(\frac{\nabla_I^2 H_{eL}}{2m_N} \right) \delta (\boldsymbol{x}_\chi - \boldsymbol{x}_n) \left[ l^{'A}_0 \left( \boldsymbol{\sigma} \cdot \frac{\vec{\nabla}_n}{i}\right) \right. \\
&\qquad \left. + \boldsymbol{l}_M' \left(\frac{\vec{\nabla}_n}{i}\right) + \boldsymbol{l}_E' \left(\boldsymbol{\sigma} \times \vec{\nabla}_n\right) \right] f
\end{aligned} 
\end{equation}
In each of the four expressions above, we can simplify each term in the commutation relation by (a) grouping the sub-terms with common derivatives (i.e. $\vec{\nabla}_n$ or $\cev{\nabla}_n$), (b) applying the operator to left or right depending on the respective term, and (c) using chain rule where appropriate. Lastly, we cancel like-terms. Some terms were independently zero. Specifically, terms that simplify to derivatives of the position-independent coefficients ($l' \cev{\nabla}_n$), where the rest of the term was not differentiated. We can then write the commutation relation as
\begin{equation}
\begin{aligned} 
\left[ H_{\chi L}, \left[ H_L, H_{e L}\right] \right]  &= \frac{1}{m_N} (\nabla_I H_{eL} \nabla_I H_{\chi L}) \\
& + \frac{2 l^{A}_0}{2m_n} \delta (\boldsymbol{x}_\chi - \boldsymbol{x}_n) 
(\boldsymbol{\sigma} \cdot \frac{\vec{\nabla}_n }{i} ) \left(\frac{-\nabla_I H_{eL}}{m_N} \right) \nabla_I \\
& + \frac{2 l^{A}_0}{2m_n} \delta (\boldsymbol{x}_\chi - \boldsymbol{x}_n) 
(\boldsymbol{\sigma} \cdot \frac{\vec{\nabla}_n }{i} ) \left(\frac{-\nabla^2_I H_{eL}}{2m_N} \right) \\
& + \frac{2 \boldsymbol{l}_M}{2 m_n} \delta (\boldsymbol{x}_\chi - \boldsymbol{x}_n) 
(\frac{\vec{\nabla}_n }{i} ) \left(\frac{-\nabla_I H_{eL}}{m_N} \right) \nabla_I  \\
& + \frac{2 \boldsymbol{l}_M}{2 m_n} \delta (\boldsymbol{x}_\chi - \boldsymbol{x}_n) 
(\frac{\vec{\nabla}_n }{i} ) \left(\frac{-\nabla^2_I H_{eL}}{2m_N} \right) \: .
\end{aligned}
\end{equation}
We can safely take $\nabla_I$, which acts on the nucleus at each lattice site, to be the same as $\nabla_n$, which acts on each nucleon. The two terms that end with a derivative acting on our initial state can not be written in terms of an elastic contributions (M, $\Delta, \Sigma', \Sigma'', \tilde{\Phi'}, \Phi''$). We remove these terms and are left with  
\begin{equation}
\begin{aligned} \label{eqn:commutationSimp}
\left[ H_{\chi L},  \left[ H_L, H_{e L}\right] \right]&  = \frac{1}{m_N} \left(\nabla_I H_{eL} \nabla_I H_{\chi L}\right) \\
& + \frac{1}{m_N}\left[ l^{A}_0  \delta (\boldsymbol{x}_\chi - \boldsymbol{x}_I) 
\left(\boldsymbol{\sigma} \cdot \frac{\vec{\nabla}_I }{i} \right) \right. \\
&\qquad \left. + \boldsymbol{l}_M \delta (\boldsymbol{x}_\chi - \boldsymbol{x}_n) 
\left(\frac{\vec{\nabla}_I }{i} \right) \right] \left(\frac{-\nabla^2_I H_{eL}}{2m_n} \right) \\
& \approx
\frac{1}{m_N} \left( \nabla_I H_{\chi L} \nabla_I H_{eL}  + \mathcal{O}\left(\frac{|\boldsymbol{k}_e +\boldsymbol{K}|}{m_n} \nabla^2_I H_{eL} \right)\right) \ .
\end{aligned} 
\end{equation}
The first term scales as $\sim  \boldsymbol{q} |\boldsymbol{k}_e + \boldsymbol{K}| H_{eL}$ whereas the remaining two terms on the second line scale as $\sim |\boldsymbol{k}_e + \boldsymbol{K}|^3/m_n H_{eL}$. For DM with mass $m_\chi \geq 10$ MeV, the DM momentum is much greater than the momentum of the excited electron ($q \gg |\boldsymbol{k}_e + \boldsymbol{K}|$). Therefore, we expect the terms in the second line of~\eqautoref{eqn:commutationSimp} to be suppressed, and we drop them in our numerical evaluations of the scattering rates. We then use the simplification, $H_{eff} = 1/\omega^2 * \left[ H_{\chi L}, \left[ H_L, H_{e L}\right] \right] \approx \frac{1}{\omega^2 m_N} (\nabla_I H_{eL} \nabla_I H_{\chi L})$ to calculate the matrix element in the next section. Note that this is only an approximation for operators with EFT coefficients described by $l_0^A$ and $l_M$; namely $\mathcal{O}_7$, $\mathcal{O}_5$, and $\mathcal{O}_8$. All other interactions can be exactly described as the first term.

\section{Elastic DM-Migdal matrix element } \label{app:Matrix}
Starting from~\eqautoref{eqn:MatrixElement}, we insert the completeness relation and use plane waves for the DM wave functions in position space,
\begin{equation*}
\begin{aligned}
\langle \boldsymbol{x}_\chi | \boldsymbol{p}_i \rangle &= e^{i \boldsymbol{p}_i \cdot \boldsymbol{x}_\chi}/\sqrt{V} \\
\langle \boldsymbol{p}_f | \boldsymbol{x}_\chi \rangle &= e^{-i \boldsymbol{p}_f \cdot \boldsymbol{x}_\chi}/\sqrt{V} \: .
\end{aligned}
\end{equation*}
Defining $\boldsymbol{q} = \boldsymbol{p}_i - \boldsymbol{p}_f$, the matrix element becomes  
\begin{equation}
\begin{aligned}
\mathcal{M}_{fi} &= \frac{1}{V} \int d^3x_L d^3x_e d^3x_\chi \langle \chi_f, n_f| \langle \lambda_f| \boldsymbol{x_L} \rangle \langle \boldsymbol{p}_e+\boldsymbol{k}| \boldsymbol{x_e} \rangle \\ & e^{i\boldsymbol{q} \cdot \boldsymbol{x}_\chi}   \frac{1}{{\omega^2 m_N} } \nabla_I H_{eL} \cdot \nabla_I H_{\chi L} | \chi_i, n_i \rangle \langle \boldsymbol{x_L}| \lambda_i \rangle \langle \boldsymbol{x_e}|\boldsymbol{p}_e \rangle \: .
\end{aligned}
\end{equation}
We then perform the $dx_\chi$ integral using the $\delta$-functions in each term of $H_{\chi L}$, which effectively set ($\boldsymbol{x}_\chi \rightarrow \boldsymbol{x}_n$) and expand $e^{i \boldsymbol{q} \cdot \boldsymbol{x}_n}$ using spherical harmonic and vector spherical harmonic identities (defined in~\autoref{app:definitions}). We then have,
\begin{equation}
\begin{aligned}
\mathcal{M}_{fi} &= \frac{4 \pi \alpha}{V^2 \omega^2  m_N} \sum_{\boldsymbol{K,K'}} \sum_{\boldsymbol{k}} \sum_I \frac{\boldsymbol{q} \cdot (\boldsymbol{k} + \boldsymbol{K'} ) \epsilon_{\boldsymbol{K,K'}}^{-1} (\boldsymbol{k}, \omega) Z(|\boldsymbol{k} + \boldsymbol{K}|)}{|\boldsymbol{k} + \boldsymbol{K}| |\boldsymbol{k} + \boldsymbol{K'}|} \\
& \times \langle \boldsymbol{p}_e+\boldsymbol{k}| e^{i (\boldsymbol{k} + \boldsymbol{K}) \cdot \boldsymbol{x_e}} | \boldsymbol{p}_e \rangle  \langle \lambda_f| e^{i (\boldsymbol{k} + \boldsymbol{K'} - \boldsymbol{q}) \cdot \boldsymbol{x_I}} | \lambda_i \rangle \\
& \times \langle \chi_f, n_f| \sum^{n,p}_N \sum_J \sqrt{4\pi} \sqrt{2J+1}\, \Big[ l_0 i^J M_{J0} + \frac{l^A_0}{m_n} i^J {M}_{J0} (\boldsymbol{\sigma} \cdot \vec{\nabla}_n) \\
&\qquad + \boldsymbol{l}_5 \left\{i^{J-1} \frac{\vec{\nabla}_n}{q} {M}_{J0} \hat{e}_0 + \frac{i^{J-2}}{\sqrt{2}} \sum_{\beta = \pm 1} \left[\beta {M}_{J0} + \frac{\vec{\nabla}_n}{q} \times \boldsymbol{M}^\beta_{J1}\right]\hat{e}^*_\beta \right\} \cdot \boldsymbol{\sigma} \\
&\qquad + \frac{\boldsymbol{l}_M}{m_n} \left\{i^{J-1} \frac{\vec{\nabla}_n}{q} {M}_{J0} \hat{e}_0 + \frac{i^{J-2}}{\sqrt{2}} \sum_{\beta = \pm 1} \left[\beta {M}_{J0} + \frac{\vec{\nabla}_n}{q} \times \boldsymbol{M}^\beta_{J1}\right]\hat{e}^*_\beta \right\} \left(\frac{\vec{\nabla}_n}{i}\right) \\
&\qquad + \frac{\boldsymbol{l}_E}{m_n} \left\{i^{J-1} \frac{\vec{\nabla}_n}{q} {M}_{J0} \hat{e}_0 + \frac{i^{J-2}}{\sqrt{2}} \sum_{\beta = \pm 1} \left[\beta {M}_{J0} + \frac{\vec{\nabla}_n}{q} \times \boldsymbol{M}^\beta_{J1}\right]\hat{e}^*_\beta \right\} \\
&\qquad\qquad\qquad \times \boldsymbol{\sigma} \times \vec{\nabla}_n \Big]| \chi_i, n_i \rangle \:.
\end{aligned}
\end{equation}
We can then write each term using the nuclear response functions,
\begin{equation}
\begin{aligned}
\mathcal{M}_{fi} &= \frac{4 \pi \alpha}{V^2 \omega^2  m_N} \sum_{\boldsymbol{K,K'}} \sum_{\boldsymbol{k}} \sum_I  \frac{\boldsymbol{q} \cdot (\boldsymbol{k} + \boldsymbol{K'} ) \epsilon_{\boldsymbol{K,K'}}^{-1} (\boldsymbol{k}, \omega) Z(|\boldsymbol{k} + \boldsymbol{K}|)}{|\boldsymbol{k} + \boldsymbol{K}| |\boldsymbol{k} + \boldsymbol{K'}|}
\\ & \langle \boldsymbol{p}_e+\boldsymbol{k}_e| e^{i (\boldsymbol{k} + \boldsymbol{K}) \cdot \boldsymbol{x_e}} | \boldsymbol{p}_e \rangle  \langle \lambda_f| e^{i (\boldsymbol{k} + \boldsymbol{K'} - \boldsymbol{q}) \cdot \boldsymbol{x_I}} | \lambda_i \rangle
\\&  \langle \chi_f, n_f | \sum^{n,p}_N \sum_J \sqrt{4\pi} \sqrt{2J+1} (i^J) \left[ l_0 M_{J0} - \boldsymbol{l}_5 \biggl\{i{\Sigma''}_{J0} \hat{e}_0 - \sum_{\beta = \pm 1} i {\Sigma'}_{J1} \frac{\hat{e}^*_\beta}{\sqrt{2}} \biggl\} \right.
\\& \left. - \frac{\boldsymbol{l}_M}{m_n} \biggl\{iq \sum_{\beta = \pm 1} \beta \Delta_{J0} \frac{\hat{e}^*_\beta}{\sqrt{2}} \biggl\} -  \frac{\boldsymbol{l}_E}{m_n} \biggl\{q\Phi''_{J0} \hat{e}_0 \biggl\} \right]| \chi_i, n_i \rangle \:.
\end{aligned}
\end{equation}
Any terms that contribute to inelastic scattering (as described in~\eqautoref{eqn:inelasticResponses}) are zero for any $J$. Therefore, in the above expression, we keep terms that contribute to elastic nuclear response functions, dropping inelastic terms. Most notably, the operator that correspond to the $l^A_0$ coefficient can not be written in terms of any elastic nuclear response functions and drops out completely. 

All DM-spin and $\boldsymbol{v^\perp_T}$ dependence is embedded into the EFT coefficients. There are further simplifications that can be made depending on which operator is chosen to be ``on'' via the $c$-couplings. Therefore, we break up our $l's$ and focus on each term proportional to the couplings $c^{(n)}_i$,
\begin{equation} \label{eqn: MatrixElementSimplified_Appendix}
\begin{aligned}
\mathcal{M}_{fi} &= \frac{4 \pi \alpha}{V^2 \omega^2  m_N} \sum_{\boldsymbol{K,K'}} \sum_{\boldsymbol{k}} \sum_I  \frac{\boldsymbol{q} \cdot (\boldsymbol{k} + \boldsymbol{K'} ) \epsilon_{\boldsymbol{K,K'}}^{-1} (\boldsymbol{k}, \omega) Z(|\boldsymbol{k} + \boldsymbol{K}|)}{|\boldsymbol{k} + \boldsymbol{K}| |\boldsymbol{k} + \boldsymbol{K'}|}
\\ & \langle \boldsymbol{p}_e+\boldsymbol{k}_e| e^{i (\boldsymbol{k} + \boldsymbol{K}) \cdot \boldsymbol{x_e}} | \boldsymbol{p}_e \rangle  \langle \lambda_f| e^{i (\boldsymbol{k} + \boldsymbol{K'} - \boldsymbol{q}) \cdot \boldsymbol{x_I}} | \lambda_i \rangle \biggl< \chi_f, n_f \Biggl | \sum^{n,p}_N \sum_J \sqrt{4\pi} \sqrt{2J+1} (i)^J 
\\& \left[ c^{(n)}_1 M_{J0} + c^{(n)}_{11} i(\frac{\boldsymbol{q}}{m_n} \cdot \boldsymbol{S_\chi}) M_{J0} - c^{(n)}_6 (\frac{\boldsymbol{q}}{m_n} \cdot \boldsymbol{S_\chi}) \frac{i|q|}{2 m_n} \Sigma''_{J0} + c^{(n)}_{10} \frac{|q|}{2 m_n} \Sigma''_{J0} \right.
\\& \left. - c^{(n)}_4 \frac{\boldsymbol{S_\chi}}{2}  \left( i{\Sigma''}_{J0} \hat{e}_0 - \sum_{\beta = \pm 1} i {\Sigma'}_{J1} \frac{\hat{e}^*_\beta}{\sqrt{2}} \right) + c^{(n)}_9 \frac{i}{2} (\frac{\boldsymbol{q}}{m_n} \times \boldsymbol{S_\chi}) \cdot \sum_{\beta= \pm} i \Sigma'_{J1} \frac{\hat{e}^*_\beta}{\sqrt{2}} \right.
\\& \left. c^{(n)}_3 \bigg\{ \frac{i}{2} (\frac{\boldsymbol{q}}{m_n} \times \boldsymbol{v^\perp_T}) \sum_{\beta= \pm} i \Sigma'_{J1} \frac{\hat{e}^*_\beta}{\sqrt{2}} - \frac{q^2}{2m^2_n} \Phi''_{J0} \bigg\} + c^{(n)}_7 \frac{\boldsymbol{v^\perp_T}}{2} \sum_{\beta= \pm} i \Sigma'_{J1} \frac{\hat{e}^*_\beta}{\sqrt{2}}  \right.
\\& \left. + c^{(n)}_5 \bigg\{ i(\frac{\boldsymbol{q}}{m_n} \times \boldsymbol{v^\perp_T}) \cdot \boldsymbol{S_\chi} M_{J0} - i(\frac{\boldsymbol{q}}{m_n} \times \boldsymbol{S_\chi}) \frac{iq}{m_n} \sum_{\beta = \pm 1} \beta {\Delta}_{J0} \frac{\hat{e}^*_\beta}{\sqrt{2}} \bigg\} \right.
\\& \left. + c^{(n)}_8 \bigg\{ (\boldsymbol{v^\perp_T} \cdot \boldsymbol{S_\chi}) M_{J0} + \boldsymbol{S_\chi} \frac{iq}{m_n} \sum_{\beta = \pm 1} \beta {\Delta}_{J0} \frac{\hat{e}^*_\beta}{\sqrt{2}} \bigg\} \right] \Biggl | \chi_i, n_i \biggl> \: .
\end{aligned}
\end{equation}
In the above expression, we have kept our model of the couplings general and keep the sum over couplings to neutrons and protons.

\section{Phonon spectrum for DM-nucleus scattering without Migdal ionization} \label{app:NuclearPhononSpectrum}

\begin{figure*}[h!]
    \centering
    \begin{subfigure}[b]{0.49\textwidth}   
        \centering 
        \includegraphics[width=\textwidth]{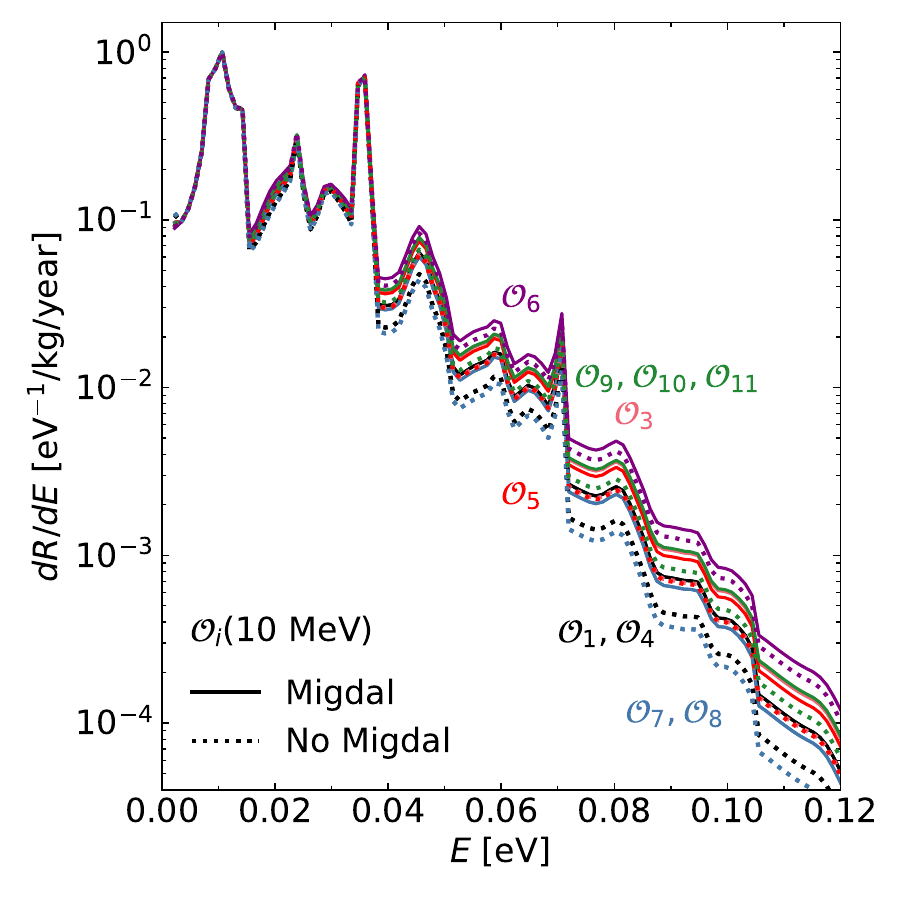}
    \end{subfigure}
    \hfill
    \begin{subfigure}[b]{0.49\textwidth}   
        \centering 
        \includegraphics[width=\textwidth]{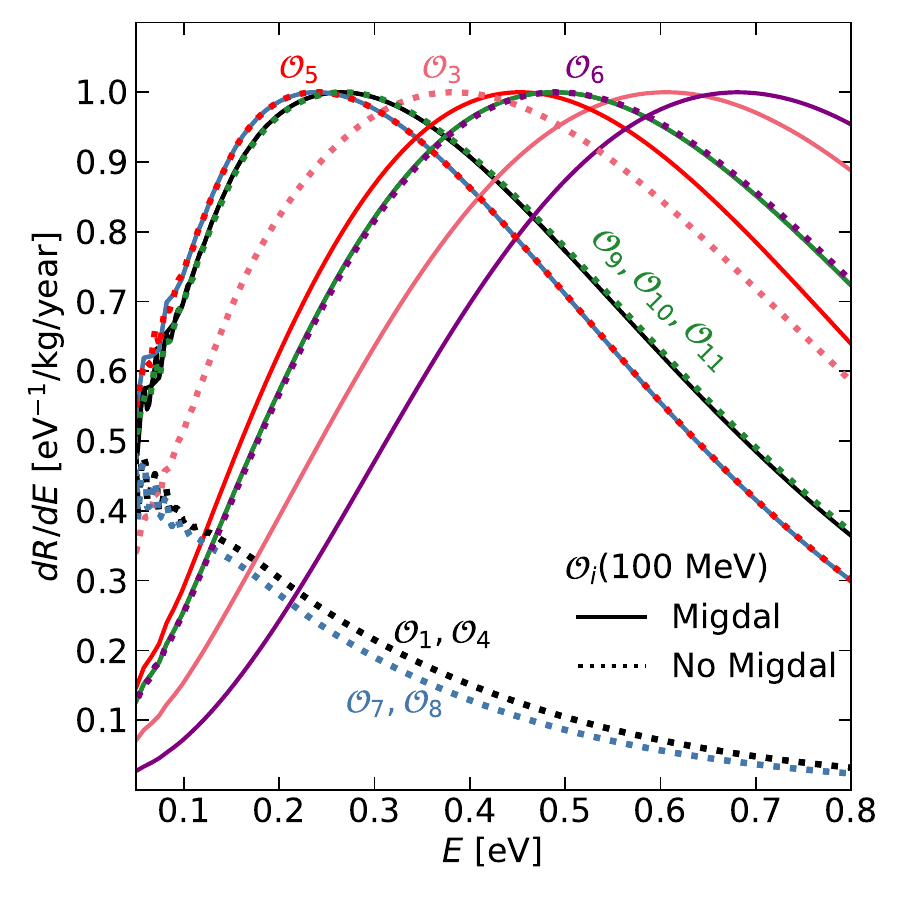}
    \end{subfigure}
    \caption[]
    {DM-Migdal and DM-nucleus scattering phonon spectrum for a germanium detector (containing 7.7\% Ge-73) for a DM spin $j_\chi = 1/2$ and a DM mass (\textbf{left}) $m_\chi = 10$~MeV and (\textbf{right}) $m_\chi = 100$~MeV. Each curve is normalized to a maximum value of 1~event/(kg-yr-eV). We show the nuclear phonon spectrum with no Migdal event as a dotted line and label the relevant operators in the appropriate colors. We show the Migdal phonon spectrum as a solid line (same as~\autoref{fig:phonon_spectrum}) for all operators with colors matching the non-Migdal spectrum. \label{fig:NuclearPhononSpectrum} } 
\end{figure*}

Detailed calculations of the phonon spectrum from spin-independent DM-nucleus scatters without an accompanying Migdal ionization are considered in~\cite{DarkElf_multiphonon}. Additionally, some spin-dependent operators are considered in~\cite{Gori:2025jzu} (note that they have a different convention in defining the reference cross section). In \autoref{fig:NuclearPhononSpectrum}, we show the characteristic shapes of the phonon spectrum with and without an accompanying Migdal ionization.  The corresponding reference cross sections for the operators with Migdal ionization are given in \autoref{tab:phonon_scale}, while the reference cross sections for the operators for DM-nucleus scatters without an accompanying Migdal ionization are given for each operator in~\autoref{tab:nuclear_scale}.  We emphasize that the event rate without a Migdal electron is much larger than the event rate with a Migdal electron for the same reference cross section. 

The grouping of operators for the phonon spectrum that is produced in events containing a Migdal ionization  (see~\autoref{fig:phonon_spectrum}) also holds true for the phonon spectrum without Migdal ionization. Therefore, at lower DM masses, $\mathcal{O}_1$ and $\mathcal{O}_4$, $\mathcal{O}_7$ and $\mathcal{O}_8$, as well as $\mathcal{O}_9$, $\mathcal{O}_{10}$, and $\mathcal{O}_{11}$ each form groups of operators that share the same operator form factor momentum and velocity dependence. As noted in the main text, the form factor $F^{(n,n')}_3$ has a complicated dependence on momentum through both $q^4$ and $q^2(v^2 - q^2/4\mu_N)$. 

Moreover, the phonon spectrum generated with accompanying Migdal ionization has an additional $\sim q^2$ momentum dependence compared to the phonon spectrum generated during DM-nucleus scattering without a Migdal signal. Therefore, for the Migdal events, $\mathcal{O}_1$ and $\mathcal{O}_4$ have the same shape as the non-Migdal events for $\mathcal{O}_9$, $\mathcal{O}_{10}$ and $\mathcal{O}_{11}$; the phonon spectra for the Migdal events generated by $\mathcal{O}_7$ and $\mathcal{O}_8$ are the same as the non-Migdal $\mathcal{O}_5$ spectrum. Lastly, the phonon spectra for the Migdal events generated by the $\mathcal{O}_9$, $\mathcal{O}_{10}$, $\mathcal{O}_{11}$ operators share the same shape as the non-Migdal $\mathcal{O}_6$ phonon spectrum. Therefore, it is hard to differentiate a Migdal event from a non-Migdal event from the phonon spectrum alone if the operator generating it is unknown.
 
As discussed in the main text, the phonon spectrum for Migdal events has dependence on $E$ through the integration bounds for momentum and velocity, which is quite weak at low DM masses where $\omega \gg E$. Although the nuclear recoil without a Migdal ionization leads to different kinematics, the integration bounds also have weak dependence on $E$ for $2E/(\mu_{\chi n} v^2)\ll 1$, which occurs at low phonon energies and small DM masses. Therefore, at low DM masses, the phonon spectrum for a nuclear recoil event with no Migdal ionization has small deviations for operators with different momentum dependence, as well as small deviations from the Migdal spectrum (see~\autoref{fig:NuclearPhononSpectrum}). Note, we must show the 10~MeV DM mass on a log scale to highlight the small deviations.

\begin{table}[h!]
\centering
\begin{tabular}{l|l|l}
Operator     & $\overline{\sigma}_n$ [$\rm{cm^2}$] ($m_\chi =10 \, \rm{MeV}$) & $\overline{\sigma}_n$ [$\rm{cm^2}$] ($m_\chi =100 \,\rm{MeV}$) \\ \cline{1-3}
$\mathcal{O}_1 = \boldsymbol{\mathbb{1}}$                     & $2.2 \times 10^{-47}$   & $3.9 \times 10^{-46}$                                       \\
$\mathcal{O}_3 = i \boldsymbol{S_n} \cdot \left( \frac{\boldsymbol{q}}{m_n} \times \boldsymbol{v^\perp} \right) $                                       & $3.2 \times 10^{-26}$                                        & $1.6 \times 10^{-26}$                                      \\
$\mathcal{O}_4 = \boldsymbol{S_\chi} \cdot \boldsymbol{S_n} $                                                                                           & $2.2 \times 10^{-41}$                                         & $3.8 \times 10^{-40}$                                       \\
$\mathcal{O}_5 = i \boldsymbol{S_\chi} \cdot \left( \frac{\boldsymbol{q}}{m_n} \times \boldsymbol{v^\perp} \right) $                                    & $7.0 \times 10^{-30}$                                         & $7.2 \times 10^{-30}$                                       \\
$\mathcal{O}_6 = \left( \boldsymbol{S_\chi} \cdot \frac{\boldsymbol{q}}{m_n} \right) \left( \boldsymbol{S_n} \cdot \frac{\boldsymbol{q}}{m_n} \right) $ & $1.1 \times 10^{-22}$                                         & $1.9 \times 10^{-24}$                                       \\
$\mathcal{O}_7 = \boldsymbol{S_n} \cdot \boldsymbol{v^\perp} $                                                                                          & $2.0 \times 10^{-35}$                                         & $2.9 \times 10^{-34}$                                       \\
$\mathcal{O}_8 = \boldsymbol{S_\chi} \cdot \boldsymbol{v^\perp} $                                                                                       & $4.3 \times 10^{-39}$                                         & $6.5 \times 10^{-38}$                                     \\
$\mathcal{O}_9 = i \boldsymbol{S_\chi} \cdot \left( \boldsymbol{S_n} \times \frac{\boldsymbol{q}}{m_n} \right)$                                         & $4.9 \times 10^{-32}$                                         & $5.8 \times 10^{-32}$                                       \\
$\mathcal{O}_{10} =  i \left( \boldsymbol{S_n} \cdot \frac{\boldsymbol{q}}{m_n} \right) $                                                               & $2.4 \times 10^{-32}$                                         & $2.9 \times 10^{-32}$                                      \\
$\mathcal{O}_{11} = i \left( \boldsymbol{S_\chi} \cdot \frac{\boldsymbol{q}}{m_n} \right) $                                                             & $5.3 \times 10^{-36}$                                         & $6.3 \times 10^{-36}$                                      
\end{tabular}
\caption{
Choice of reference cross section for the phonon spectrum without Migdal ionization for each operator in~\autoref{fig:NuclearPhononSpectrum} at two different DM masses. Note, for $\mathcal{O}_{1}$ we assume 100\% of the detector material can interact (rather than 7.7\% for Ge-73) and include the contribution from protons. \label{tab:nuclear_scale}}
\end{table}

\bibliographystyle{JHEP} 
\bibliography{spinDepMigdal.bib}

\providecommand{\href}[2]{#2}\begingroup\raggedright\begin{thebibliography}{100}

\bibitem{CRESST-III-2021}
M.~Stahlberg, \emph{{Probing Low-Mass DarkMatter with CRESST-III - Data Analysis and First Results}}, Ph.D. thesis, Vienna, Tech. U., 2021.
\newblock 10.34726/hss.2021.45935.

\bibitem{DarkSide-50-2023}
{\scshape DarkSide-50 Collaboration} collaboration, \emph{Search for low-mass dark matter wimps with 12 ton-day exposure of darkside-50}, \href{https://doi.org/10.1103/PhysRevD.107.063001}{\emph{Phys. Rev. D} {\bfseries 107} (2023) 063001}.

\bibitem{SuperCDMS:2023geu}
{\scshape SuperCDMS} collaboration, \emph{{First Measurement of the Nuclear-Recoil Ionization Yield in Silicon at 100~eV}}, \href{https://doi.org/10.1103/PhysRevLett.131.091801}{\emph{Phys. Rev. Lett.} {\bfseries 131} (2023) 091801} [\href{https://arxiv.org/abs/2303.02196}{{\ttfamily 2303.02196}}].

\bibitem{XENONnT2025}
{\scshape XENON Collaboration} collaboration, \emph{Wimp dark matter search using a 3.1 tonne-year exposure of the xenonnt experiment}, \href{https://doi.org/10.1103/msw4-t342}{\emph{Phys. Rev. Lett.} {\bfseries 135} (2025) 221003}.

\bibitem{LZ:2025igz}
{\scshape LZ} collaboration, \emph{Searches for light dark matter and evidence of coherent elastic neutrino-nucleus scattering of solar neutrinos with the lux-zeplin (lz) experiment},  \href{https://arxiv.org/abs/2512.08065}{{\ttfamily 2512.08065}}.

\bibitem{PandaX4T_2025}
{\scshape PandaX Collaboration} collaboration, \emph{Search for light dark matter with 259 days of data in pandax-4t}, \href{https://doi.org/10.1103/rtnh-jn8s}{\emph{Phys. Rev. Lett.} {\bfseries 135} (2025) 211001}.

\bibitem{Fan2010}
J.~Fan, M.~Reece and L.-T.~Wang, \emph{Non-relativistic effective theory of dark matter direct detection}, \href{https://doi.org/10.1088/1475-7516/2010/11/042}{\emph{Journal of Cosmology and Astroparticle Physics} {\bfseries 2010} (2010) 042}.

\bibitem{Fitzpatrick:2013}
A.L.~Fitzpatrick, W.C.~Haxton, E.~Katz, N.~Lubbers and Y.~Xu, \emph{The effective field theory of dark matter direct detection}, \href{https://doi.org/10.1088/1475-7516/2013/02/004}{\emph{JCAP} {\bfseries 02} (2013) 004} [\href{https://arxiv.org/abs/arXiv:1203.3542}{{\ttfamily arXiv:1203.3542}}].

\bibitem{Anand2013}
N.~Anand, A.L.~Fitzpatrick and W.C.~Haxton, \emph{Weakly interacting massive particle-nucleus elastic scattering response}, \href{https://doi.org/10.1103/PhysRevC.89.065501}{\emph{Phys. Rev. C} {\bfseries 89} (2014) 065501}.

\bibitem{SuperCDMS:2015lcz}
{\scshape SuperCDMS} collaboration, \emph{Dark matter effective field theory scattering in direct detection experiments}, \href{https://doi.org/10.1103/PhysRevD.91.092004}{\emph{Phys. Rev. D} {\bfseries 91} (2015) 092004} [\href{https://arxiv.org/abs/1503.03379}{{\ttfamily 1503.03379}}].

\bibitem{Gazda2016}
D.~Gazda, R.~Catena and C.~Forss\'en, \emph{Ab initio nuclear response functions for dark matter searches}, \href{https://doi.org/10.1103/PhysRevD.95.103011}{\emph{Phys. Rev. D} {\bfseries 95} (2017) 103011}.

\bibitem{DarkSide-50:2020swd}
{\scshape DarkSide-50} collaboration, \emph{Effective field theory interactions for liquid argon target in darkside-50 experiment}, \href{https://doi.org/10.1103/PhysRevD.101.062002}{\emph{Phys. Rev. D} {\bfseries 101} (2020) 062002} [\href{https://arxiv.org/abs/2002.07794}{{\ttfamily 2002.07794}}].

\bibitem{LUX:2020oan}
{\scshape LUX} collaboration, \emph{Effective field theory analysis of the first lux dark matter search}, \href{https://doi.org/10.1103/PhysRevD.103.122005}{\emph{Phys. Rev. D} {\bfseries 103} (2021) 122005} [\href{https://arxiv.org/abs/2003.11141}{{\ttfamily 2003.11141}}].

\bibitem{Griffin2021}
S.M.~Griffin, K.~Inzani, T.~Trickle, Z.~Zhang and K.M.~Zurek, \emph{Extended calculation of dark matter-electron scattering in crystal targets}, \href{https://doi.org/10.1103/PhysRevD.104.095015}{\emph{Phys. Rev. D} {\bfseries 104} (2021) 095015}.

\bibitem{SuperCDMS:2022crd}
{\scshape SuperCDMS} collaboration, \emph{Effective field theory analysis of cdmslite run 2 data},  \href{https://arxiv.org/abs/2205.11683}{{\ttfamily 2205.11683}}.

\bibitem{PhysRevD.111.095033}
J.-H.~Liang, Y.~Liao, X.-D.~Ma and H.-L.~Wang, \emph{Systematic investigation on vector dark matter-nucleus scattering in effective field theories}, \href{https://doi.org/10.1103/PhysRevD.111.095033}{\emph{Phys. Rev. D} {\bfseries 111} (2025) 095033}.

\bibitem{CDEX:2025mcb}
{\scshape CDEX} collaboration, \emph{Constraints on dark matter boosted by supernova shock within the effective field theory framework from the cdex-10 experiment}, \href{https://doi.org/10.1103/kxmk-7zfk}{\emph{Phys. Rev. D} {\bfseries 112} (2025) 092011} [\href{https://arxiv.org/abs/2504.03559}{{\ttfamily 2504.03559}}].

\bibitem{Essig:2011nj}
R.~Essig, J.~Mardon and T.~Volansky, \emph{Direct detection of sub-gev dark matter}, \href{https://doi.org/10.1103/PhysRevD.85.076007}{\emph{Phys. Rev. D} {\bfseries 85} (2012) 076007} [\href{https://arxiv.org/abs/1108.5383}{{\ttfamily 1108.5383}}].

\bibitem{Migdal:1939}
A.B.~Migdal, \emph{Ionizatsiya atomov pri yadernykh reaktsiyakh}, {\emph{Soviet Physics JETP} {\bfseries 9} (1939) 1163}.

\bibitem{Migdal:1978}
A.B.~Migdal and L.E.~Ballentine, \emph{Qualitative methods in quantum theory}, \href{https://doi.org/10.1063/1.2995000}{\emph{Physics Today} {\bfseries 31} (1978) 60}.

\bibitem{Ibe:2017yqa}
M.~Ibe, W.~Nakano, Y.~Shoji and K.~Suzuki, \emph{Migdal effect in dark matter direct detection experiments}, \href{https://doi.org/10.1007/JHEP03(2018)194}{\emph{JHEP} {\bfseries 03} (2018) 194} [\href{https://arxiv.org/abs/1707.07258}{{\ttfamily 1707.07258}}].

\bibitem{Knapen:2017ekk}
S.~Knapen, T.~Lin, M.~Pyle and K.M.~Zurek, \emph{{Detection of Light Dark Matter With Optical Phonons in Polar Materials}}, \href{https://doi.org/10.1016/j.physletb.2018.08.064}{\emph{Phys. Lett. B} {\bfseries 785} (2018) 386} [\href{https://arxiv.org/abs/1712.06598}{{\ttfamily 1712.06598}}].

\bibitem{Essig2016}
R.~Essig, M.~Fern{\'a}ndez-Serra, J.~Mardon, A.~Soto, T.~Volansky and T.-T.~Yu, \emph{Direct detection of sub-gev dark matter with semiconductor targets}, \href{https://doi.org/10.1007/JHEP05(2016)046}{\emph{Journal of High Energy Physics} {\bfseries 2016} (2016) 46}.

\bibitem{DarkELF2021_dielectric}
S.~Knapen, J.~Kozaczuk and T.~Lin, \emph{Dark matter-electron scattering in dielectrics}, \href{https://doi.org/10.1103/PhysRevD.104.015031}{\emph{Phys. Rev. D} {\bfseries 104} (2021) 015031}.

\bibitem{darkELF2022}
S.~Knapen, J.~Kozaczuk and T.~Lin, \emph{python package for dark matter scattering in dielectric targets}, \href{https://doi.org/10.1103/PhysRevD.105.015014}{\emph{Phys. Rev. D} {\bfseries 105} (2022) 015014}.

\bibitem{Singal2023}
C.E.~Dreyer, R.~Essig, M.~Fernandez-Serra, A.~Singal and C.~Zhen, \emph{Fully ab-initio all-electron calculation of dark matter-electron scattering in crystals with evaluation of systematic uncertainties}, \href{https://doi.org/10.1103/PhysRevD.109.115008}{\emph{Phys. Rev. D} {\bfseries 109} (2024) 115008}.

\bibitem{Giffin:2025hdx}
P.~Giffin, B.~Lillard, P.~Munbodh and T.-T.~Yu, \emph{Simplified spin dependence in dark matter direct detection},  \href{https://arxiv.org/abs/2511.10764}{{\ttfamily 2511.10764}}.

\bibitem{Tiffenberg:2017aac}
{\scshape SENSEI} collaboration, \emph{{Single-electron and single-photon sensitivity with a silicon Skipper CCD}}, \href{https://doi.org/10.1103/PhysRevLett.119.131802}{\emph{Phys. Rev. Lett.} {\bfseries 119} (2017) 131802} [\href{https://arxiv.org/abs/1706.00028}{{\ttfamily 1706.00028}}].

\bibitem{Crisler:2018gci}
{\scshape SENSEI} collaboration, \emph{Sensei: First direct-detection constraints on sub-gev dark matter from a surface run}, \href{https://doi.org/10.1103/PhysRevLett.121.061803}{\emph{Phys. Rev. Lett.} {\bfseries 121} (2018) 061803} [\href{https://arxiv.org/abs/1804.00088}{{\ttfamily 1804.00088}}].

\bibitem{SENSEI:2019ibb}
{\scshape SENSEI} collaboration, \emph{Sensei: Direct-detection constraints on sub-gev dark matter from a shallow underground run using a prototype skipper-ccd}, \href{https://doi.org/10.1103/PhysRevLett.122.161801}{\emph{Phys. Rev. Lett.} {\bfseries 122} (2019) 161801} [\href{https://arxiv.org/abs/1901.10478}{{\ttfamily 1901.10478}}].

\bibitem{SENSEI:2020dpa}
{\scshape SENSEI} collaboration, \emph{Sensei: Direct-detection results on sub-gev dark matter from a new skipper-ccd}, \href{https://doi.org/10.1103/PhysRevLett.125.171802}{\emph{Phys. Rev. Lett.} {\bfseries 125} (2020) 171802} [\href{https://arxiv.org/abs/2004.11378}{{\ttfamily 2004.11378}}].

\bibitem{SENSEI:2021hcn}
{\scshape SENSEI} collaboration, \emph{{SENSEI: Characterization of Single-Electron Events Using a Skipper Charge-Coupled Device}}, \href{https://doi.org/10.1103/PhysRevApplied.17.014022}{\emph{Phys. Rev. Applied} {\bfseries 17} (2022) 014022} [\href{https://arxiv.org/abs/2106.08347}{{\ttfamily 2106.08347}}].

\bibitem{CDEX_electron}
{\scshape CDEX Collaboration} collaboration, \emph{Constraints on sub-gev dark matter--electron scattering from the cdex-10 experiment}, \href{https://doi.org/10.1103/PhysRevLett.129.221301}{\emph{Phys. Rev. Lett.} {\bfseries 129} (2022) 221301}.

\bibitem{SENSEI:2024yyt}
{\scshape SENSEI} collaboration, \emph{{SENSEI at SNOLAB: Single-Electron Event Rate and Implications for Dark Matter}}, \href{https://doi.org/10.1103/PhysRevLett.134.161002}{\emph{Phys. Rev. Lett.} {\bfseries 134} (2025) 161002} [\href{https://arxiv.org/abs/2410.18716}{{\ttfamily 2410.18716}}].

\bibitem{SENSEI:2023zdf}
{\scshape SENSEI} collaboration, \emph{First direct-detection results on sub-gev dark matter using the sensei detector at snolab}, \href{https://doi.org/10.1103/PhysRevLett.134.011804}{\emph{Phys. Rev. Lett.} {\bfseries 134} (2025) 011804} [\href{https://arxiv.org/abs/2312.13342}{{\ttfamily 2312.13342}}].

\bibitem{SENSEI:2025qvp}
{\scshape SENSEI} collaboration, \emph{{SENSEI: A Search for Diurnal Modulation in sub-GeV Dark Matter Scattering}},  \href{https://arxiv.org/abs/2510.20889}{{\ttfamily 2510.20889}}.

\bibitem{DAMIC-M:2023hgj}
{\scshape DAMIC-M} collaboration, \emph{{Search for Daily Modulation of MeV Dark Matter Signals with DAMIC-M}}, \href{https://doi.org/10.1103/PhysRevLett.132.101006}{\emph{Phys. Rev. Lett.} {\bfseries 132} (2024) 101006} [\href{https://arxiv.org/abs/2307.07251}{{\ttfamily 2307.07251}}].

\bibitem{DAMICm2023}
{\scshape DAMIC-M Collaboration} collaboration, \emph{First constraints from damic-m on sub-gev dark-matter particles interacting with electrons}, \href{https://doi.org/10.1103/PhysRevLett.130.171003}{\emph{Phys. Rev. Lett.} {\bfseries 130} (2023) 171003}.

\bibitem{DAMIC-M:2025ltz}
{\scshape DAMIC-M} collaboration, \emph{{Daily Modulation Constraints on Light Dark Matter with DAMIC-M}},  \href{https://arxiv.org/abs/2511.13962}{{\ttfamily 2511.13962}}.

\bibitem{DAMIC-M2025}
{\scshape DAMIC-M} collaboration, \emph{{Probing benchmark models of hidden-sector dark matter with DAMIC-M}}, \href{https://doi.org/10.1103/2tcc-bqck}{\emph{Phys. Rev. Lett.} (2025) } [\href{https://arxiv.org/abs/2503.14617}{{\ttfamily 2503.14617}}].

\bibitem{SuperCDMS_electron}
R.~Agnese, T.~Aralis, T.~Aramaki, I.J.~Arnquist, E.~Azadbakht, W.~Baker et~al., \emph{First dark matter constraints from a supercdms single-charge sensitive detector}, \href{https://doi.org/10.1103/PhysRevLett.121.051301}{\emph{Phys. Rev. Lett.} {\bfseries 121} (2018) 051301}.

\bibitem{SuperCDMS:2020ymb}
{\scshape SuperCDMS} collaboration, \emph{Constraints on low-mass, relic dark matter candidates from a surface-operated supercdms single-charge sensitive detector}, \href{https://doi.org/10.1103/PhysRevD.102.091101}{\emph{Phys. Rev. D} {\bfseries 102} (2020) 091101} [\href{https://arxiv.org/abs/2005.14067}{{\ttfamily 2005.14067}}].

\bibitem{SuperCDMS2025}
{\scshape SuperCDMS} collaboration, \emph{Light dark matter constraints from supercdms hvev detectors operated underground with an anticoincidence event selection}, \href{https://doi.org/10.1103/PhysRevD.111.012006}{\emph{Phys. Rev. D} {\bfseries 111} (2025) 012006} [\href{https://arxiv.org/abs/2407.08085}{{\ttfamily 2407.08085}}].

\bibitem{SuperCDMS_electron_2025}
{\scshape SuperCDMS} collaboration, \emph{Search for low-mass electron-recoil dark matter using a single-charge sensitive supercdms-hvev detector}, \href{https://doi.org/10.1103/5lnp-6mng}{\emph{Phys. Rev. D} {\bfseries 113} (2026) 032001} [\href{https://arxiv.org/abs/2509.03608}{{\ttfamily 2509.03608}}].

\bibitem{EDELWEISS2020}
{\scshape EDELWEISS Collaboration} collaboration, \emph{First germanium-based constraints on sub-mev dark matter with the edelweiss experiment}, \href{https://doi.org/10.1103/PhysRevLett.125.141301}{\emph{Phys. Rev. Lett.} {\bfseries 125} (2020) 141301}.

\bibitem{Catena:2019gfa}
R.~Catena, T.~Emken, N.A.~Spaldin and W.~Tarantino, \emph{Atomic responses to general dark matter-electron interactions}, \href{https://doi.org/10.1103/PhysRevResearch.2.033195}{\emph{Phys. Rev. Res.} {\bfseries 2} (2020) 033195} [\href{https://arxiv.org/abs/1912.08204}{{\ttfamily 1912.08204}}].

\bibitem{Catena2021}
R.~Catena, T.~Emken, M.~Matas, N.A.~Spaldin and E.~Urdshals, \emph{Crystal responses to general dark matter-electron interactions}, .

\bibitem{Liang:2024ecw}
J.-H.~Liang, Y.~Liao, X.-D.~Ma and H.-L.~Wang, \emph{A systematic investigation on dark matter-electron scattering in effective field theories},  \href{https://arxiv.org/abs/2406.10912}{{\ttfamily 2406.10912}}.

\bibitem{Krnjaic:2024bdd}
G.~Krnjaic, D.~Rocha and T.~Trickle, \emph{The non-relativistic effective field theory of dark matter-electron interactions},  \href{https://arxiv.org/abs/2407.14598}{{\ttfamily 2407.14598}}.

\bibitem{Vergados:2005dpd}
J.D.~Vergados and H.~Ejiri, \emph{The role of ionization electrons in direct neutralino detection}, \href{https://doi.org/10.1016/j.physletb.2004.11.085}{\emph{Phys. Lett. B} {\bfseries 606} (2005) 313} [\href{https://arxiv.org/abs/hep-ph/0401151}{{\ttfamily hep-ph/0401151}}].

\bibitem{Moustakidis:2005gx}
C.C.~Moustakidis, J.D.~Vergados and H.~Ejiri, \emph{Direct dark matter detection by observing electrons produced in neutralino-nucleus collisions}, \href{https://doi.org/10.1016/j.nuclphysb.2005.08.033}{\emph{Nucl. Phys. B} {\bfseries 727} (2005) 406} [\href{https://arxiv.org/abs/hep-ph/0507123}{{\ttfamily hep-ph/0507123}}].

\bibitem{Essig:2019xkx}
R.~Essig, J.~Pradler, M.~Sholapurkar and T.-T.~Yu, \emph{Relation between the migdal effect and dark matter-electron scattering in isolated atoms and semiconductors}, \href{https://doi.org/10.1103/PhysRevLett.124.021801}{\emph{Phys. Rev. Lett.} {\bfseries 124} (2020) 021801} [\href{https://arxiv.org/abs/1908.10881}{{\ttfamily 1908.10881}}].

\bibitem{Baxter:2019pnz}
D.~Baxter, Y.~Kahn and G.~Krnjaic, \emph{Electron ionization via dark matter-electron scattering and the migdal effect}, \href{https://doi.org/10.1103/PhysRevD.101.076014}{\emph{Phys. Rev. D} {\bfseries 101} (2020) 076014} [\href{https://arxiv.org/abs/1908.00012}{{\ttfamily 1908.00012}}].

\bibitem{Li:2022acp}
J.~Li, L.~Su, L.~Wu and B.~Zhu, \emph{Spin-dependent sub-gev inelastic dark matter-electron scattering and migdal effect. part i. velocity independent operator}, \href{https://doi.org/10.1088/1475-7516/2023/04/020}{\emph{JCAP} {\bfseries 04} (2023) 020} [\href{https://arxiv.org/abs/2210.15474}{{\ttfamily 2210.15474}}].

\bibitem{Adams:2022zvg}
D.~Adams, D.~Baxter, H.~Day, R.~Essig and Y.~Kahn, \emph{Measuring the migdal effect in semiconductors for dark matter detection}, \href{https://doi.org/10.1103/PhysRevD.107.L041303}{\emph{Phys. Rev. D} {\bfseries 107} (2023) L041303}.

\bibitem{Dolan2017}
M.J.~Dolan, F.~Kahlhoefer and C.~McCabe, \emph{Directly detecting sub-gev dark matter with electrons from nuclear scattering}, .

\bibitem{XENON:2019zpr}
{\scshape XENON} collaboration, \emph{Search for light dark matter interactions enhanced by the migdal effect or bremsstrahlung in xenon1t}, \href{https://doi.org/10.1103/PhysRevLett.123.241803}{\emph{Phys. Rev. Lett.} {\bfseries 123} (2019) 241803} [\href{https://arxiv.org/abs/1907.12771}{{\ttfamily 1907.12771}}].

\bibitem{DarkSide:2022dhx}
{\scshape DarkSide} collaboration, \emph{Search for dark-matter{\textendash}nucleon interactions via migdal effect with darkside-50}, \href{https://doi.org/10.1103/PhysRevLett.130.101001}{\emph{Phys. Rev. Lett.} {\bfseries 130} (2023) 101001} [\href{https://arxiv.org/abs/2207.11967}{{\ttfamily 2207.11967}}].

\bibitem{COSINE-100:2021poy}
{\scshape COSINE-100} collaboration, \emph{Searching for low-mass dark matter via the migdal effect in cosine-100}, \href{https://doi.org/10.1103/PhysRevD.105.042006}{\emph{Phys. Rev. D} {\bfseries 105} (2022) 042006} [\href{https://arxiv.org/abs/2110.05806}{{\ttfamily 2110.05806}}].

\bibitem{LZ_Migdal}
N.F.~Bell, P.~Cox, M.J.~Dolan, J.L.~Newstead and A.C.~Ritter, \emph{Exploring light dark matter with the migdal effect in hydrogen-doped liquid xenon}, \href{https://doi.org/10.1103/PhysRevD.109.L091902}{\emph{Phys. Rev. D} {\bfseries 109} (2024) L091902}.

\bibitem{Knapen2021}
S.~Knapen, J.~Kozaczuk and T.~Lin, \emph{Migdal effect in semiconductors}, \href{https://doi.org/10.1103/PhysRevLett.127.081805}{\emph{Physical Review Letters} {\bfseries 127} (2021) }.

\bibitem{SuperCDMS:2022kgp}
{\scshape SuperCDMS} collaboration, \emph{A search for low-mass dark matter via bremsstrahlung radiation and the migdal effect in supercdms},  \href{https://arxiv.org/abs/2203.02594}{{\ttfamily 2203.02594}}.

\bibitem{EDELWEISS:2022ktt}
{\scshape EDELWEISS} collaboration, \emph{Search for sub-gev dark matter via the migdal effect with an edelweiss germanium detector with nbsi transition-edge sensors}, \href{https://doi.org/10.1103/PhysRevD.106.062004}{\emph{Phys. Rev. D} {\bfseries 106} (2022) 062004} [\href{https://arxiv.org/abs/2203.03993}{{\ttfamily 2203.03993}}].

\bibitem{CDEX2022}
{\scshape CDEX Collaboration} collaboration, \emph{Studies of the earth shielding effect to direct dark matter searches at the china jinping underground laboratory}, \href{https://doi.org/10.1103/PhysRevD.105.052005}{\emph{Phys. Rev. D} {\bfseries 105} (2022) 052005}.

\bibitem{Berghaus2023}
K.V.~Berghaus, A.~Esposito, R.~Essig and M.~Sholapurkar, \emph{The migdal effect in semiconductors for dark matter with masses below $\sim$ 100 mev}, \href{https://doi.org/10.1007/JHEP01(2023)023}{\emph{Journal of High Energy Physics} {\bfseries 2023} (2023) }.

\bibitem{Esposito:2025iry}
A.~Esposito and A.~Rocchi, \emph{The migdal effect in solid crystals and the role of non-adiabaticity},  \href{https://arxiv.org/abs/2505.08864}{{\ttfamily 2505.08864}}.

\bibitem{EDELWEISS_surface}
{\scshape EDELWEISS Collaboration} collaboration, \emph{Searching for low-mass dark matter particles with a massive ge bolometer operated above ground}, \href{https://doi.org/10.1103/PhysRevD.99.082003}{\emph{Phys. Rev. D} {\bfseries 99} (2019) 082003}.

\bibitem{CDEX:2019hzn}
{\scshape CDEX} collaboration, \emph{Constraints on spin-independent nucleus scattering with sub-gev weakly interacting massive particle dark matter from the cdex-1b experiment at the china jinping underground laboratory}, \href{https://doi.org/10.1103/PhysRevLett.123.161301}{\emph{Phys. Rev. Lett.} {\bfseries 123} (2019) 161301} [\href{https://arxiv.org/abs/1905.00354}{{\ttfamily 1905.00354}}].

\bibitem{CDEX:2020tkb}
{\scshape CDEX} collaboration, \emph{First experimental constraints on wimp couplings in the effective field theory framework from cdex}, \href{https://doi.org/10.1007/s11433-020-1666-8}{\emph{Sci. China Phys. Mech. Astron.} {\bfseries 64} (2021) 281011} [\href{https://arxiv.org/abs/2007.15555}{{\ttfamily 2007.15555}}].

\bibitem{Kang:2018rad}
S.~Kang, S.~Scopel, G.~Tomar and J.-H.~Yoon, \emph{On the sensitivity of present direct detection experiments to wimp\textendash{}quark and wimp\textendash{}gluon effective interactions: A systematic assessment and new model\textendash{}independent approaches}, \href{https://doi.org/10.1016/j.astropartphys.2019.07.001}{\emph{Astropart. Phys.} {\bfseries 114} (2020) 80} [\href{https://arxiv.org/abs/1810.00607}{{\ttfamily 1810.00607}}].

\bibitem{TOMAR2023102851}
G.~Tomar, S.~Kang and S.~Scopel, \emph{Low-mass extension of direct detection bounds on wimp-quark and wimp-gluon effective interactions using the migdal effect}, \href{https://doi.org/https://doi.org/10.1016/j.astropartphys.2023.102851}{\emph{Astroparticle Physics} {\bfseries 150} (2023) 102851} [\href{https://arxiv.org/abs/2210.00199}{{\ttfamily 2210.00199}}].

\bibitem{Liang:2024tef}
J.-H.~Liang, Y.~Liao, X.-D.~Ma and H.-L.~Wang, \emph{Comprehensive constraints on fermionic dark matter-quark tensor interactions in direct detection experiments},  \href{https://arxiv.org/abs/2401.05005}{{\ttfamily 2401.05005}}.

\bibitem{Cirelli2013}
M.~Cirelli, E.D.~Nobile and P.~Panci, \emph{Tools for model-independent bounds in direct dark matter searches}, \href{https://doi.org/10.1088/1475-7516/2013/10/019}{\emph{Journal of Cosmology and Astroparticle Physics} {\bfseries 2013} (2013) 019}.

\bibitem{KANG201950}
S.~Kang, S.~Scopel, G.~Tomar and J.~Yoon, \emph{Present and projected sensitivities of dark matter direct detection experiments to effective wimp-nucleus couplings}, \href{https://doi.org/https://doi.org/10.1016/j.astropartphys.2019.02.006}{\emph{Astroparticle Physics} {\bfseries 109} (2019) 50}.

\bibitem{Tovey200017}
D.~Tovey, R.~Gaitskell, P.~Gondolo, Y.~Ramachers and L.~Roszkowski, \emph{A new model-independent method for extracting spin-dependent cross section limits from dark matter searches}, \href{https://doi.org/https://doi.org/10.1016/S0370-2693(00)00846-7}{\emph{Physics Letters B} {\bfseries 488} (2000) 17}.

\bibitem{Ramani2019}
H.~Ramani and G.~Woolley, \emph{Spin-dependent light dark matter constraints from mediators},  \href{https://arxiv.org/abs/1905.04319}{{\ttfamily 1905.04319}}.

\bibitem{Wang:2021oha}
W.~Wang, K.-Y.~Wu, L.~Wu and B.~Zhu, \emph{Direct detection of spin-dependent sub-gev dark matter via migdal effect}, \href{https://doi.org/10.1016/j.nuclphysb.2022.115907}{\emph{Nucl. Phys. B} {\bfseries 983} (2022) 115907} [\href{https://arxiv.org/abs/2112.06492}{{\ttfamily 2112.06492}}].

\bibitem{PICASSO2017}
E.~Behnke, M.~Besnier, P.~Bhattacharjee, X.~Dai, M.~Das, A.~Davour et~al., \emph{Final results of the picasso dark matter search experiment}, \href{https://doi.org/https://doi.org/10.1016/j.astropartphys.2017.02.005}{\emph{Astroparticle Physics} {\bfseries 90} (2017) 85}.

\bibitem{AnandProc}
N.~Anand, A.L.~Fitzpatrick and W.C.~Haxton, \emph{Model-independent analyses of dark-matter particle interactions},  \href{https://arxiv.org/abs/arXiv:1405.6690 [nucl-th]}{{\ttfamily arXiv:1405.6690 [nucl-th]}}.

\bibitem{Fan:2010gt}
J.~Fan, M.~Reece and L.-T.~Wang, \emph{Non-relativistic effective theory of dark matter direct detection}, \href{https://doi.org/10.1088/1475-7516/2010/11/042}{\emph{JCAP} {\bfseries 11} (2010) 042} [\href{https://arxiv.org/abs/1008.1591}{{\ttfamily 1008.1591}}].

\bibitem{Catena_2015}
R.~Catena and B.~Schwabe, \emph{Form factors for dark matter capture by the sun in effective theories}, \href{https://doi.org/10.1088/1475-7516/2015/04/042}{\emph{Journal of Cosmology and Astroparticle Physics} {\bfseries 2015} (2015) 042}.

\bibitem{Trickle2022}
T.~Trickle, Z.~Zhang and K.M.~Zurek, \emph{Effective field theory of dark matter direct detection with collective excitations}, \href{https://doi.org/10.1103/PhysRevD.105.015001}{\emph{Physical Review D} {\bfseries 105} (2022) }.

\bibitem{Z_effective}
P.J.~Brown, A.G.~Fox, E.N.~Maslen, M.A.~O’Keefe and B.T.M.~Willis, \emph{Intensity of diffracted intensities}, {\emph{International Tables for Crystallography} (2006) }.

\bibitem{Gresham2014}
M.I.~Gresham and K.M.~Zurek, \emph{Effect of nuclear response functions in dark matter direct detection}, \href{https://doi.org/10.1103/PhysRevD.89.123521}{\emph{Phys. Rev. D} {\bfseries 89} (2014) 123521}.

\bibitem{XENON:2018voc}
{\scshape XENON} collaboration, \emph{Dark matter search results from a one ton-year exposure of xenon1t}, \href{https://doi.org/10.1103/PhysRevLett.121.111302}{\emph{Phys. Rev. Lett.} {\bfseries 121} (2018) 111302} [\href{https://arxiv.org/abs/1805.12562}{{\ttfamily 1805.12562}}].

\bibitem{CDMS2018}
{\scshape SuperCDMS Collaboration} collaboration, \emph{Low-mass dark matter search with cdmslite}, \href{https://doi.org/10.1103/PhysRevD.97.022002}{\emph{Phys. Rev. D} {\bfseries 97} (2018) 022002}.

\bibitem{SuperCDMS2018}
{\scshape SuperCDMS Collaboration} collaboration, \emph{Results from the super cryogenic dark matter search experiment at soudan}, \href{https://doi.org/10.1103/PhysRevLett.120.061802}{\emph{Phys. Rev. Lett.} {\bfseries 120} (2018) 061802}.

\bibitem{PICO2016}
{\scshape PICO Collaboration} collaboration, \emph{Dark matter search results from the pico-60 ${\mathrm{cf}}_{3}\mathrm{I}$ bubble chamber}, \href{https://doi.org/10.1103/PhysRevD.93.052014}{\emph{Phys. Rev. D} {\bfseries 93} (2016) 052014}.

\bibitem{PICO2019}
{\scshape PICO Collaboration} collaboration, \emph{Dark matter search results from the complete exposure of the pico-60 ${\mathrm{c}}_{3}{\mathrm{f}}_{8}$ bubble chamber}, \href{https://doi.org/10.1103/PhysRevD.100.022001}{\emph{Phys. Rev. D} {\bfseries 100} (2019) 022001}.

\bibitem{CRESST_II_2016}
G.~Angloher, A.~Bento, C.~Bucci, L.~Canonica, X.~Defay, A.~Erb et~al., \emph{Results on light dark matter particles with a low-threshold cresst-ii detector}, \href{https://doi.org/10.1140/epjc/s10052-016-3877-3}{\emph{The European Physical Journal C} {\bfseries 76} (2016) 25}.

\bibitem{CRESST_II_data}
G.~Angloher, P.~Bauer, A.~Bento, C.~Bucci, L.~Canonica, X.~Defay et~al., \emph{Description of cresst-ii data},  \href{https://arxiv.org/abs/1701.08157}{{\ttfamily 1701.08157}}.

\bibitem{DAMA2008}
R.~Bernabei, P.~Belli, A.~Bussolotti, F.~Cappella, R.~Cerulli, C.~Dai et~al., \emph{The dama/libra apparatus}, \href{https://doi.org/https://doi.org/10.1016/j.nima.2008.04.082}{\emph{Nuclear Instruments and Methods in Physics Research Section A: Accelerators, Spectrometers, Detectors and Associated Equipment} {\bfseries 592} (2008) 297}.

\bibitem{DarkSide2018}
{\scshape DarkSide Collaboration} collaboration, \emph{Low-mass dark matter search with the darkside-50 experiment}, \href{https://doi.org/10.1103/PhysRevLett.121.081307}{\emph{Phys. Rev. Lett.} {\bfseries 121} (2018) 081307}.

\bibitem{Bishara:2017pfq}
F.~Bishara, J.~Brod, B.~Grinstein and J.~Zupan, \emph{From quarks to nucleons in dark matter direct detection}, \href{https://doi.org/10.1007/JHEP11(2017)059}{\emph{JHEP} {\bfseries 11} (2017) 059} [\href{https://arxiv.org/abs/1707.06998}{{\ttfamily 1707.06998}}].

\bibitem{Emken:2019tni}
T.~Emken, R.~Essig, C.~Kouvaris and M.~Sholapurkar, \emph{Direct detection of strongly interacting sub-gev dark matter via electron recoils}, \href{https://doi.org/10.1088/1475-7516/2019/09/070}{\emph{JCAP} {\bfseries 09} (2019) 070} [\href{https://arxiv.org/abs/1905.06348}{{\ttfamily 1905.06348}}].

\bibitem{Chen2023}
Y.~Chen, B.~Fornal, P.~Sandick, J.~Shu, X.~Xue, Y.~Zhao et~al., \emph{Earth shielding and daily modulation from electrophilic boosted dark particles}, \href{https://doi.org/10.1103/PhysRevD.107.033006}{\emph{Phys. Rev. D} {\bfseries 107} (2023) 033006}.

\bibitem{GreshamArXiv}
M.I.~Gresham and K.M.~Zurek, \emph{Effect of nuclear response functions in dark matter direct detection}, \href{https://doi.org/10.1103/PhysRevD.89.123521}{\emph{Phys. Rev. D} {\bfseries 89} (2014) 123521}.

\bibitem{BSM_exchanges2017}
J.A.~Dror, R.~Lasenby and M.~Pospelov, \emph{Dark forces coupled to nonconserved currents}, \href{https://doi.org/10.1103/PhysRevD.96.075036}{\emph{Phys. Rev. D} {\bfseries 96} (2017) 075036}.

\bibitem{Gori:2025jzu}
S.~Gori, S.~Knapen, T.~Lin, P.~Munbodh and B.~Suter, \emph{Spin-dependent scattering of sub-gev dark matter: Models and constraints}, \href{https://doi.org/10.1103/1r9r-n4sv}{\emph{Phys. Rev. D} {\bfseries 112} (2025) 075019} [\href{https://arxiv.org/abs/2506.11191}{{\ttfamily 2506.11191}}].

\bibitem{EXCEED_DM}
T.~Trickle, \emph{Extended calculation of electronic excitations for direct detection of dark matter}, \href{https://doi.org/10.1103/PhysRevD.107.035035}{\emph{Phys. Rev. D} {\bfseries 107} (2023) 035035}.

\bibitem{Hochberg:2025dom}
Y.~Hochberg, D.~Novko, R.~Ovadia and A.~Politano, \emph{{Unconventional Materials for Light Dark Matter Detection}},  \href{https://arxiv.org/abs/2507.07164}{{\ttfamily 2507.07164}}.

\bibitem{Klos2013}
P.~Klos, J.~Men\'endez, D.~Gazit and A.~Schwenk, \emph{Large-scale nuclear structure calculations for spin-dependent wimp scattering with chiral effective field theory currents}, \href{https://doi.org/10.1103/PhysRevD.88.083516}{\emph{Phys. Rev. D} {\bfseries 88} (2013) 083516}.

\bibitem{Dimitrov1995}
V.I.~Dimitrov, J.~Engel and S.~Pittel, \emph{Scattering of weakly interacting massive particles from $^{73}\mathrm{Ge}$}, \href{https://doi.org/10.1103/PhysRevD.51.R291}{\emph{Phys. Rev. D} {\bfseries 51} (1995) R291}.

\bibitem{DarkElf_multiphonon}
B.~Campbell-Deem, S.~Knapen, T.~Lin and E.~Villarama, \emph{Dark matter direct detection from the single phonon to the nuclear recoil regime}, \href{https://doi.org/10.1103/PhysRevD.106.036019}{\emph{Phys. Rev. D} {\bfseries 106} (2022) 036019}.

\end{thebibliography}\endgroup

\end{document}